\documentclass[aps, pra, reprint, tightenlines, letterpaper, amsmath, amssymb, preprintnumbers, floatfix, longbibliography, nofootinbib,superscriptaddress]{revtex4-2}
\usepackage{tabularx}
\usepackage{dsfont}
\usepackage{extarrows}
\usepackage{amsmath}
\usepackage{amssymb}
\usepackage{physics}
\usepackage{tabu}
\usepackage{tabularx}
\usepackage{bm}
\usepackage{graphicx}
\usepackage{placeins}
\usepackage{textgreek}
\usepackage{multirow}
\usepackage{makecell}
\usepackage{soul}
\usepackage{diagbox}
\usepackage{multirow}
\usepackage{listings}

\usepackage[normalem]{ulem}

\makeatletter

\usepackage[hypertexnames=false]{hyperref}
\hypersetup{
    colorlinks=true,       
    linkcolor=blue,          
    citecolor=blue,        
    filecolor=blue,      
    urlcolor=blue           
}

\newcolumntype{Y}{>{\centering\arraybackslash}X}

\makeatletter
\pretocmd\frontmatter@thefootnote{\color{black}}{}{}
\makeatother

\usepackage{orcidlink}

\makeatletter  
\renewcommand\onecolumngrid{
\do@columngrid{one}{\@ne}%
\def\set@footnotewidth{\onecolumngrid}
\def\footnoterule{\kern-6pt\hrule width 1.5in\kern6pt}%
}
\makeatother    

\begin{document}

\begin{figure}
  \vskip -1.cm
  \leftline{\includegraphics[width=0.15\textwidth]{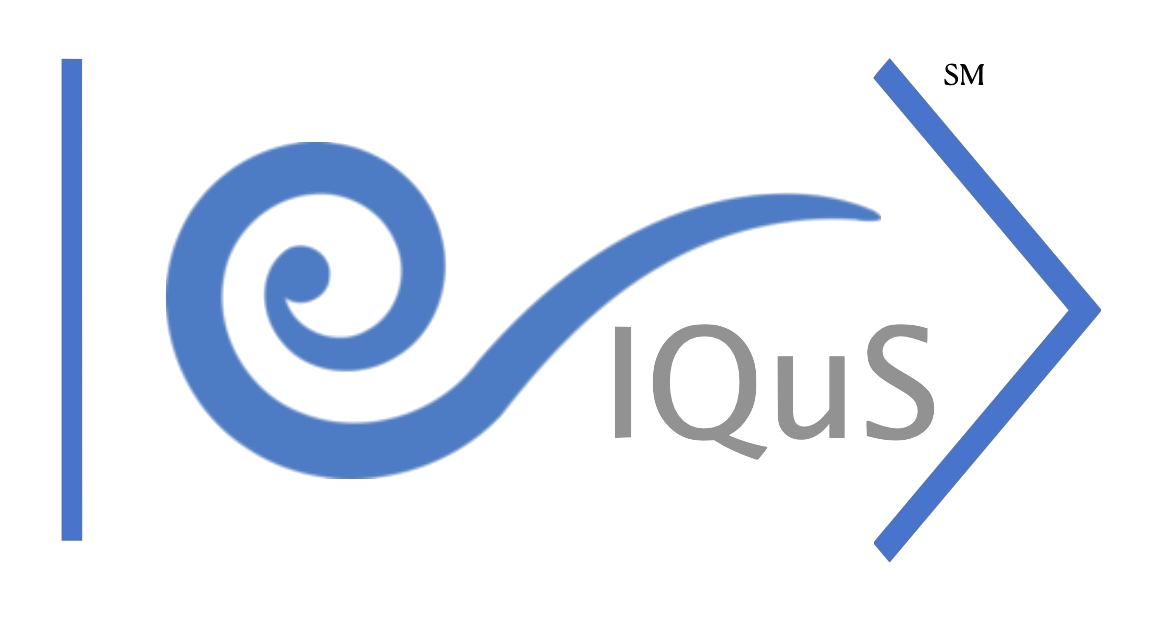}}
  \vskip -1.cm
\end{figure}

\title{Digital quantum simulations of scattering in quantum field theories using W states}

\author{Roland C. Farrell\,\orcidlink{0000-0001-7189-0424}}
\email{rolandf@caltech.edu}
\affiliation{Institute for Quantum Information and Matter, California Institute of Technology}
\affiliation{Department of Physics, California Institute of Technology}
\author{Nikita A. Zemlevskiy\,\orcidlink{0000-0002-0794-2389}}
\email{zemlni@uw.edu}
\affiliation{InQubator for Quantum Simulation (IQuS), Department of Physics,
University of Washington}
\author{Marc Illa\,\orcidlink{0000-0003-3570-2849}}
\email{marcilla@uw.edu}
\affiliation{InQubator for Quantum Simulation (IQuS), Department of Physics,
University of Washington}
\author{John Preskill\,\orcidlink{0000-0002-2421-4762}}
\email{preskill@caltech.edu}
\affiliation{Institute for Quantum Information and Matter, California Institute of Technology}
\affiliation{Department of Physics, California Institute of Technology}
\affiliation{AWS Center for Quantum Computing}

\preprint{IQuS@UW-21-099}
\date{\today}

\begin{abstract}
\noindent
High-energy particle collisions can convert energy into matter through the inelastic production of new particles.
Quantum computers are an ideal platform for simulating the out-of-equilibrium dynamics of collisions and the formation of subsequent many-particle states.
In this work, evidence for inelastic particle production is observed in one-dimensional Ising field theory using IBM's quantum computers.
The scattering experiment is performed on 104 qubits of {\tt ibm\_marrakesh} and uses up to 5,589 two-qubit gates to access the post-collision dynamics.
An outgoing heavy particle produced in the collision is identified from the skewness of the measured energy density. 
Integral to this computation is a new quantum algorithm for preparing the initial state (wavepackets) of a quantum field theory scattering simulation.
This method efficiently prepares wavepackets by extending recent protocols for creating W states with mid-circuit measurement and feedforward. 
The required circuit depth is independent of wavepacket size and spatial dimension, representing a superexponential improvement over previous methods.
Our wavepacket preparation algorithm can be applied to a wide range of lattice models and is demonstrated in one-dimensional Ising field theory, scalar field theory, the Schwinger model and two-dimensional Ising field theory.

\end{abstract}

\maketitle


\begingroup
\renewcommand*\addcontentsline[3]{}  
\section{Introduction}
\label{sec:intro}
\endgroup
\pagenumbering{arabic}
\setcounter{page}{1}
\noindent
Quantum computers promise to be powerful tools able to simulate the dynamics of strongly-coupled quantum many-body systems~\cite{Benioff1980,Feynman1982,Feynman1986,Lloyd1073}.
In many instances, these simulations are beyond the capabilities of even the most powerful supercomputers~\cite{Troyer_2005}.
Quantum simulations of Quantum Field Theories (QFTs) would enable first-principles studies of the behavior of matter present in extreme astrophysical environments, high-energy particle collisions and the early universe~\cite{Bauer:2022hpo,Catterall:2022wjq,Humble:2022klb,Banuls:2019bmf,Bauer:2023qgm,DiMeglio:2023nsa,Bauer:2025nzf}.
This pinnacle of {\it ab initio} scientific investigation, along with advances in quantum technologies, has motivated the first simulations of QFTs on quantum devices~\cite{Li:2024lrl,
Chai:2023qpq,
Cochran:2024rwe,
Schuster:2023klj,
Angelides:2023noe,
Guo:2024tnb,
Zhu:2024dvz,
Martinez:2016yna,
Kokail:2018eiw,
Meth:2023wzd,
Atas:2021ext,
ARahman:2021ktn,
Illa:2022jqb,
ARahman:2022tkr,
Atas:2022dqm,
Kavaki:2024ijd,
Than:2024zaj,
Lewis:2025xtu,
Nguyen:2021hyk,
Davoudi:2024wyv,
Mueller:2024mmk,
De:2024smi,
Klco:2018kyo,
Klco:2019evd,
Ciavarella:2021nmj,
Ciavarella:2021lel,
Ciavarella:2023mfc,
Turro:2024pxu,
Farrell:2022wyt,
Farrell:2022vyh,
Alexandrou:2025vaj,
Crippa:2024hso,
Klco:2019xro}.
Notably, quantum simulations of QFTs have been among the first to surpass the capabilities of exact statevector simulations~\cite{Farrell:2023fgd,Farrell:2024fit,Zemlevskiy:2024vxt,Ciavarella:2024lsp,Ciavarella:2024fzw,Gonzalez-Cuadra:2024xul,Gyawali:2024hrz,Yang:2020yer,Hayata:2024fnh}, and are strong candidates for a quantum advantage.

A fundamental process in QFTs is scattering, in which particles 
travel toward each other, collide, and eventually form an asymptotic state. 
The evolution of the quantum state throughout the collision encodes information about the interactions, particle content and mechanisms that govern dynamics in the theory.
At the ultra-relativistic energies accessed in particle collider experiments, kinetic energy is converted into matter through the production of many particles.
Predicting the abundances of particles produced in these processes, as well as their collective dynamics, is a forefront theoretical problem.
In this work, evidence for particle production is observed for the first time in a quantum simulation of inelastic scattering.
A key element of this simulation is a new algorithm for preparing the initial state, i.e., single-particle wavepackets. 
The algorithm has two steps:

\vspace{0.75em}
\begin{itemize}
    \item[1.] Prepare a state $| W(k_0)\rangle$ that establishes the spatial profile, momentum content and quantum numbers of the target wavepacket, but has contributions from multi-particle states.
    \item[2.] Project $|W(k_0)\rangle$ onto single-particle eigenstates by minimizing the energy with symmetry-preserving quantum circuits. 
    The energy minimum corresponds to the target wavepacket.
\end{itemize}

The initial state $|W(k_0)\rangle$ is similar to the W state~\cite{Dur:2000zz} that has been well studied in quantum information science due to its robust entanglement structure and applications to quantum communication, sensing and optimization~\cite{Agrawal2006,Wang2007,Liu2011,Li_2007,Joo_2003,Wang_2020,9259949,Catalano:2024bdh}.
Recent work has shown that W states can be prepared in constant circuit depth using mid-circuit measurement and feedforward (MCM-FF) ~\cite{Piroli:2024ckr,Buhrman:2023rft,Piroli:2021fjn,Yu:2024szp}.
Inspired by techniques from Refs.~\cite{Smith:2022nbd, Cruz_2019}, we improve the W state preparation protocol in Ref.~\cite{Piroli:2024ckr} by removing the need for ancillas. 
To the best of our knowledge, this is the first ancilla-free protocol for preparing W states in constant depth.
A straightforward generalization of this method is used to efficiently initialize $|W(k_0)\rangle$ in the first step of the wavepacket preparation algorithm.

The symmetry-preserving circuits that minimize the energy in step 2 are found using the variational algorithm ADAPT-VQE~\cite{Grimsley:2018wnd} running on classical computers.
In the future, ADAPT-VQE could be performed efficiently on a quantum computer.
Our wavepacket preparation algorithm can be applied to a wide range of lattice models.
Furthermore, the required circuit depth is independent of wavepacket size and only scales with the correlation length.
This is a significant improvement over previous methods, which either incur an exponentially-scaling classical computing overhead~\cite{Zemlevskiy:2024vxt,Farrell:2024fit} or require a circuit depth that scales polynomially with the wavepacket size~\cite{Davoudi:2024wyv,Chai:2023qpq,Jordan:2011ci,Hite:2025pvb,Turco:2023rmx,Turco:2025jot}.

We demonstrate the utility of this algorithm by constructing wavepacket preparation circuits in one-dimensional Ising field theory, scalar field theory, the Schwinger model, and in two-dimensional Ising field theory. 
In one-dimensional Ising field theory, the circuits that prepare wavepackets on lattices with 100+ qubits are determined using a Matrix Product State (MPS) simulator.
These wavepacket preparation circuits are then used to initialize a scattering simulation on 104 qubits of IBM's quantum computer {\tt ibm\_marrakesh}.
The application of up to 45 steps of Trotterized time evolution (5,589 two-qubit gates) evolves the system well beyond the collision.
The paths of the particles throughout the scattering process are identified from measurements of the energy density.
The creation of a slow-moving heavy particle during the collision causes the energy density in the post-collision state to be skewed.
The skewness is extracted from our quantum simulations and is evidence for inelastic particle production.

Sections \ref{sec:WPsummary},~\ref{sec:qsim} and~\ref{sec:discussion} constitute the main text of this paper and are designed to be read sequentially.
Section~\ref{sec:WPsummary} provides a high-level overview of the wavepacket preparation algorithm.
Section~\ref{sec:qsim} presents results from using {\tt ibm\_marrakesh} to simulate inelastic scattering in one-dimensional Ising field theory.
Section~\ref{sec:discussion} concludes and discusses prospects for a near-term quantum advantage in simulations of scattering.
The Methods section supports the results presented in the main text.
In particular, the constant-depth circuit that prepares $|W(k_0)\rangle$ is given in Methods~\ref{sec:WKprep}.
Circuits that prepare wavepackets in one-dimensional Ising field theory, scalar field theory, the Schwinger model, and two-dimensional Ising field theory are determined using an exact statevector simulator in Methods~\ref{sec:WPQFTcircs}.
In Methods~\ref{sec:qcirc_scatt}, circuits that prepare wavepackets on large lattices in one-dimensional Ising field theory are constructed using a MPS circuit simulator, and in Methods~\ref{sec:csimscatt} they are used to simulate scattering with MPS.
Details about the experiments performed on {\tt ibm\_marrakesh}, including the error mitigation techniques used, are given in Methods~\ref{sec:qsimDetails}.
Methods~\ref{sec:skew} shows how inelastic effects can be detected from the asymmetry in the outgoing energy density and explains our method for quantifying this with skewness.
The appendices provide additional information and expand on discussions in Methods.

\begingroup
\renewcommand*\addcontentsline[3]{}  
\section{Overview of the Wavepacket Preparation Algorithm}
\label{sec:WPsummary}
\endgroup
\noindent
A quantum simulation of particle scattering begins with the preparation of single-particle wavepackets.
Our method prepares wavepackets by minimizing the energy starting from a suitable initial state.
The circuits that minimize the energy could be determined from running variational algorithms on a quantum computer.
For one-dimensional systems these variational algorithms can instead be implemented using a MPS circuit simulator.
Importantly, the energy minimization circuits can be made hardware-efficient by adapting the ansatze to the connectivity and native gate set of the target quantum computer.
Therefore, this method is particularly advantageous for simulations on near-term quantum computers that are limited by circuit depth.
For simplicity, this section focuses on preparing wavepackets in one-dimensional lattice QFTs.

\begin{figure*}
    \centering
     \includegraphics[width=0.85\linewidth]{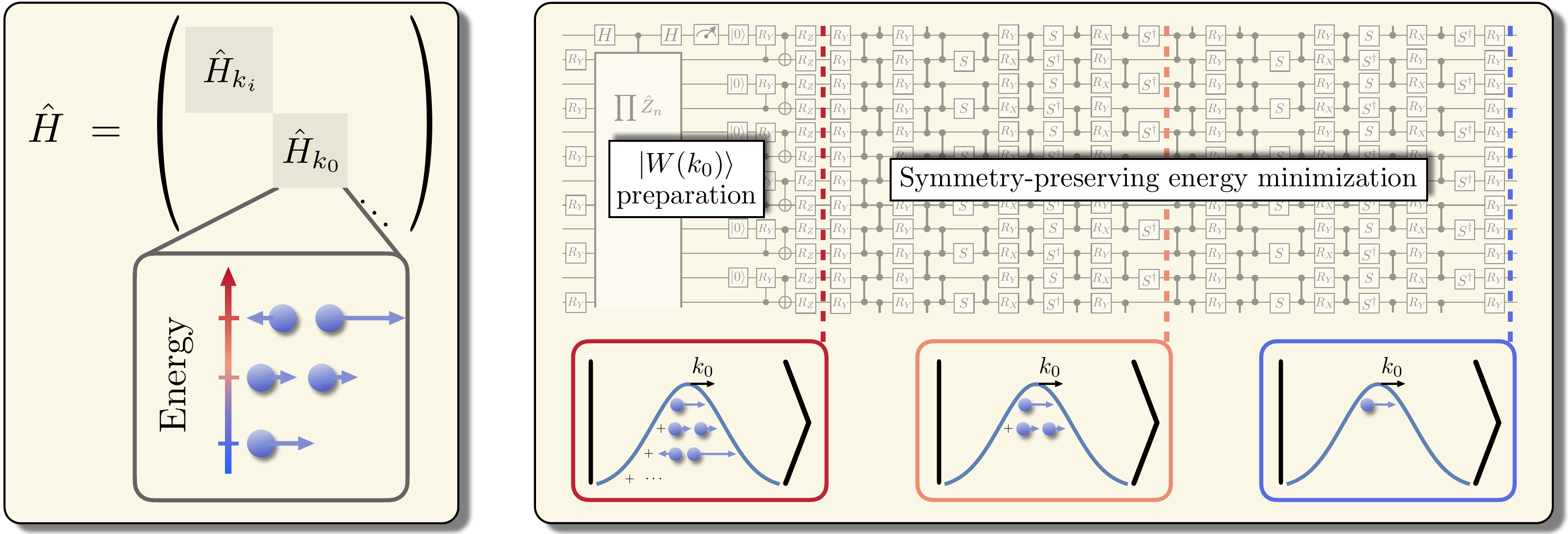 }
    \caption{\textit{Wavepacket preparation using W states and symmetry-preserving energy minimization.} Left: due to translational invariance, the Hamiltonian decomposes into blocks $\hat{H}_{k}$, each with definite momentum $k$.
    The lowest-energy state in each block corresponds to the single-particle momentum eigenstate $|\psi_k\rangle$.
    Higher-energy states can be interpreted as multiple particles with total momentum $k$ (or heavier single-particle states not shown).
    Right: quantum circuits that prepare the target wavepacket $|\psi_{\text{wp}}\rangle$.
    First, the state $|W(k_0)\rangle$ is prepared with a constant-depth circuit using MCM-FF (step 1).
    This initial state has the momentum content of the target wavepacket but incorrectly contains higher-energy components.
    Next, the energy is minimized using circuits that are translationally invariant, real, and conserve other system-specific symmetries (step 2).
    These constraints preserve the momentum content of $|W(k_0)\rangle$ while projecting the wavefunction onto the desired single-particle states.
    The energy minimum corresponds to $|\psi_{\text{wp}}\rangle$.}
\label{fig:WPOverview}
\end{figure*}
A wavepacket is a superposition of single-particle eigenstates that is localized around a position $x_0$ and a momentum $k_0$.
A wavepacket with a small spread in momentum space $\sigma_{}$ will have a large spread in position space $\sim\!\sigma_{}^{-1}$, and will propagate for a long time as a localized particle.
The wavepacket wavefunction is
\begin{equation}
\vert \psi_{\text{wp}} \rangle  \ = \ {\cal N}\sum_k e^{-i k x_0}\, e^{-(k_0 - k)^2/(4\sigma_{}^2)} \vert \psi_k \rangle \ ,
\label{eq:psiWPFull}
\end{equation}
where ${\cal N}$ is a normalization factor and $|\psi_k\rangle$ is the lightest single-particle eigenstate with momentum $k$.\footnote{A consistent phase convention is necessary when defining the single-particle eigenstates $|\psi_k\rangle$, see App.~\ref{app:ExactWP}.} 
The sum runs over $k \in 2 \pi n/L$, where $L$ is the number of lattice sites and $n$ is an integer such that $k\in (-\pi, \pi]$.
A wavepacket that propagates to the right (left) will have the bulk of its amplitude in eigenstates with $k>0$ ($k<0$).

Consider a translationally-invariant system with periodic boundary conditions (PBCs).
Its Hamiltonian $\hat{H}$ is block diagonal, with each block labeled by a momentum $k$.
If the Hamiltonian has a mass gap, the ground state and first excited state of the $k=0$ block correspond to the vacuum and the lightest particle at rest, respectively.
The other $|\psi_k\rangle$ are the lowest-energy states of the $k\neq0$ blocks.
The block-diagonal structure of the Hamiltonian is illustrated in the left panel of Fig.~\ref{fig:WPOverview}.
These properties of the Hamiltonian motivate a variational quantum algorithm for preparing $|\psi_{\text{wp}}\rangle$. 
As a warm up, first consider the task of preparing $|\psi_k\rangle$. 

The state $|\psi_k\rangle$ with $k\neq 0$ can be prepared in two steps.
First, some state $|k\rangle$ that is in a definite momentum block of the Hamiltonian (i.e., an eigenstate under spatial translations, $e^{-i \hat{k} n}|k\rangle = e^{-i k n}|k\rangle$) is initialized.
One simple choice for $|k\rangle$ is the superposition of all states related by translation to $|00...001\rangle$ weighted by the appropriate phase. 
This assumes a basis where $|00...001\rangle$ has the quantum numbers of the target single-particle state.\footnote{For example, in scalar field theory, single-particle states are odd under the $Z_2$ symmetry that takes $\phi\to-\phi$.}
If one lattice site maps to one qubit, this state would be
\begin{align}
|k\rangle \ = \ \frac{1}{\sqrt{L}}\sum_{n=0}^{L-1} e^{i k n}|2^n\rangle \ ,
\label{eq:psik0}
\end{align}
where the state is specified by its binary value on $L$ bits, e.g., $|2^1\rangle = |00...010\rangle$.
This state is in the correct $k$ block of the Hamiltonian and has the long-range entanglement inherent to momentum eigenstates~\cite{Gioia:2021xtp}, but needs to be rotated to the lowest-energy state.
A circuit that approximately implements this rotation $|\psi_k\rangle \approx \hat{U}(\vec{\theta}_{\star}) |k\rangle$ can be determined by minimizing the energy,
\begin{align}
\hat{U}(\vec{\theta}_{\star}) \ = \ \text{argmin} \langle \psi_{\text{ansatz}} |\hat{U}(\vec{\theta})^{\dagger}\hat{H} \hat{U}(\vec{\theta})|\psi_{\text{ansatz}}\rangle \ ,
\label{eq:UThetastar}
\end{align}
where $|\psi_{\text{ansatz}}\rangle = |k\rangle$ and
$\hat{U}(\vec{\theta})$ is a translationally-invariant unitary operator $e^{i \hat{k} n} \hat{U}(\vec{\theta}) e^{-i \hat{k} n} = \hat{U}(\vec{\theta})$ that depends on variational parameters $\vec{\theta}$.
This circuit must also preserve the other quantum numbers of $|\psi_k\rangle$, and translational invariance ensures that the resulting state remains in the correct $k$ block of the Hamiltonian.
This forms the basis for a variational algorithm, and circuits that prepare $|k\rangle$ and minimize the energy can readily be constructed.

The generalization of this strategy to prepare $|\psi_{\text{wp}}\rangle$ begins with the initialization of some state $|W(k_0)\rangle$, that has the correct amplitude and phase in each $k$ block of the Hamiltonian.
A simple choice is a wavepacket built from $|k\rangle$,
\begin{align}
\vert W(k_0) \rangle  \ &= \ {\cal N}\sum_k e^{-i k x_0}\, e^{-(k_0 - k)^2/(4\sigma_{}^2)} \vert k \rangle \ \nonumber \\ &=\ \sum_n e^{i\phi_n}c_n |2^n\rangle \ , 
\label{eq:psiWP0}
\end{align}
where the magnitude of $c_n$ follows a Gaussian centered at $n=x_0$.
The special case of a uniform superposition is the W state~\cite{Dur:2000zz}.\footnote{The use of W states for preparing wavepackets in analog quantum simulations was recently proposed in Ref.~\cite{Bennewitz:2024ixi}.}
Using techniques from Refs.~\cite{Cruz_2019,Smith:2022nbd} we improve the W state preparation circuits in Ref.~\cite{Piroli:2024ckr}, and generalize them to prepare $\vert W(k_0) \rangle$.
The circuit is given in Methods~\ref{sec:WKprep} and utilizes MCM-FF to prepare $\vert W(k_0) \rangle$ in two-qubit gate depth 7. 
The depth is independent of lattice geometry, system size and target wavepacket size. 

A simplified description of the circuit is the following:
first, single-qubit rotations produce the state $|000...\rangle  + \sqrt{\delta}|W(k_0)\rangle \ + \ \delta|X\rangle $, where $|X\rangle$ contains states with two or more $1$s in their binary representations, and $\delta$ is an input parameter.
Next, the parity $\prod \hat{Z}_n$ (and only the parity) is measured using the constant-depth circuit from Ref.~\cite{Piroli:2024ckr}. 
Post-selecting on odd parity prepares $|W(k_0)\rangle$ with infidelity ${\cal I} = {\cal O}(\delta^2)$ and succeeds with probability $p_{\text{success}} \geq 0.43\delta$ (assuming $\delta\le 1$).
For large wavepackets, the success probability and infidelity are independent of wavepacket size (see Methods~\ref{sec:WKprep}).
The state $|W(k_0)\rangle$ has the momentum content of the target wavepacket but incorrectly contains components with two or more particles, as well as heavier single-particle components.
This is illustrated by the first step in the right panel of Fig.~\ref{fig:WPOverview}.

After preparing $|W(k_0)\rangle$, the wavefunction in each $k$ block is rotated to the lowest-energy state while preserving its amplitude ($e^{-(k_0 - k)^2/(4\sigma_{}^2)}$) and phase ($e^{-i k x_0}$).
The circuit that approximately performs this rotation $|\psi_{\text{wp}}\rangle \approx \hat{U}(\vec{\theta}_{\star}) |W(k_0)\rangle$ is again determined from minimizing the energy in Eq.~\eqref{eq:UThetastar}.
Now, $|\psi_{\text{ansatz}}\rangle = |W(k_0)\rangle$, and $\hat{U}(\vec{\theta})$ is translationally invariant, real, and conserves the other quantum numbers of $|W(k_0)\rangle$. 
Translational invariance preserves the amplitude of the initial state in each $k$ block and reality preserves the phase.
At a practical level, $\hat{U}(\vec{\theta})$ can be constructed from translationally-invariant and imaginary operators $\hat{O}_j$ as $\hat{U}(\vec{\theta})= \prod_j e^{i \theta_j \hat{O}_j}$.
Each of the terms $e^{i \theta_j \hat{O}_j}$ is then converted to a quantum circuit via Trotterization.
Operators $\hat{O}_j$ generated from the Lie algebra of the Hamiltonian are particularly effective at minimizing the energy and guide the circuit design in this work.
The energy minimization effectively projects out all unwanted multi-particle and heavy single-particle components of the $|W(k_0)\rangle$ wavefunction, leaving only the target single-particle wavepacket $|\psi_{\text{wp}}\rangle$.
This is illustrated by the second step in the right panel of Fig.~\ref{fig:WPOverview}.

This wavepacket preparation method can be applied to a wide range of lattice models in any finite spatial dimension.
The required circuit depth is independent of wavepacket volume, provided that the qubit connectivity matches that of the target lattice.
In the next section, this algorithm is used to initialize simulations of scattering in one-dimensional Ising field theory on IBM's quantum computers.
In Methods~\ref{sec:WPQFTcircs}, it is applied to the preparation of wavepackets in one-dimensional scalar field theory, the Schwinger model, and two-dimensional Ising field theory.

\vspace{2em}

\begingroup
\renewcommand*\addcontentsline[3]{}  
\section{Quantum Simulations of Scattering in One-Dimensional Ising Field Theory}
\label{sec:qsim}
\endgroup
\noindent
The physics of the Ising model illustrates the complexity of quantum phenomena that can emerge from simple interactions.
This has made it a target for quantum simulations, with recent demonstrations of
many-body localization~\cite{Shtanko:2023tjn}, discrete time crystals~\cite{Shinjo:2024vci}, Majorana edge modes~\cite{Mi:2022egw} and string breaking~\cite{De:2024smi}. 
Our work focuses on the field theory that emerges when the Ising model Hamiltonian,
\begin{align}
\hat{H}  \ &=\  -  \sum_{n=0}^{L-1}\left [ \frac{1}{2}\left (\hat{Z}_{n-1}\hat{Z}_{n}+\hat{Z}_n\hat{Z}_{n+1}\right ) +  g_x \hat{X}_n + g_z\hat{Z}_n \right ] \nonumber \\
&\equiv \ \sum_{n=0}^{L-1}\hat{H}_n \ , \label{eq:HIFT}
\end{align}
is tuned to criticality: $g_x\to1,\,g_z\to0$ and $L\to \infty$.
The energy density $\hat{H}_n$ will be an important observable and is defined in the second line.
The physics at the critical point is described by the conformal field theory (CFT) of a free, massless Majorana fermion.
A family of Ising QFTs are reached by tuning to criticality, but keeping the scaling invariant ratio
\begin{equation}
\eta_{\text{latt}} \ = \ \frac{g_x-1}{\vert g_z\vert^{\frac{D-\Delta_{\epsilon}}{D-\Delta_\sigma}}} \ = \ \frac{g_x-1}{\vert g_z\vert^{8/15}}
\end{equation}
held fixed.
The second equality uses the spacetime dimension $D=2$ and scaling dimensions of the relevant CFT deformations $\Delta_{\epsilon} = 1$ and $\Delta_{\sigma} = 1/8$.
For generic values of $\eta_{\text{latt}}$, the field theory describes massive interacting particles and is non-integrable.
The lattice spacing is set to 1 throughout this work and all lengths are expressed in lattice units. 
The Hamiltonian $\hat H$ is dimensionless; hence time is also dimensionless, and the operator $e^{-i\hat H t}$ describes evolution for dimensionless time $t$.
More information on Ising field theory can be found in Refs.~\cite{Zamolodchikov:1989hfa,Zamolodchikov:1989fp,Jha:2024jan}.

Dynamical simulations of scattering in Ising field theory were recently performed using MPS methods in Ref.~\cite{Jha:2024jan}.
By changing the energy of the initial state, clear distinctions between elastic scattering, scattering near a resonance, and particle production via inelastic scattering were observed.
This work showcased the wealth of information that can be accessed with real-time simulations of collisions.

\begin{figure}
    \centering
    \includegraphics[width=\linewidth]{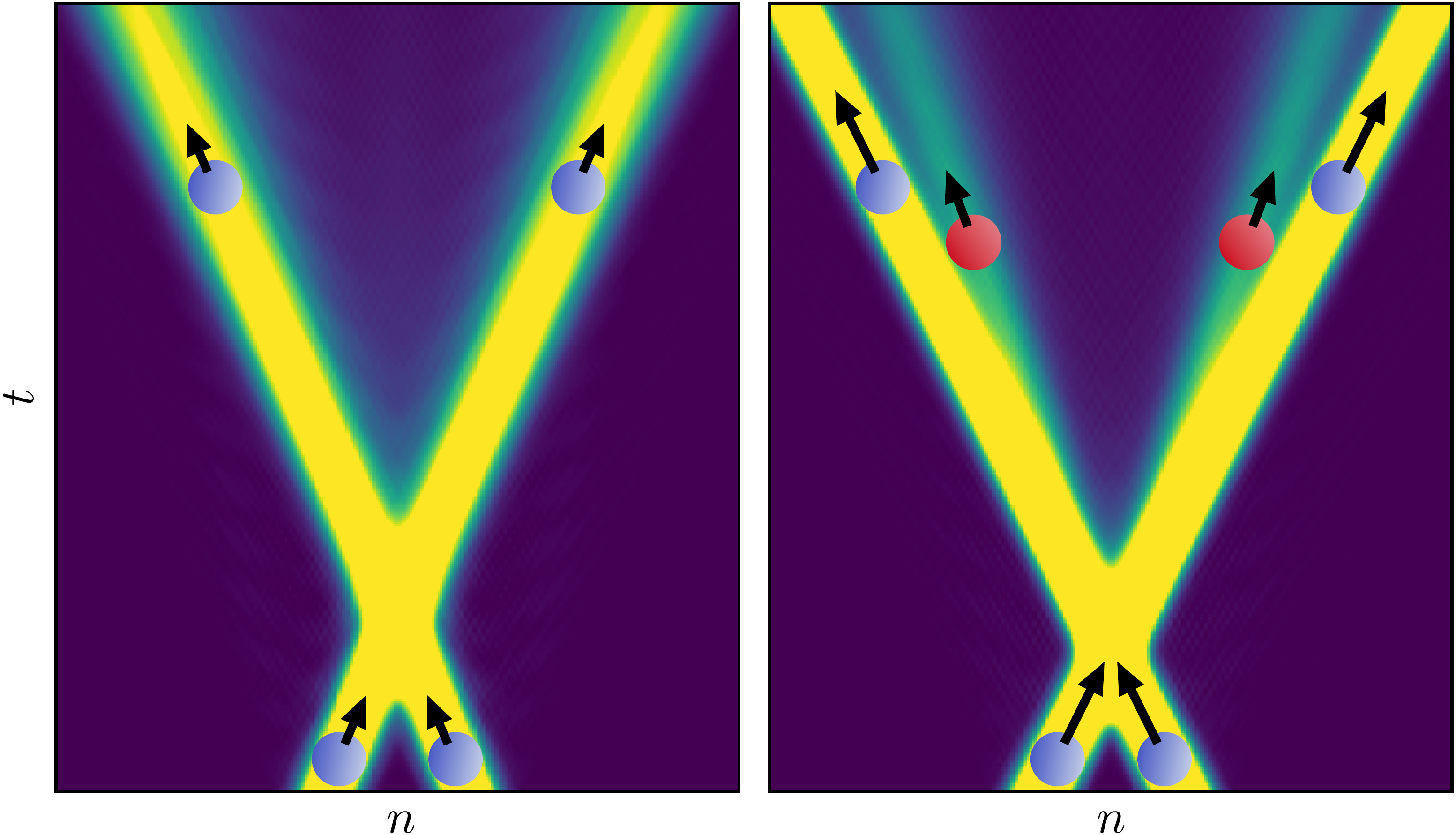}
    \caption{\textit{Low-energy vs.\ high-energy collisions.} A heatmap of the energy density throughout the scattering process as a function of lattice position $n$ and time $t$.
    Propagating particles are identified as beams of energy with a constant velocity.
    Left: the elastic scattering of two light $|1\rangle$ particles (blue), $11\to11$. Right: the inelastic process $11\to12$ that produces a heavy $|2\rangle$ particle (red).}
\label{fig:elastic_vs_inelastic}
\end{figure}

Our quantum simulations use $g_x=1.25$ and $g_z=0.15$ where there are two stable particles: one light particle $|1\rangle$ with mass $m_1=1.59$ and one heavy particle $|2\rangle$ with mass $m_2=2.98$.
The initial state for our scattering simulations consists of two wavepackets of $|1\rangle$ particles that are separated in space.
As time goes on, they travel toward each other, collide, and separate. 
For low center-of-mass energies $E_{\text{tot}}$, only the elastic process $11\to 11$ shown in the left panel of Fig.~\ref{fig:elastic_vs_inelastic} is kinematically allowed. 
The lowest-energy inelastic process is $11\to 12$ where a single heavy $|2\rangle$ particle is produced during the collision. The observation of this process is the goal of our quantum simulations.
This inelastic channel opens up for energies above the particle production threshold, $E_{\text{tot}}>E_{\text{thr}}=m_1+m_2 = 4.57$, and can be detected from additional outgoing tracks in the energy density.
This is seen in the right panel of Fig.~\ref{fig:elastic_vs_inelastic}, which depicts the production of the heavy $|2\rangle$ particle. In the kinematic regime illustrated by the right panel, three different scattering outcomes occur in superposition: elastic scattering, inelastic scattering with the heavier particle moving right, and inelastic scattering with the heavier particle moving left. Hence four outgoing tracks are visible, even though there are only two outgoing particles for each outcome. 

For energies $E_{\text{tot}}\gg E_{\text{thr}}$ there are additional inelastic processes that produce many particles.
Such processes generate significant entanglement and are challenging to simulate with MPS methods.
Simulations of these ultra-high-energy collisions are also beyond the capabilities of current quantum hardware.
At the energies accessed in our quantum simulations all inelastic channels besides $11\to12$ can be neglected.\footnote{The quantum simulations in this work are at $E_{\text{tot}}=3.46m_1$; hence the three-body channel $11\to 111$ is kinematically allowed, but has a much smaller branching ratio than the $11\to 12$ channel~\cite{Jha:2024jan}.} 
See App.~\ref{app:kinematics} for details.

\vspace{1em}

\begin{figure}
    \centering
    \includegraphics[width=\linewidth]{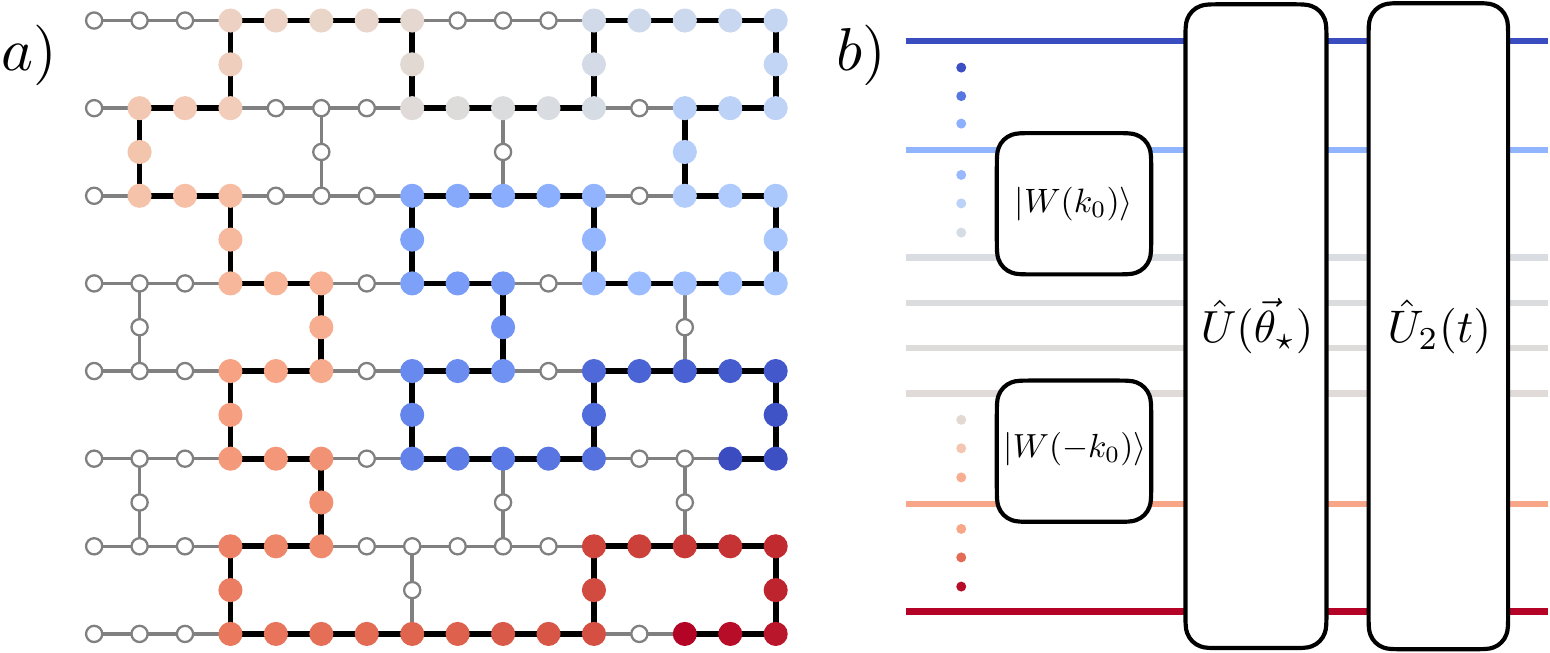}
    \caption{a) The $L=104$ qubit layout used on {\tt ibm\_marrakesh}. 
    b) The structure of the quantum circuits used to simulate scattering in Ising field theory. The colors indicate the qubits used in the lattice-to-device mapping.
    The energy minimization circuit $\hat{U}(\vec{\theta}_*)$ creates wavepackets when acting on $|W(\pm k_0)\rangle$.
    The circuit $\hat{U}_2(t)$ implements time evolution with second-order Trotterization.}
\label{fig:marrakesh_layout}
\end{figure}

\begingroup
\renewcommand*\addcontentsline[3]{}  
\subsection{Inelastic scattering on IBM's quantum computers}
\endgroup
\noindent 
In this section, inelastic scattering is simulated on $L=104$ qubits of {\tt ibm\_marrakesh} using the lattice-to-qubit mapping shown in Fig.~\ref{fig:marrakesh_layout}a).
A lattice with open boundary conditions (OBCs) is chosen to avoid noisy gates.
This is discussed in Methods~\ref{sec:qsimDetails}, along with the modifications to the simulation protocols that are necessary to accommodate OBCs.
The first step of the wavepacket preparation algorithm outlined in Sec.~\ref{sec:WPsummary} is the initialization of $|W(k_0)\rangle$.
Methods~\ref{sec:WKprep} shows that this can be achieved with a constant-depth circuit using MCM-FF.
This approach is currently not optimal on IBM's {\tt heron r2} quantum computers,\footnote{This is in part due to readout being $\sim\!3\times$ noisier and $\sim \!38\times$ slower than two-qubit gates (2584ns vs 68ns), and also the current incompatibility of MCM-FF with dynamical decoupling and native $R_{ZZ}(\theta)$-gates. Improvements are expected in future versions of the hardware and software stack.}
and instead $|W(k_0)\rangle$ is initialized with the unitary circuit in the left panel of Fig.~\ref{fig:IsingWPCircs}.
Wavepackets are then prepared by applying symmetry-preserving circuits $\hat{U}(\vec{\theta}_{\star})$ that minimize the energy.
These circuits are determined using ADAPT-VQE running on the MPS circuit simulator Quimb~\cite{gray2018quimb}.
Information on wavepacket preparation is given in Methods~\ref{sec:IsingStatevector} and Methods~\ref{sec:qcirc_scatt}.
The structure of the circuit that prepares two wavepackets is shown in Fig.~\ref{fig:marrakesh_layout}b).
The initial state is chosen so the wavepackets have support on $d\gtrsim21$ sites, and the total energy is $E_{\text{tot}}/E_{\text{thr}}=1.2$. These parameters give the strongest signal of inelastic scattering in the least circuit depth (see Methods~\ref{sec:qsimDetails}).
\begin{figure*}
    \centering
    \includegraphics[width=\linewidth]{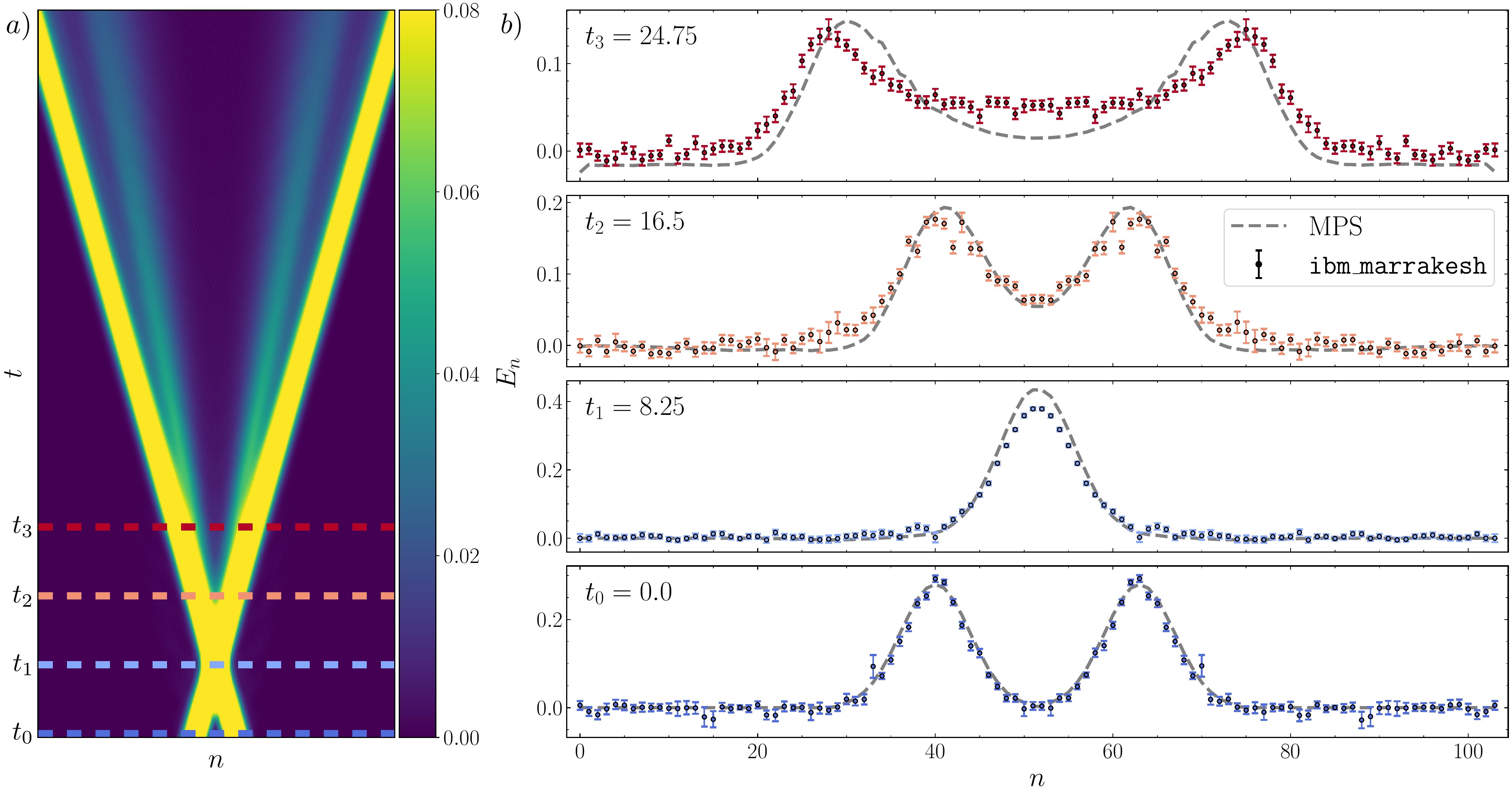}
    \caption{\textit{Simulations of inelastic particle production in one-dimensional Ising field theory.} a) The energy density $E_n$ throughout the scattering process obtained with a MPS circuit simulator on a $L=256$ lattice. 
    b) Results from simulations of scattering using $L=104$ qubits of {\tt ibm\_marrakesh} for a selection of times depicted in a).
    The wavepackets are initialized at $t_0=0$ and collide around $t_1=8.25$.
    Outgoing particles begin to form at $t_2=16.5$ and $t_3=24.75$. The asymmetry of the energy density in each ``bump" signals the formation of the $|2\rangle$ particle produced in the inelastic process $11\to12$.
    The y-axis range at each time is different to clearly show the features of the energy density.
    The initial wavepacket parameters are $k_0=0.32\pi$ and $\sigma_{}=0.13$, and each wavepacket is supported on $d \gtrsim 21$ sites.
    A Trotter step size of $\delta t=1/16$ is used to evolve the system in a) and $\delta t=0.55$ is used in b).}
\label{fig:ibm_results}
\end{figure*}
Once established, the wavepackets are time evolved with a second-order Trotterized unitary $\hat{U}_2(t)$.
A Trotter step size of $\delta t=0.55$ is chosen to balance the Trotter error and circuit depth (see App.~\ref{app:systematics}).
For each simulation time, the vacuum-subtracted energy density
\begin{align}
E_n \ = \ \langle\psi_{\text{2wp}}|\hat{H}_n(t) |\psi_{\text{2wp}}\rangle \ - \ \langle\psi_{\text{vac}}|\hat{H}_n(t) |\psi_{\text{vac}}\rangle 
\label{eq:vacsubEn}
\end{align}
is measured. 
Here, $\hat{H}_n(t) = \hat{U}_2(t)^{\dagger}\hat{H}_n \hat{U}_2(t)$ with $\hat{H}_n$ defined in Eq.~\eqref{eq:HIFT}, $|\psi_{\text{2wp}}\rangle$ is the prepared two-wavepacket state and $|\psi_{\text{vac}}\rangle$ is the prepared vacuum.
The circuits that prepare $|\psi_{\text{2wp}}\rangle$ and $|\psi_{\text{vac}}\rangle$ have very similar structure and are explained in Methods~\ref{sec:qsimDetails}.

MPS calculations of the energy density throughout the scattering process are shown in Fig.~\ref{fig:ibm_results}a).\footnote{MPS simulations are performed using CuPy \cite{cupy} for GPU acceleration.} 
Two pairs of particle tracks emerge from the collision region.
The majority of the energy is in the outer pair of tracks, which correspond to the trajectories of the light $|1\rangle$ particles.
The inner pair of tracks has a lower velocity compared to the elastic trajectories, and represent the heavy $|2\rangle$ particles produced in the inelastic process $11\to12$.
Schematically, this process is
\begin{align}
|1(k_0) \,  1(-k_0)\rangle \ \to \ |1(k_0')\,  2(-k_0')\rangle\ + \ 1\leftrightarrow2 \ ,
\label{eq:inelastic_scattering_wavefunction}
\end{align}
where $k_0'$ is the momentum of outgoing particles.
The kinematics are such that the trajectories of the outgoing $|1\rangle$ particles overlap with the elastic tracks and cannot be distinguished in Fig.~\ref{fig:ibm_results}a) (see Methods~\ref{sec:csimscatt} for details).
It is shown in App.~\ref{app:kinematics} that the identification of the outgoing tracks with $|1\rangle$ and $|2\rangle$ particles is consistent with their velocities predicted from the single-particle dispersion relations.
We emphasize that the elastic and inelastic process occur in superposition, and there are no components of the wavefunction with four particles.

The energy density measured on {\tt ibm\_marrakesh} is shown for a selection of simulation times in Fig.~\ref{fig:ibm_results}b).
Two distinct wavepackets initialized at $t_0=0$ propagate toward each other and collide around $t_1=8.25$.
Later simulation times probe the formation and evolution of the post-collision state.
At $t_2=16.5$ the outgoing wavepackets begin to separate, and by $t_3=24.75$ asymptotic particles begin to form.
Up to $t_2=16.5$, there is good agreement between the energy density determined from {\tt ibm\_marrakesh} (data) and MPS (gray dashed line).
Significant systematic errors develop in the quantum data at $t_3=24.75$ due to the larger circuit volume.
For $t>t_3$, the asymptotic $|2\rangle$ particle produced in the $11\to12$ process becomes identifiable as a separate ``bump" in the energy density that propagates at a lower velocity.
Unfortunately, the cumulative effects of noise made simulations beyond $t=24.75$ impossible on {\tt ibm\_marrakesh}.

\begin{figure}
    \centering
    \includegraphics[width=\linewidth]{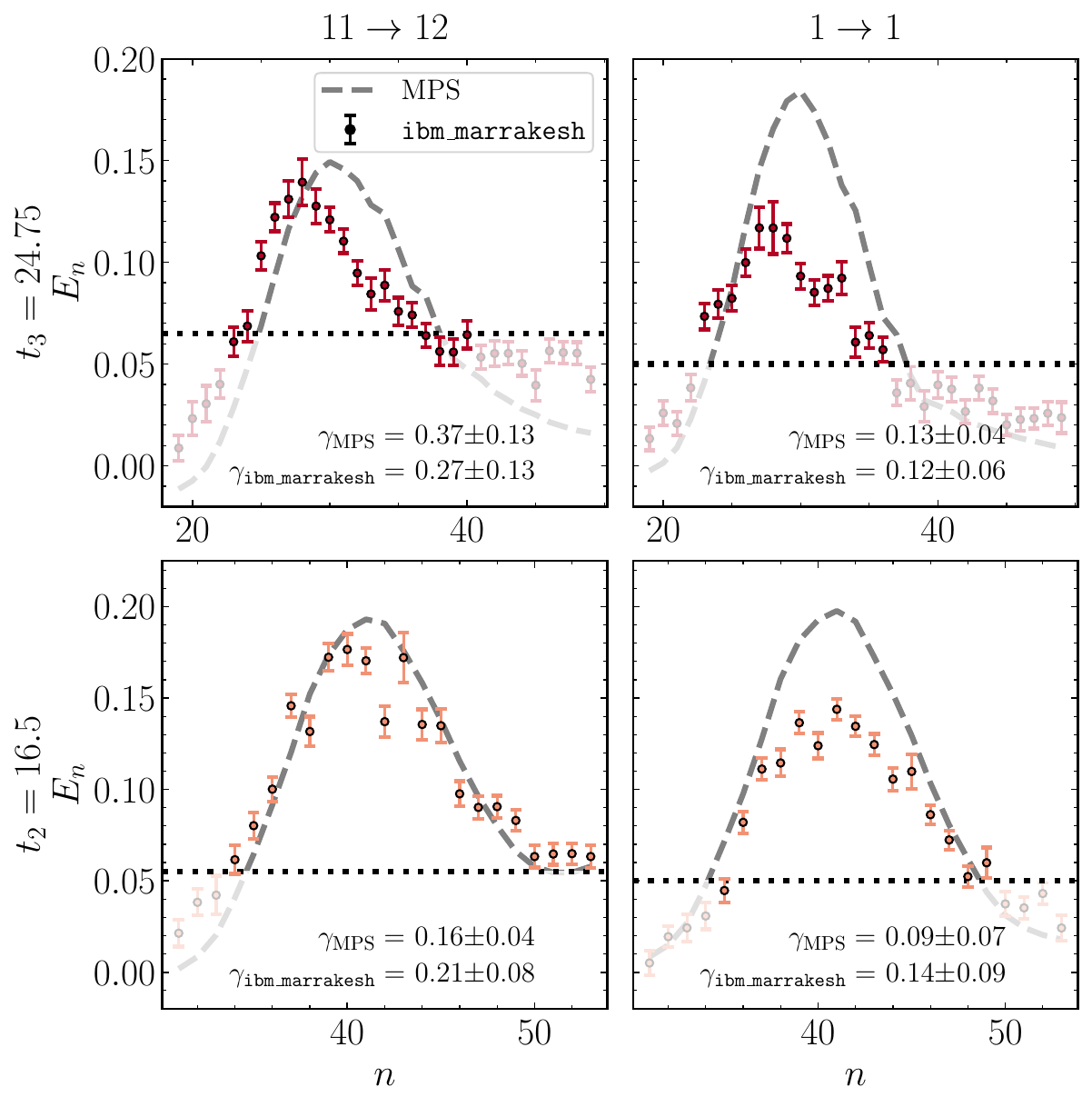}
    \caption{\textit{Asymmetry in the post-collision and single-particle energy densities.} 
    The regions of positive energy density the left half of the lattice in the $11\to12$ scattering process (left column) and in the $1\to1$ process of single particle propagation (right column).
    Results at times $t_2$ and $t_3$ obtained from {\tt ibm\_marrakesh} and MPS are shown.
    The skewness of the energy density, $\gamma$, is computed by considering points in an interval determined by an energy cutoff (black dotted line).
    The uncertainty in $\gamma$ comes from varying the energy cutoff as described in Methods~\ref{sec:skew}, and from statistical error.
    The skewness is increased in the $11\to12$ energy density compared to $1\to1$ due to the inelastic production of the heavy $|2\rangle$ particle.
    The simulation parameters are the same as in Fig.~\ref{fig:ibm_results}b).}
\label{fig:1wp_vs_2wp}
\end{figure}

Despite the device noise, we observe evidence for the inelastic production of the heavy $|2\rangle$ particle.
After the collision, the energy density of the outgoing particles on each side of the collision becomes skewed toward the center of the lattice.
This is due to the presence of the $|2\rangle$ particle in the wavefunction that travels slower, and is absent for scattering below inelastic threshold, as shown in Methods~\ref{sec:skew}.
We compute the skewness, $\gamma$, from the third moment of the energy density of each region of outgoing particles as described in Methods~\ref{sec:skew}.
The left column of Fig.~\ref{fig:1wp_vs_2wp} shows the energy density of the bumps after the collision, and gives $\gamma$ obtained from both MPS and {\tt ibm\_marrakesh}.
The quantum results of the post-collision state at $t_2$ are right-skewed ($\gamma>0$) and in agreement with MPS.
The right-skewness increases at $t_3$ due to the emergence of the heavy $|2\rangle$ particle.
Some of this skewness is due to systematic errors in the quantum data that increase $E_n$ between the outgoing particles.
To isolate skewness due to inelastic effects from systematic errors, additional simulations of single wavepacket propagation are performed.
The energy density of a single wavepacket at times $t_2$ and $t_3$ are shown in the right column of Fig.~\ref{fig:1wp_vs_2wp}.
The MPS simulations of single wavepacket propagation still exhibit positive skewness due to boundary effects as discussed in Methods~\ref{sec:skew}.
The skewness of the post-collision state is $0.6\sigma$ larger than the single-particle state at $t_2$, and $1\sigma$ larger at $t_3$.
This provides evidence that the post-collision skewness is, at least partially, due to inelastic particle production.

The quantum resources used for these simulations are detailed in Methods~\ref{sec:qsimDetails}.
The latest simulation time corresponds to $n_T=45$ Trotter steps and requires 5,589 two-qubit gates with a two-qubit gate depth of 130. 
With such a large quantum volume, effective error mitigation is crucial for recovering reliable results.
A full description of our error mitigation strategy is provided in Methods~\ref{sec:qsimDetails}, with certain aspects highlighted here.
Pauli twirling~\cite{Wallman:2015uzh} shapes the noise into a stochastic Pauli channel.
For each simulation time, circuits that evolve both $|\psi_{\text{2wp}}\rangle$ and $|\psi_{\text{vac}}\rangle $ are run.
The time evolution of the vacuum is used to learn how the Pauli noise channel affects local observables.
Noise-free expectation values in the scattering simulations are then estimated using the learned noise model.
This method, known as Operator Decoherence Renormalization (ODR)~\cite{Farrell:2023fgd}, is an extension of Refs.~\cite{Urbanek:2021oej,ARahman:2022tkr}.

Two unforeseen features of the device noise were encountered during the simulations.
First, the Pauli noise channel is very asymmetric, and $\langle\hat{Z}_n\rangle$ measurements are effectively $\sim\!3\times$ noisier than $\langle\hat{X}_n\rangle$ and $\langle\hat{Z}_n\hat{Z}_{n+1}\rangle$ measurements (see Methods~\ref{sec:qsimDetails} for details on device noise characteristics).
The source of this asymmetry is a mystery, but could be due to incomplete twirling of the (non-Clifford) $R_{ZZ}$ gate.
Additionally, we observed that one faulty single- or two-qubit gate could degrade the whole simulation.
This was not noticed in previous simulations of wavepacket dynamics on IBM's quantum computers performed by the authors that utilized similar circuit depths~\cite{Zemlevskiy:2024vxt,Farrell:2024fit}.
We attribute this increased sensitivity to the larger simulation light cone of $c t_{\text{max}}\sim\!40$ qubits reached in this work, compared to $c t_{\text{max}}\sim\! 8$ qubits in our previous works.
This effectively allows the noise to contaminate a larger region of the device in the same circuit depth.
To reduce these effects, IBM's calibration data was used to carefully avoid high-error gates in the lattice-to-qubit mapping.
The simulation light-cone also explains the increase in systematic errors at $t_3$ compared to $t_2$, despite the circuit depth only increasing by 30. 
This is because the simulation light-cone at $t_3$ encapsulates approximately 2.5$\times$ more two-qubit gates than at $t_2$.
These observations highlight that the effects of circuit noise on the results of a quantum simulation depend heavily on the physical process being simulated.
\vspace{2em}

\begingroup
\renewcommand*\addcontentsline[3]{}  
\section{Discussion}
\label{sec:discussion}
\endgroup
\noindent
The quantum simulations in this work are the first to probe the non-equilibrium QFT dynamics generated in the wake of particle collisions.
This is a significant improvement over previous quantum simulations of scattering that were constrained to early times and low energies.
Advancing beyond this regime required sophisticated strategies for managing device errors that are amplified by a large simulation light cone $ct_\text{max}$.
Algorithmic errors and quantum resource requirements must be balanced in an optimal simulation that operates within the correct regime of length scales.
Our simulations of Ising field theory used: system size $L=104$, particle propagation distance $vt_{\text{max}}\!\sim40$, wavepacket size $d\!\sim\!21$ and correlation length $\xi\propto1/m\sim0.6$ (where $m$ is the mass of the lightest particle).
Even with these minimal parameters, observing evidence of inelastic particle production required the full capability of state-of-the-art quantum hardware.

Our simulations are far from the continuum limit of $\xi\to\infty$ (in lattice units),
and are not designed to make precise predictions about the underlying QFT.
Instead, our work demonstrates that simulations far from the continuum can still be informative and capture features of non-equilibrium processes in QFTs.
The use of quantum computers to explore quantum many-body dynamics, without the goal of making precise quantitative predictions, aligns with the capabilities expected in the pre-fault-tolerant era. 
Precision simulations of scattering are possible in one-dimensional QFTs using MPS techniques~\cite{Pichler:2015yqa,Van_Damme_2021,Rigobello:2021fxw,Vovrosh:2022bpj,Belyansky:2023rgh,Papaefstathiou:2024zsu,Milsted:2020jmf,Su:2024uuc,Jha:2024jan,Barata:2025hgx}.
For example, the Ising field theory simulations in Ref.~\cite{Jha:2024jan} utilized an $L,\,vt_{\text{max}},\,d,\,\xi$, and inverse time step $1/\delta t$ that were about $10\times$ larger than are used in this work.
The smaller systematic errors in their simulations allowed individual inelastic channels to be isolated and enabled robust predictions of the underlying QFT.
Future quantum simulations, likely incorporating some form of error correction, may be able to access ultra-high-energy scattering where
the production of highly entangled, many-particle states renders MPS methods unreliable. 

The wavepacket preparation algorithm  developed in this work is particularly advantageous in higher dimensions.
By utilizing MCM-FF, wavepackets in any finite spatial dimension can be prepared with a circuit depth that is independent of the wavepacket size, provided that the qubits share the connectivity of the target lattice.
In contrast, existing methods have a circuit depth that scales polynomially in the spatial {\it volume} of the wavepacket.
This scaling is often worse if all-to-all connectivity is not assumed. 
The speedup from using MCM-FF arises because some long-range correlations in quantum states can be built 
with constant-depth circuits that are supplemented with local operations and classical communication.
Efficient algorithms allocate the minimal amount of two-qubit gates needed to create the requisite entanglement, which is then distributed throughout the system with MCM-FF.
The key insight is that the long-range entanglement inherent to momentum plane waves~\cite{Gioia:2021xtp} is also present in W states~\cite{Gioia:2023adm}.
Therefore, wavepackets can be built from W states by only modifying correlations localized over $\sim\! \xi$ sites. 
Finding a basis where this observation is useful was guided by the symmetries and hierarchies in length scales characteristic of the target system.

The efficient preparation of wavepackets is a prerequisite for achieving a quantum advantage in simulations of scattering. 
An advantage could be realized in two dimensions, where no dynamical simulations of QFT scattering have ever been performed.
The underlying assumption of classical hardness is supported by empirical evidence that simulations of dynamics in two dimensions are challenging using classical computers~\cite{Haghshenas:2025euj,Park:2025vyf,Dziarmaga:2022via,PRXQuantum.6.020302}.
Such simulations would unlock the {\it ab initio} study of many phenomena absent in one-dimensional systems.
For example, in two dimensions, spatial rotations on the circle allow single particles to have spin, pairs of particles to have anyonic statistics, and multiple particles to have relative orbital angular momentum.
Dynamical studies of these effects in QFTs are now possible~\cite{Iqbal:2023wvm,Iqbal:2024drh,Xu:2024guv,Minev:2024pmj,Andersen:2022xmz} due to the emergence of flagship quantum computers with native two-dimensional connectivity~\cite{GoogleQuantumAIandCollaborators:2024efv,Gao:2024fik,DeCross:2024tmi,Chen:2023erd,Rodriguez:2024bhh,Muniz:2024dna}.
These advancements have the potential to open a new frontier of scientific exploration that is powered by quantum computing.

\vspace{1em}

\begingroup
\renewcommand*\addcontentsline[3]{}  
\begin{acknowledgements}
\noindent
We would like to thank Ivan Chernyshev, David Simmons-Duffin, Johnnie Gray, Ash Milsted, Martin Savage and Federica Surace for helpful discussions. 
We also thank Ilan Rosen, Brendan Saxberg and Abhinav Kandala for sharing methods for improvements in $R_{ZZ}$ gate calibrations in our simulations.
RF and JP acknowledge support from the U.S. Department of Energy QuantISED program through the theory consortium “Intersections of QIS and Theoretical Particle Physics” at Fermilab, from the U.S. Department of Energy, Office of Science, Accelerated Research in Quantum Computing, Quantum Utility through Advanced Computational Quantum Algorithms (QUACQ), and from the Institute for Quantum Information and Matter, an NSF Physics Frontiers Center
(PHY-2317110). 
RF additionally acknowledges support from a Burke Institute prize fellowship.
NZ acknowledges support provided by the DOE, Office of Science, Office of Nuclear Physics, InQubator for Quantum Simulation (IQuS)
under Award Number DOE (NP) Award DE-SC0020970 via the program on Quantum Horizons: QIS Research and Innovation for Nuclear Science.
NZ is also supported by the Department of Physics
and the College of Arts and Sciences
at the University of Washington. 
MI acknowledges support provided by the Quantum Science Center (QSC),
which is a National Quantum Information Science Research Center of the U.S.\ Department of Energy.
JP acknowledges funding provided by the
U.S. Department of Energy Office of High Energy Physics (DE-SC0018407), the U.S. Department of Energy, Office of Science, Accelerated Research in Quantum Computing, Fundamental Algorithmic Research toward Quantum Utility (FAR-Qu), and the U.S. Department of Energy, Office of Science, National Quantum Information Science Research Centers, Quantum Systems Accelerator.
The computations presented in this work were conducted in the Resnick High Performance Computing Center,
a facility supported by the Resnick Sustainability Institute at Caltech and also enabled by the use of advanced computational, storage and networking infrastructure provided by the Hyak supercomputer system at the University of Washington.
RF and NZ acknowledge the use of IBM Quantum Credits for this work. 
The views expressed are those of the authors, and do not reflect the official policy or position of IBM or the IBM Quantum team.
This research used resources of the National Energy Research
Scientific Computing Center, a DOE Office of Science User Facility
supported by the Office of Science of the U.S. Department of Energy
under Contract No. DE-AC02-05CH11231 using NERSC award
NERSC DDR-ERCAP0034353.

\end{acknowledgements}
\endgroup

\clearpage
\newpage{}

\onecolumngrid
\begingroup
\hypersetup{linkcolor=black}
\tableofcontents
\endgroup

\clearpage

\section*{Methods}
\label{sec:methods}
\setcounter{section}{0}
\renewcommand\thesection{}
\renewcommand\thesubsection{\Alph{subsection}}
\noindent
Throughout this paper the qubits are labeled $|q_{L-1}\dots q_2 q_1 q_0\rangle$ (little-endian notation) and the top wire in every circuit corresponds to $q_0$.
The system size is $L$ and all systems have PBCs (with $q_L=q_0$) unless otherwise stated.
The wavepacket size is $d$, the mass of the lightest particle is $m$, $|\psi_{\text{wp}}\rangle$ is the exact wavepacket, $|\psi_{\text{ansatz}}\rangle$ is the approximately-prepared wavepacket and $|\psi_{\text{vac}}\rangle$ is the vacuum.

\subsection{A constant-depth circuit for initializing \texorpdfstring{$|W(k_0)\rangle$}{}}
\label{sec:WKprep}
\noindent
The efficient preparation of the state in Eq.~\eqref{eq:psiWP0} on $d$ sites,
\begin{equation}
\vert W(k_0) \rangle  \ =\ \sum_{n=0}^{d-1} e^{i\phi_n}c_n |2^n\rangle ,
\label{eq:psiWP02}
\end{equation}
is an important primitive for the preparation of wavepackets.
For $\phi_n = 0$ and $c_n=1/\sqrt{d}$  this reduces to the W state.
\begin{figure}
    \centering
    \includegraphics[width=\linewidth]{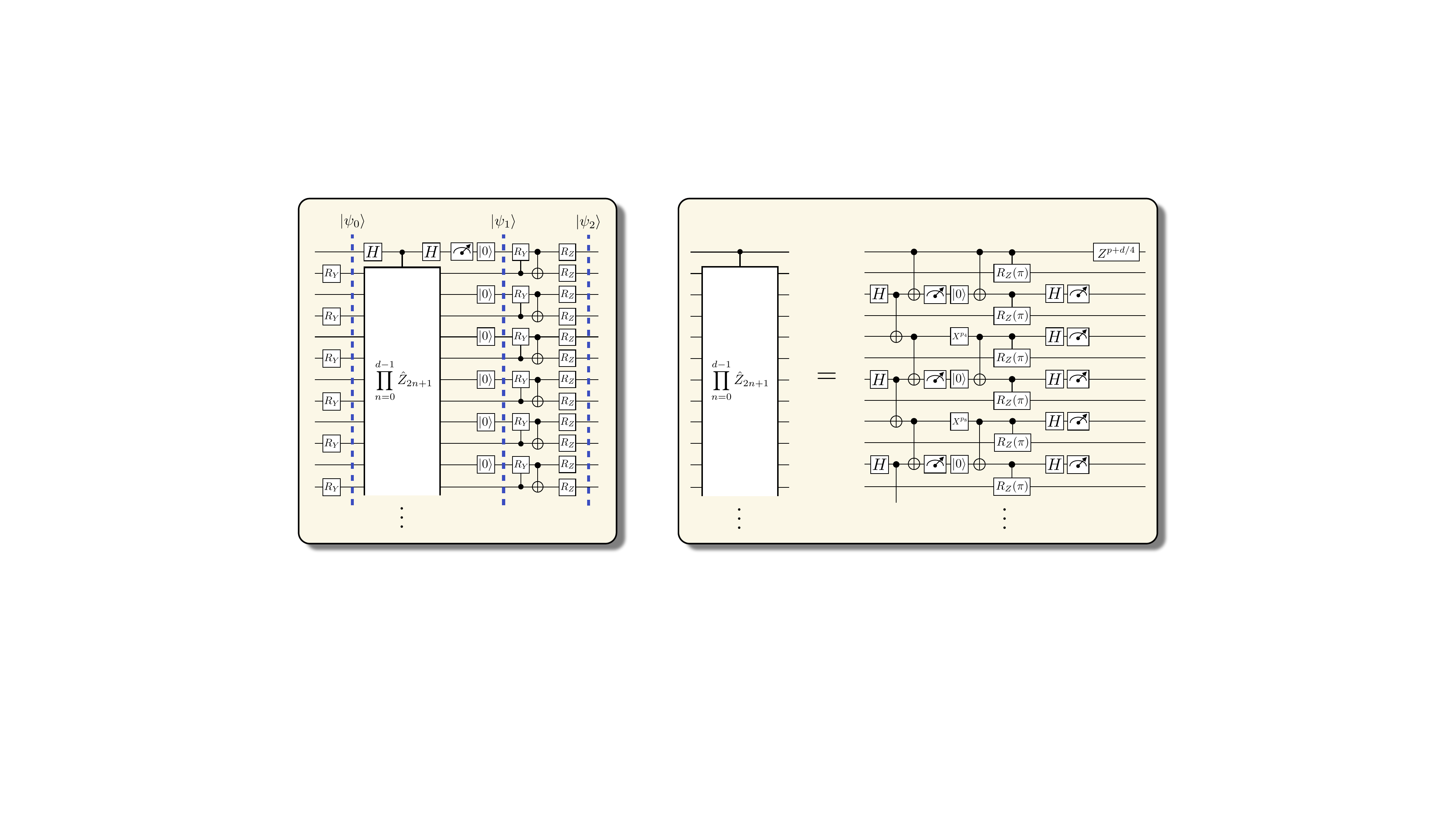}
    \caption{\textit{Circuits that prepare $|W(k_0)\rangle$ in constant depth using MCM-FF.} Left: a circuit that prepares $|W(k_0)\rangle$ in Eq.~\eqref{eq:psiWP02}.
    Right: the circuit from Ref.~\cite{Piroli:2024ckr} that implements the global controlled unitary in constant depth.}
    \label{fig:ConstantDepth}
\end{figure}
The authors of Ref.~\cite{Piroli:2024ckr} presented an ingenious circuit for preparing the W state in constant depth that requires $d$ ancilla qubits.
Their method begins by initializing $|\psi_0\rangle = (\sqrt{1-\delta/d}|0\rangle +\sqrt{\delta/d}|1\rangle )^{\otimes d}$ on the system qubits. Then the parity $\prod \hat{Z}_n$ is measured using a constant-depth circuit.
Odd parity occurs with $p_{\text{success}}\geq0.43\delta$ and prepares the W state with infidelity ${\cal I} ={\cal O}(\delta^2)$.
We generalize their method to prepare $|W(k_0)\rangle$ for arbitrary $c_n$ and $\phi_n$. 
Additionally, we remove the ancillas and provide exact expressions for $p_{\text{success}}$ and ${\cal I}$.

The circuit that prepares $|W(k_0)\rangle$ is shown in Fig.~\ref{fig:ConstantDepth}.
The states $|\psi_0\rangle,\,|\psi_1\rangle,\,|\psi_2\rangle$ are labeled at intermediate stages of the circuit and are described here.
First, $q_0$ is prepared in $|+\rangle$ and
\begin{align}
|\psi_0\rangle \ = \ \bigotimes_{n=0}^{d/2} \left [\cos{\left ( \frac{\theta_{2n+1}}{2}\right ) }|0\rangle \ + \ \sin{\left ( \frac{\theta_{2n+1}}{2}\right ) }|1\rangle \right ] \ , \ \ \sin{\left ( \frac{\theta_{2n+1}}{2}\right ) } \ = \ \sqrt\delta \sqrt{c_{2n}^2 + c_{2n+1}^2} \ ,
\label{eq:psi0}
\end{align}
is prepared on the odd-numbered qubits by applying single-qubit $R_Y$ rotations, $\prod_n e^{-i \theta_{2n+1} \hat{Y}_{2n+1} /2}$.
Then, the parity is kicked back to qubit $q_0$ by applying $\prod_{n=0}^{d-1}\hat{Z}_{2n+1}$ controlled on $q_0$.
This controlled-parity operation is implemented with the constant-depth circuit developed in Ref.~\cite{Piroli:2024ckr}, and shown in the right panel of Fig.~\ref{fig:ConstantDepth}.
Several details about this circuit are described here. The second layer of CNOTs acts on pairs of qubits $q_{n}q_{n+2}$ with $n=0,4,\ldots,d-6$.
The $p_n$ in the first round of MCM-FF are the sums of the outcomes on all measured qubits $q_m$ with $m<n$ and
the $p$ in the second round of MCM-FF is the sum of all measurement outcomes. 
After applying the controlled $\prod_{n=0}^{d-1}\hat{Z}_{2n+1}$ and a Hadamard on qubit $q_0$, the state of $q_0$ is correlated with the parity, i.e., $|\psi\rangle = |\text{even parity}\rangle|0\rangle + |\text{odd parity}\rangle|1\rangle$.

Next, $q_0$ is measured and outcomes are post-selected on odd parity, which occurs with success probability
\begin{align}
p_{\text{success}}  &= \frac{1}{2} - \frac{1}{2}\prod_{n=0}^{d/2 - 1}\left [1-2\delta(c_{2n}^2+c_{2n+1}^2)\right ] \nonumber \\
&=  \delta  -  \delta^2\left [1 - \sum_{n=0}^{d/2-1}(c_{2n}^2 + c_{2n+1}^2)^2 \right ]  +  \frac{2}{3}\delta^3 \left [1- 3 \sum_{n=0}^{d/2-1}(c_{2n}^2 + c_{2n+1}^2)^2+2\sum_{n=0}^{d/2-1}(c_{2n}^2 + c_{2n+1}^2)^3  \right ]  +  {\cal O}(\delta^4) \nonumber \\
&\geq \frac{1}{2}\left (1-e^{-2\delta} \right ) \ \geq \ 0.43 \delta \ .
\label{eq:psuccess}
\end{align}
This expression has been simplified by using the normalization condition $\sum_n c_n^2 = 1$.
The lower bound of  $p_{\text{success}}$ is derived in App.~\ref{app:psuccess} for $\delta \leq 1$.
The even-numbered qubits are reset to $|0\rangle$ and the state after a successful measurement outcome is
\begin{align}
|\psi_1\rangle \ = \ \sum_{n=0}^{d/2-1}  \sqrt{c_{2n}^2+c_{2n+1}^2}\,  |2^{2n+1}\rangle  \ + \ {\cal O}(\delta)\ .
\end{align}
For a given pair of neighboring qubits, this state has concentrated the probability of finding either in $|1\rangle$ to the odd-numbered qubits.
Inspired by Ref.~\cite{Smith:2022nbd}, the final step spreads this probability to the even sites, previously used as ancillas.
This is done using the controlled-$R_Y$ CNOT sequence from Ref.~\cite{Cruz_2019}, with rotation angles
\begin{align}
\tan{\left (\frac{\theta_{2n}}{2} \right )} \ = \ \frac{c_{2n}}{c_{2n+1}} \ .
\end{align}
Lastly, the phases in $|W(k_0)\rangle$ are added with single-qubit $R_Z$ rotations, $\prod_n e^{-i \phi_n \hat{Z}_n /2}$.
The final state $|\psi_2\rangle$ is an approximation to $| W(k_0)\rangle$ with infidelity
\begin{align}
{\cal I} \ &= \ 1 \ - \ |\langle W(k_0)|\psi_2\rangle|^2 \ = \ 1 \ - \ \frac{1}{p_{\text{success}}} \sum_{n=0}^{d/2 - 1}\delta(c_{2n}^2 + c_{2n+1}^2)\prod_{\ell\neq n}\left[1 - \delta (c_{2\ell}^2 + c_{2\ell+1}^2)\right] \nonumber \\
&= \ \frac{\delta^2}{6}\left [1 -3\sum_{n=0}^{d/2-1}(c^2_{2n} + c_{2n+1}^2)^2 +2\sum_{n=0}^{d/2-1}(c^2_{2n} + c_{2n+1}^2)^3 \right ] \ + \ {\cal O}(\delta^3) \ .
\end{align}
Note that expanding ${\cal I}$ to the first non-zero order required $p_{\text{success}}$ in Eq.~\eqref{eq:psuccess} to third order.
Since $\sum_n c_n^2 = 1$, it is reasonable to take $c_n^2 = {\cal O}(d^{-1})$, and both $p_{\text{success}}$ and ${\cal I}$ are independent of $d$ for large $d$.

\subsection{Quantum circuits for preparing wavepackets in lattice quantum field theories}
\label{sec:WPQFTcircs}
\noindent
The algorithm outlined in Sec.~\ref{sec:WPsummary} can prepare wavepackets in a wide range of lattice QFTs.
This section provides specific applications to one-dimensional Ising field theory, $\lambda \hat{\phi}^4$ scalar field theory and the Schwinger model, as well as two-dimensional Ising field theory.
Circuits that prepare wavepackets on system sizes with $\leq 28$ qubits are found using the {\tt qiskit} statevector simulator~\cite{Javadi-Abhari:2024kbf}, and the prepared state is benchmarked against the exact wavepacket determined from exact diagonalization.
Our exact diagonalization methods are explained in App.~\ref{app:ExactWP}.
In Methods~\ref{sec:csimscatt}, the circuits that prepare wavepackets in one-dimensional Ising field theory on a $L=256$ lattice are determined using the Quimb MPS circuit simulator.

MCM-FF is inefficient on classical circuit simulators because each measurement outcome creates a new branch in the circuit that must be computed separately. 
In the worst case, this causes the simulation cost to scale exponentially with the number of measurements. 
Therefore, all simulations that use classical computing in this work initialize $|W(k_0)\rangle$ with the unitary circuit in the left panel of Fig.~\ref{fig:IsingWPCircs}.\footnote{The wavepackets prepared in this section using statevector simulators have the wavepacket size $d$ equal to the system size $L$.}
Additional circuits for preparing $|W(k_0)\rangle$ that reduce the circuit depth by utilizing beyond-linear connectivity and/or MCM-FF are given in App.~\ref{app:WP0prep}.
The approximate wavepacket is then prepared by variationally minimizing the energy using circuits that are translationally invariant, real and preserve the other system-specific symmetries.
We perform this optimization using ADAPT-VQE, which provides an efficient framework for symmetry-aware optimization. 
ADAPT-VQE is a greedy algorithm that builds the structure of the ansatz circuits layer by layer.
At each step in the algorithm, the circuit layer that is most effective at minimizing the energy is identified from a pool of symmetry-preserving parameterized circuits $\{ \hat{U}(\theta) \}$.
This circuit layer is appended to the ansatz circuit and the values of the variational parameters $\vec{\theta}$ are optimized to minimize the energy.
The key ingredient in ADAPT-VQE is the choice of circuit pool, or equivalently an operator pool $\{ \hat{O}\}$ from which the circuits are generated, i.e., $\hat{U}_j(\theta) = e^{i \theta_j \hat{O}_j}$.
Guided by symmetries, an operator pool inspired by the Lie algebra of the Hamiltonian will be used throughout this work.
Operator pools inspired by the Hamiltonian algebra have previously been used to great success in Refs.~\cite{Farrell:2023fgd,Farrell:2024fit,Zemlevskiy:2024vxt,Gustafson:2024bww,Ciavarella:2024lsp,VanDyke:2022ffj,Farrell:2024mgu}.
A full description of ADAPT-VQE is given in App.~\ref{app:57Adapt}.

\subsubsection{One-dimensional Ising field theory}
\label{sec:IsingStatevector}
\begin{figure}
    \centering
    \includegraphics[width=\linewidth]{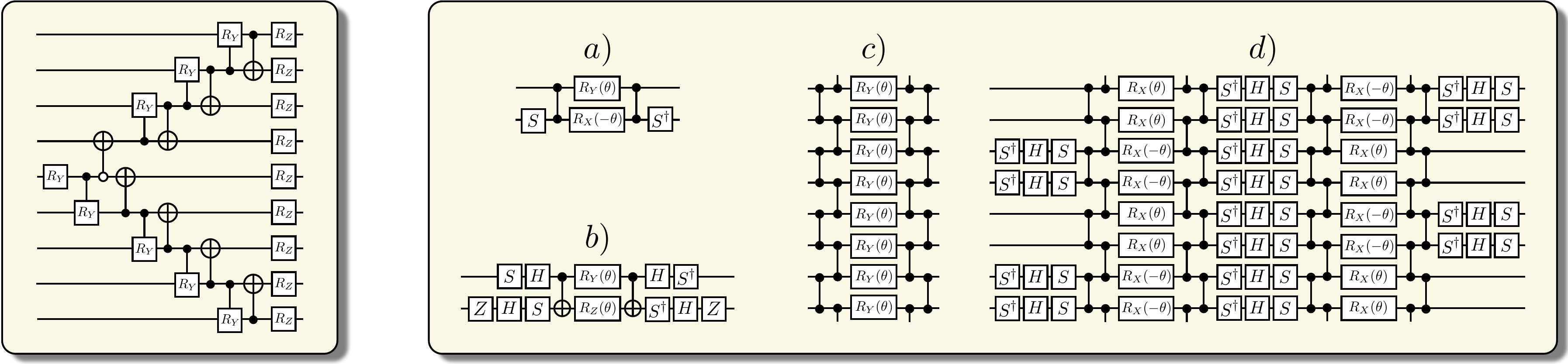}
    \caption{\textit{Circuits used to prepare wavepackets in one-dimensional Ising field theory. } Left: a unitary circuit that prepares $|W(k_0)\rangle$ in Eq.~\eqref{eq:psiWP02} across 9 sites.
    The rotation angles are given in Eq.~\eqref{eq:WP0angles}.
    Right: circuit elements used in ADAPT-VQE to prepare wavepackets in Ising field theory. They implement the unitary evolution of the operators in Eq.~\eqref{eq:opPool}.
    a) implements $\exp{-i\frac{\theta}{2}\left (\hat{Y}\hat{Z}+\hat{Z}\hat{Y} \right )}$ and 
    b) implements $\exp{-i\frac{\theta}{2}\left (\hat{Y}\hat{X}  + \hat{X}\hat{Y} \right )}$.
    Examples of circuits that implement  $\exp{-i\frac{\theta}{2}\sum \hat{Z}\hat{Y}\hat{Z}}$ and $\exp{-i\frac{\theta}{2}\sum \left (\hat{Z}\hat{X}\hat{Y} + \hat{Y}\hat{X}\hat{Z}\right )}$ across 8 qubits with PBCs are shown in c) and d), respectively.}
    \label{fig:IsingWPCircs}
\end{figure}

\noindent
In one-dimensional Ising field theory, one lattice site maps to one qubit and the initial state for preparing wavepackets is the $|W(k_0)\rangle$ in Eq.~\eqref{eq:psiWP02}.
A unitary circuit that prepares $|W(k_0)\rangle$ is shown in the left panel of Fig.~\ref{fig:IsingWPCircs}.
This circuit is most efficient for odd wavepacket sizes $d$. The $R_Y$ angles are given by recursively solving the equations,
\begin{align}
\left [\sin\left (\frac{\theta_{\eta}}{2}\right )\right ]^2 \ &= \ \sum_{i=\eta}^{d-1} \, c_i^2  \ , \nonumber \\[4pt]
\cos{\left (\frac{\theta_{\eta+j+1}}{2}\right )}\prod_{i=0}^{j} \sin{\left (\frac{\theta_{\eta+i} }{2}\right )} \ &= \ c_{\eta+j} \ \ , \ \ j\in[0,1,\ldots,\eta-1] \ , \nonumber \\[4pt] 
\cos{\left (\frac{\theta_{\eta}}{2}\right )}\cos{\left (\frac{\theta_{\eta-j-1}}{2}\right )}\prod_{i=2}^{j} \sin{\left (\frac{\theta_{\eta-i} }{2}\right )} \ &= \ c_{\eta-j} \ \ , \ \ j\in[1,2,\ldots,\eta-1] \ ,
\label{eq:WP0angles}
\end{align}
where $\eta\equiv(d-1)/2$. The $R_Z$ rotation angles are the $\phi_n$ from Eq.~\eqref{eq:psiWP02}.
The circuit depth scales linearly with wavepacket size,
\begin{equation}
\text{CNOT depth: } \ 2\lfloor d/2 \rfloor+2 \ ,
\end{equation}
and is likely optimal assuming linear connectivity and no MCM-FF.

The next step in the wavepacket preparation algorithm is to apply
translationally-invariant and real circuits that minimize the energy.
This is achieved by using ADAPT-VQE to optimize
parameterized circuits $\{e^{i \theta \hat{O}}\}$ constructed from a symmetry-preserving pool of operators $\{\hat{O}\}_{1d \ \text{Ising}}$. 
Enforcing reality and hermiticity constrains all pool operators to be imaginary and anti-symmetric, and the union of all unique operators generated from $i\sum_n[\hat{H}, \hat{X}_n ]$ and $i\sum_n [ \hat{H}, [ \hat{H}, [\hat{H}, \hat{X}_n ] ] ]$ is found to be effective,
\begin{align}
\{ {\hat O}\}_{1d \ \text{Ising}}  =  \sum_{n=0}^{L-1} \bigg \{ \hat{Y}_n 
 ,   \hat{Z}_n\hat{Y}_{n+1}\hat{Z}_{n+2} , \left (\hat{Y}_n\hat{Z}_{n+1}+ \hat{Z}_n\hat{Y}_{n+1}\right )  ,   \left (\hat{Y}_n\hat{X}_{n+1}+ \hat{X}_n\hat{Y}_{n+1}\right )  ,  \left (\hat{Z}_n\hat{X}_{n+1}\hat{Y}_{n+2}+ \hat{Y}_n\hat{X}_{n+1}\hat{Z}_{n+2}\right )  \bigg \} \ .
\label{eq:opPool}
\end{align}
Circuits implementing unitary evolution generated by the operators in the pool are given in the right panel of Fig.~\ref{fig:IsingWPCircs}.
The construction of the circuits is based on the methods in Ref.~\cite{Chernyshev:2025jyw}, and details can be found in App.~\ref{app:qcircs}.
The CNOT depth corresponding to the unitary evolution with respect to each pool operator in Eq.~\eqref{eq:opPool} is $\{0,4,4,4,8\}$

\begin{figure}
    \centering
    \includegraphics[width=\linewidth]{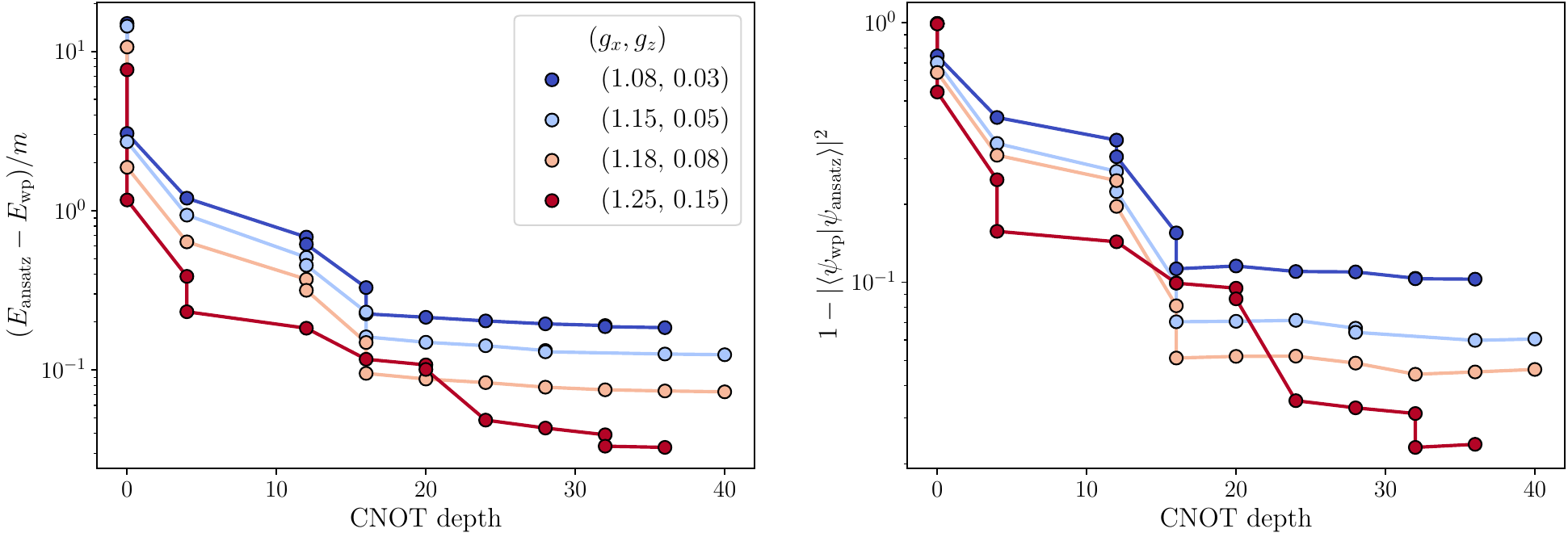}
    \caption{\textit{Quality of prepared wavepackets in one-dimensional Ising field theory.} 
    The deviation in energy (left) and infidelity (right) of the prepared wavepacket as a function of CNOT depth.
    The CNOT depth is monotonically related to the step of ADAPT-VQE, and corresponds to the additional depth after $|W(k_0)\rangle$ preparation.
    Results are shown for wavepacket parameters $\sigma_{}=0.13,\,k_0=0.36\pi$ and $L=28$ for various couplings $g_x,\,g_z$.}
    \label{fig:IsingADADPT_results}
\end{figure}

To assess the quality of the ADAPT-VQE-prepared wavepacket $|\psi_{\text{ansatz}}\rangle$, we compare it to the exact wavepacket  $|\psi_{\text{wp}}\rangle$ by computing the deviation of the energy $(E_{\text{wp}}-E_{\text{ansatz}})/m$ in units of the mass gap $m$, and the infidelity $1-|\langle\psi_{\text{wp}}|\psi_{\text{ansatz}}\rangle|^2$.
Figure~\ref{fig:IsingADADPT_results} shows these two quantities as functions of CNOT depth, obtained from up to 12 steps of ADAPT-VQE for $L=28$ and $\sigma_{}=0.13,\,k_0=0.36\pi$.
Results are shown for four sets of couplings that approach the field-theory limit of vanishing mass gap $\{(g_x,g_z,m)\}=\{(1.25,0.15,1.6),(1.18,0.08,1.1),(1.15,0.05,0.8),(1.08,0.03,0.7)\}$.
The circuits are optimized to minimize the energy (left plot), which improves the overlap with the target wavepacket (right plot).
An infidelity $<0.12$ is reached by depth 16 for all couplings considered.

A plateau in the convergence is observed around the 6th step (depth 16) of ADAPT-VQE for all couplings except $g_x=1.25,g_z=0.15$.
In App.~\ref{app:57Adapt} it is shown that these plateaus can be overcome by expanding the operator pool to include terms in the Lie algebra of the Hamiltonian that correspond to 5th- and 7th-order commutators.
In general, the convergence to the target wavepacket degrades with decreasing mass gap (increasing correlation length).
This is expected as deeper circuits are generically needed to build out longer correlations.
Recently, it has been demonstrated that variational quantum algorithms that utilize MCM-FF have a more favorable optimization landscape~\cite{Deshpande:2024kpt}, and are able to more efficiently prepare states with long-range correlations~\cite{Alam:2024mit,Niu:2024oxx,Yan:2024xev}.
This is an exciting direction where further improvements are expected.
The sequence of operators and variational parameters corresponding to the 8th step of ADAPT-VQE in Fig.~\ref{fig:IsingADADPT_results} are given in App.~\ref{app:ADAPTparam}.

\subsubsection{One-dimensional scalar field theory}
\label{sec:scalar}
\noindent
Historically, scalar field theory has served as a sandbox for understanding aspects of QFTs, with many applications to (beyond) Standard Model physics.
It describes the dynamics of the Higgs Boson~\cite{PhysRevLett.13.508}, as well as particles that emerge from the spontaneous breaking of exact and approximate symmetries, such as the pseudoscalar mesons~\cite{Gasser:1983yg} and the axion~\cite{PhysRevLett.40.279,PhysRevLett.40.223}.
Quantum simulations of scattering in scalar field theory were first laid out in two seminal papers~\cite{Jordan:2011ci,Jordan:2012xnu}.
These works gave protocols for simulating scattering in scalar field theory on quantum computers with an exponential quantum advantage.
Scattering in one-dimensional scalar field theory has also been shown to be BQP-complete~\cite{Jordan_2018}, meaning that any efficient computation on a quantum computer can be mapped to scattering in scalar field theory with a polynomial overhead.

There have been numerous efforts and proposals toward quantum simulations of scattering in scalar field theory \cite{Klco:2019xro,Klco:2019yrb,Klco:2020aud,Kurkcuoglu:2021dnw,Macridin:2021uwn,Liu:2021otn,Abel:2025zxb,Illa:2022jqb, Li:2022ped,Hardy:2024ric,Kreshchuk:2023btr,Briceno:2023xcm,Turco:2023rmx}.
Recently, the first such simulation was performed on 120 qubits of IBM's quantum computers~\cite{Zemlevskiy:2024vxt}.
That work used variational optimization in every stage of the quantum simulation to minimize the circuit depth.
Building off the techniques in Refs.~\cite{Farrell:2023fgd,Farrell:2024fit}, scalable circuits that prepared wavepackets were found by minimizing the infidelity with the exact wavepacket using classical computers.
One challenge was that there were no physics-informed constraints on the structure of the wavepacket preparation circuits, making the optimization difficult.
Additionally, because it relied on minimizing the infidelity, the wavepacket size was constrained by the limitations of exact statevector simulations.
The wavepacket preparation method described in this section overcomes both of these difficulties by:
1) relying on a minimization of a local observable (the energy), not the global infidelity, and 2) only having to optimize over quantum circuits whose structure is heavily constrained by symmetries.
This section begins with a background for scalar field theory on the lattice.

The Hamiltonian for $\lambda \hat{\phi}^4$ scalar field theory on a one-dimensional lattice is
\begin{align}
\hat{H} \ &= \ \sum_{n=0}^{L-1}\left (\frac{1}{2}m_0^2 \hat{\phi}_n^2 \ + \ \frac{1}{2}\hat{\Pi}_n^2 \ + \ \frac{1}{2}\left (  \hat{\phi}_{n+1}-\hat{\phi}_n \right )^2 \ +\ \frac{\lambda}{4!}\hat{\phi}_n^4\right ) \nonumber \\
&\equiv\ \sum_{n=0}^{L-1}\left (\hat{H}_{n}^{(\text{h.o.})} \ - \ \hat{\phi}_{n}\hat{\phi}_{n+1}  \ +\  \frac{\lambda}{4!}\hat{\phi}_n^4\right ) \ ,
\end{align}
where $m_0$ is the bare mass and $\lambda$ is the coupling strength.
For later use, the second line has separated out the single-site harmonic oscillator Hamiltonian,
\begin{align}
\hat{H}_n^{(\text{h.o.})} \ = \ \frac{1}{2}\left (m_0^2 + 2 \right )\hat{\phi}_n^2 + \frac{1}{2} \hat{\Pi}_n^2 \ .
\label{eq:Hscalar1}
\end{align}
The bosonic field operator $\hat{\phi}_n$ and its conjugate momentum $\hat{\Pi}_n$ satisfy the canonical commutation relations $[ \hat{\phi}_n, \hat{\Pi}_m ]=i\delta_{n,m}$.
This Hamiltonian has a $Z_2$ symmetry that takes $\hat{\phi}_n \to -\hat{\phi}_n$ (and $\hat{\Pi}_n \to -\hat{\Pi}_n$). The vacuum is even under $Z_2$, whereas the lowest-energy single-particle states are odd.

The bosonic nature of scalar particles means the Hilbert space on each lattice site is infinite-dimensional and identical to that of a quantum harmonic oscillator. To simulate this theory on a quantum computer, the local Hilbert space must be made finite.
This requires choosing a basis to represent the fields, and then truncating the number of basis states.
Two bases will be used in this work: the harmonic oscillator basis composed of eigenstates of $\hat{H}^{(\text{h.o.})}$, $\vert E_j\rangle^{(\text{h.o.})}$,
and the field basis where $\hat{\phi} \vert \phi_j\rangle = \phi_j\vert \phi_j\rangle$.
To minimize confusion, $n$ will be used as a position space index, and $j$ will be used as an index labeling states in the local Hilbert space at each lattice site $n$.
Once a basis is chosen, the local Hilbert space is truncated to $2^{n_q}$ states and the degrees of freedom of each site are mapped onto $n_q$ qubits. 

The prescription for digitizing scalar field theory in the field basis was described in Refs.~\cite{Jordan:2011ci, Jordan:2012xnu}.
The eigenvalues of the field operator are restricted to the interval $[-\phi_{\text{max}},\phi_{\text{max}}]$ and uniformly sampled at intervals $\delta_\phi$,
\begin{align}
\phi_j = -\phi_{\text{max}} + j \delta_{\phi} \ \ , \ \ \delta_{\phi} = \frac{2 \phi_{\text{max}}}{2^{n_q}-1} \ \ , \ \ j\in[0,2^{n_q}-1] \ .
\end{align}
The basis where $\hat{\Pi}$ is diagonal is related to the $\phi$-basis by a site-wise Fourier transform. 
By the Nyquist-Shannon sampling theorem, field-space digitization errors are minimized when the eigenvalues of $\hat{\Pi}$, $\Pi_j$ are also sampled uniformly,
\begin{align}
\Pi_j \ = \ -\frac{\pi}{\delta_{\phi}} + \left (j+\frac{1}{2} \right )\frac{2 \pi}{2^{n_q} \delta_{\phi}} \ \ , \ \ j\in[0,2^{n_q}-1] \ .
\label{eq:Pij}
\end{align}
In the field basis, there is a simple mapping from fields to spin operators:
\begin{align}
\hat{\phi}_n \ = \ - \frac{\phi_{\text{max}}}{2^{n_q} - 1} \sum_{j=0}^{n_q - 1}2^j \hat{Z}_{n_q  n + j} \ ,
\end{align}
and
\begin{align}
\hat{\Pi}_n \ = \ \hat{V}^{\dagger}\left (- \frac{\pi}{2^{n_q} \delta_{\phi}} \sum_{j=0}^{n_q - 1}2^j \hat{Z}_{n_q  n + j} \right )\hat{V} \ ,
\end{align}
where $\hat{V}$ is the $2^{n_q}\! \times \! 2^{n_q}$ symmetric Fourier transform~\cite{Klco:2018zqz} matrix $V_{kj} = \frac{1}{2^{n_q/2}}e^{i {\bf k} {\bf j}}$, with ${\bf k} = {\bf \Pi}\delta\phi$ and ${\bf \Pi}$ defined in Eq.~\eqref{eq:Pij}.
Importantly, the Fourier transform can be implemented with ${\cal O}(n_q)$ circuit depth~\cite{Klaver:2024vkw}, making this construction efficient for quantum simulation. 
There is one free parameter, $\phi_{\text{max}}$, which is chosen to minimize the field digitization errors.
This can be done by ensuring that the maximum energy of the $\hat{\phi}$ and $\hat{\Pi}$ terms in the single-site Hamiltonian $\hat{H}^{(\text{h.o.})}+\lambda/4! \,\hat{\phi}^4$, are equal.
This constraint gives an optimal $\phi_{\text{max}}$ that is the solution to the cubic equation\footnote{This was determined numerically in Refs.~\cite{Klco:2018zqz,Zemlevskiy:2024vxt}, and analytically for $\lambda=0$ in Ref.~\cite{Bauer:2021gek}.}
\begin{align}
\frac{\pi(2^{n_q}-1)^2}{2^{n_q +1}} \ = \ \phi_{\text{max}}^2\sqrt{(m_0^2+2)+\frac{\lambda}{12}\phi_{\text{max}}^2} \ .
\label{eq:phimax}
\end{align}
For $\lambda=0$ this solution also preserves the  symplectic symmetry of the harmonic oscillator that takes $\hat{\phi} \leftrightarrow  \hat{\Pi}/\sqrt{m_0^2 +2}$.
In the rest of this section, $n_q=3$ is used.

The first step in the wavepacket preparation algorithm outlined in Sec.~\ref{sec:WPsummary} is to initialize a state with the correct amplitude and phase in each momentum block of the Hamiltonian.
For a single-particle wavepacket in scalar field theory, this initial state should be odd under $Z_2$.
Therefore, it is convenient to work in the basis of harmonic oscillator eigenstates $|E_j\rangle^{(\text{h.o.})}$.
The vacuum $|E_0\rangle^{(\text{h.o.})} =  |0\rangle^{\otimes n_q}$ is $Z_2$-even and the first excited state $|E_1\rangle^{(\text{h.o.})} =  |0\rangle^{\otimes n_q -1}|1\rangle$ is $Z_2$-odd.
A $Z_2$-odd state with momentum $k$ is constructed using a straightforward generalization of Eq.~\eqref{eq:psik0}, 
\begin{align}
\vert k\rangle^{(\text{h.o.})} \ =\  \frac{1}{\sqrt{L}}\sum_{n=0}^{L-1}e^{i k n}\, \left (|E_0\rangle^{(\text{h.o.})}\right )^{\otimes L-n-1}\left (|E_1\rangle^{(\text{h.o.})}\right )\left (|E_0\rangle^{(\text{h.o.})}\right )^{\otimes n} \ ,
\label{eq:psik1}
\end{align}
and superpositions of $\vert k \rangle^{(\text{h.o.})}$ can be combined into a wavepacket,
\begin{align}
| W(k_0)\rangle^{(\text{h.o.})} \ = \ {\cal N}\sum_ke^{-i k x_0}\, e^{-(k_0 - k)^2/(4\sigma_{}^2)} |k\rangle^{(\text{h.o.})} \ = \ \sum_{n=0}^{L-1} e^{i\phi_n}c_n |000\rangle^{\otimes L-1-n}|001\rangle|000\rangle^{\otimes n}\ .
\label{eq:Wk0Scalar}
\end{align}
The second equality is for $n_q=3$ and has emphasized that $|W(k_0)\rangle^{(\text{h.o.})}$ is simply a generalized version of the initial state used to prepare wavepackets in Ising field theory, Eq.~\eqref{eq:psiWP02}. 
This state can be prepared using the $|W(k_0)\rangle$ preparation circuits constructed previously.

\begin{figure}
    \centering
    \includegraphics[width=\linewidth]{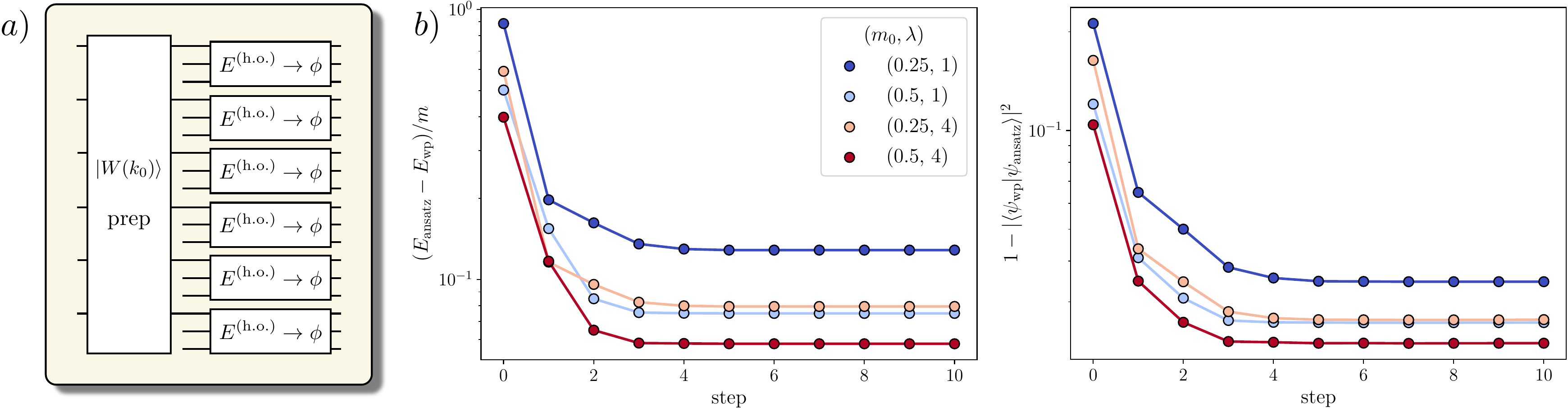}
    \caption{\textit{Wavepacket preparation in scalar field theory.} a) A circuit that prepares the initial state of the wavepacket preparation algorithm $|W(k_0)\rangle^{(\phi)}$. b) The deviation of the energy (left) and infidelity (right) of the wavepackets prepared with up to 10 steps of ADAPT-VQE. Results for $\sigma_{}=0.35,\,k_0=0.44\pi$ and $L=9$ are shown for $n_q=3$ (27 qubits).}
    \label{fig:scalarWP}
\end{figure}

The remainder of the wavepacket preparation is more efficient in the field basis.
While an efficient circuit that implements the $E^{(\text{h.o.})} \to \phi$ change of basis is not known, it can be found by exactly synthesizing the corresponding $2^{n_q}$-dimensional unitary into gates.
Since a modest $n_q$ is expected to be sufficient for many observables of interest~\cite{Klco:2018zqz}, this single-site change of basis during state preparation will likely not be the limiting factor in quantum simulations of scattering.
An example of a circuit preparing a wavepacket across 6 sites is shown in Fig.~\ref{fig:scalarWP}a).
First $|W(k_0)\rangle^{(\text{h.o.})}$ is prepared in the harmonic oscillator basis.
Then the circuits implementing the $E^{(\text{h.o.})}\to\phi$ change of basis are applied locally to each lattice site.
For $n_q=3$, the change of basis circuit found by BQSkit~\cite{osti_1785933} has CNOT depth 11 (compare to depth 6 for the symmetric Fourier transform). 
The result is $|W(k_0)\rangle^{(\phi)}$ in the field basis.
%


With $|W(k_0)\rangle^{(\phi)}$ initialized, the next step is to minimize the energy using ADAPT-VQE equipped with a pool of translationally-invariant, real and $Z_2$-symmetric operators.
Inspired by the Lie algebra of the Hamiltonian, an operator pool similar to that used in Ref.~\cite{Zemlevskiy:2024vxt} is found to be effective,
\begin{align}
\{ \hat{O} \}_{\text{scalar}} \ &= \ \{ \hat{O}_{\phi\Pi} (s)\} \ = \ \left \{ -\frac{i}{2} \left [\sum_{n=0}^{L-1} \hat{\phi}_n \hat{\phi}_{n+s}\, , \  \sum_{n=0}^{L-1} \hat{\Pi}_{n}^2\right ] \right \} \nonumber \\
&\to \ \left \{ \sum_{n=0}^{L-1}\left (\hat{\phi}_n\hat{\Pi}_{n+s} + \hat{\Pi}_n \hat{\phi}_{n+s} \right )  \right \} \ .
\end{align}
The second line has assumed the fields satisfy the canonical commutation relations, which is only approximately true with digitized fields.
However, the pool operators still have the desired symmetry properties.
In the field basis, $\hat{\phi}_n$ is real and $\hat{\Pi}_n$ is imaginary, and therefore $\exp\left (i \theta \hat{O}_{\phi\Pi}\right )$ is real.
These operators build correlations over $s$ spatial sites with $s=0,1,2,\ldots,\lfloor L/2\rfloor$. 
The unitary evolution with respect to these operators is implemented by Trotterizing and switching from the $\phi$-basis to the $\Pi$-basis using the quantum Fourier transform.
For $\hat{O}_{\phi\Pi}(0)$, we instead find the matrix of eigenvectors $\hat{U}$ and eigenvalues $\hat{D}$ of $\left (\hat{\phi}_n\hat{\Pi}_{n} + \hat{\Pi}_n \hat{\phi}_{n}\right )$, and then implement $U^{\dagger} e^{i \theta \hat{D}} U$. 
BQSkit's unitary circuit synthesis finds a CNOT depth 10 implementation of $\hat{U}$ and $\hat{U}^{\dagger}$, and $e^{i\theta \hat{D}}$ is a product of diagonal rotations.

The ADAPT-VQE results for preparing wavepackets with $\sigma_{}=0.35$, $k_0=0.44\pi$ and $L=9$ (27 qubits) are shown in Fig.~\ref{fig:scalarWP}b).
A selection of couplings $\{(m_0,\lambda,m)\}=\{(0.5,4,0.94),(0.25,4,0.85),(0.5,1,0.66),(0.25,1,0.52)\}$ are chosen, with $m$ the mass gap.
For these couplings, Eq.~\eqref{eq:phimax} gives $\phi_{\text{max}}=\{2.21,2.23,2.41,2.45\}$.
Again, minimizing the energy rapidly decreases the infidelity with the target wavepacket.
It is surprising how well  $|W(k_0)\rangle^{(\phi)}$ (the state at step 0) approximates the exact wavepacket.
This state is a superposition over tensor products between different spatial sites, yet it still has high overlap with the exact wavepacket for all couplings tested.
Convergence is faster for smaller mass gaps, and all couplings reach an infidelity $<0.05$ by the second step of ADAPT-VQE.

\subsubsection{The Schwinger model}
\label{eq:SchwingerWP}
\noindent
The Schwinger model describes electrons, positrons and photons interacting in one  dimension, and is often studied as a toy model for Quantum Chromodynamics (QCD).
Like QCD, the Schwinger model is a lattice gauge theory that exhibits confinement and a chiral condensate.
It has emerged as a popular testbed for quantum simulation algorithms on quantum devices~\cite{Farrell:2023fgd,
Farrell:2024fit,
Klco:2018kyo,
Nguyen:2021hyk,
deJong:2021wsd,
Pomarico:2023png,
Zhou:2021kdl,
Zhang:2023hzr,
Mil:2019pbt,
Lu:2018pjk,
Mueller:2022xbg,
Riechert:2021ink,
Yang:2020yer,
Kokail:2018eiw,
Martinez:2016yna,
Guo:2024tnb,
Angelides:2023noe,
Schuster:2023klj}.
Recently, a wavepacket was prepared in the Schwinger model on 112 qubits of IBM's quantum computers~\cite{Farrell:2024fit}.
Wavepacket preparation circuits were first determined on small system sizes by minimizing the infidelity with the exact wavepacket using classical computers.
The SC-ADAPT-VQE algorithm~\cite{Farrell:2023fgd} was then used to scale these circuits up and prepare wavepackets on 112 qubits.
Similar to the discussion in the previous section on scalar field theory, this approach has limitations related to wavepacket scalability and circuit optimization.
The wavepacket preparation method described here overcomes both of these problems.

In this work, the Schwinger model is discretized onto a staggered lattice with fermions (antifermions) occupying even- (odd-) numbered sites, and the gauge fields occupying the links between them.
This maps $L$ spatial sites to $2L$ staggered sites.
Gauss's law constrains the allowed states of the gauge fields and fermions. 
With OBCs, explicit gauge field degrees of freedom can be completely removed from the theory, leaving a system of fermions interacting through a linear Coulomb potential.
With PBCs, there is a single mode of the gauge field whose dynamics is not constrained.
To keep translational invariance manifest, it is convenient to parameterize this mode by the average electric field $\hat{{\cal E}} = \sum_n \hat{{\cal E}}_n/(2L)$, and its conjugate parallel transporter $\hat{U}^{2L} = \prod_n \hat{U}_n$, such that $[\hat{{\cal E}},\hat{U}^{2L} ] = \hat{U}$~\cite{Dempsey:2022nys}.
After performing the Jordan-Wigner (JW) mapping from fermionic to spin operators the Hamiltonian is,
\begin{align}
\hat{H} \ = \ &\frac{m_0}{2}\sum_{n=0}^{2L-1}\left [ (-1)^n \hat{Z}_n + \hat{I}\right ] \  - \ \frac{g^2}{2}\sum_{n=0}^{2L-1}\sum_{s=1}^{L}\left (s - \frac{s^2}{2L} \right )\left (1-\frac{\delta_{s,L}}{2} \right )\hat{Q}_n\hat{Q}_{n+s} \nonumber \ + \ g^2 L \hat{{\cal E}}^2 \\
&+ \ \frac{1}{2}\sum_{n=0}^{2L-2}\left (\hat{\sigma}_n^+ \hat{U} \hat{\sigma}_{n+1}^- + {\rm h.c.} \right ) \ + \ \frac{1}{2}(-1)^{L + 1}\left (\hat{\sigma}^+_{2L-1}\hat{U}\hat{\sigma}^-_{0} +{\rm h.c.} \right ) \ .
\end{align}
The bare fermion mass is $m_0$, the electric charge is $g$ and the staggered electric charge operator is $\hat{Q}_n = -\frac{1}{2}[(-1)^n +\hat{Z}_n ]$. 
The $(-1)^{L +1}$ in the hopping term between sites $2L-1$ and $0$ comes from the JW mapping.\footnote{The extra sign can be verified by comparing the spectrum of a JW-mapped free fermion with PBCs to the exact result.} 

The zero mode of the gauge field is bosonic and its Hilbert space is formally infinite.
Like for scalar field theory, this Hilbert space is made finite by choosing a basis and truncating to some number of states.
In this section, we will truncate the Hilbert space to one state, i.e., will not consider gauge field dynamics.
This can be improved by allocating some number of qubits to represent the gauge field.
With this truncation, the Hamiltonian only contains operators acting on staggered fermion sites,\footnote{The Hamiltonian in Eq.~\eqref{eq:SchwingerH} is equivalent to that used in Ref.~\cite{Zache:2018cqq} with the right choice of initial conditions.}
\begin{align}
\hat{H} \ \to \ &\frac{m_0}{2}\sum_{n=0}^{2L-1}\left [ (-1)^n \hat{Z}_n + \hat{I}\right ] \  - \ \frac{g^2}{2}\sum_{n=0}^{2L-1}\sum_{s=1}^{L}\left (s - \frac{s^2}{2L} \right )\left (1-\frac{\delta_{s,L}}{2} \right )\hat{Q}_n\hat{Q}_{n+s} \nonumber \\
&+ \ \frac{1}{2}\sum_{n=0}^{2L-2}\left (\hat{\sigma}_n^+  \hat{\sigma}_{n+1}^- + {\rm h.c.} \right ) \ + \ \frac{1}{2}(-1)^{L + 1}\left (\hat{\sigma}^+_{2L-1}\hat{\sigma}^-_{0} +{\rm h.c.} \right ) \ .
\label{eq:SchwingerH}
\end{align}
This is the Hamiltonian that will be used in the rest of this section.

The first step for preparing wavepackets is to initialize a state with the momentum content and quantum numbers of the target wavepacket. 
In the Schwinger model, due to confinement, the single-particle eigenstates are charge-neutral ``hadrons".
As a result, the initial state must also be charge-neutral, not composed of individual electron ($e^{-}$) or positron ($e^+$) excitations.
One place to start is at strong coupling (SC), $g\to\infty$, where hadrons are tightly-bound pairs of $e^+ e^-$ on neighboring staggered sites.
The wavefunction for a plane wave of hadrons at SC is,
\begin{align}
|k\rangle^{(\text{SC})} \ = \ &\frac{1}{\sqrt{2L}}\Bigg [ \sum_{\ell=0}^{L-1}e^{i k \ell}\, |01\rangle^{\otimes L-\ell-1}|10\rangle |01\rangle^{\otimes \ell} \nonumber \\
&- e^{ik/2} \left ( \sum_{\ell=0}^{L-2}e^{i k \ell}\, |01\rangle^{\otimes L-\ell-2}|00\rangle|11\rangle|01\rangle^{\otimes \ell}  +   (-1)^{L+1}e^{ik \left (L-1\right )}|11\rangle |01\rangle^{\otimes L-2}\vert 00\rangle\right ) \Bigg ] \ .
\end{align}
The sum index $\ell$ emphasizes that these sums are over spatial sites.
The state in the first line represents a plane wave of $|e^+ e^{-}\rangle$, with the $e^+$ on the higher-numbered staggered site.
The states in the second line are related by charge conjugation, and are plane waves of $|e^- e^+\rangle$ with the $e^-$ on the higher-numbered staggered site.
At all couplings, the lightest hadron corresponds to a massive photon and is therefore odd under charge conjugation.
This is why there is a $(-1)$ between the sums in the first and second lines.
On a staggered lattice, charge conjugation is realized by a global spin flip followed by a translation by one staggered site.
One staggered site is half of a spatial site, and this translation gives the $e^{i k/2}$ relative phase between the sums in the first and second line.
Finally, when the $|e^-e^+\rangle$ is translated across sites $0$ and $2L-1$, there is an additional factor of $(-1)^{L+1}$ due to the kinetic term in the JW mapping.

\begin{figure}
    \centering
    \includegraphics[width=\linewidth]{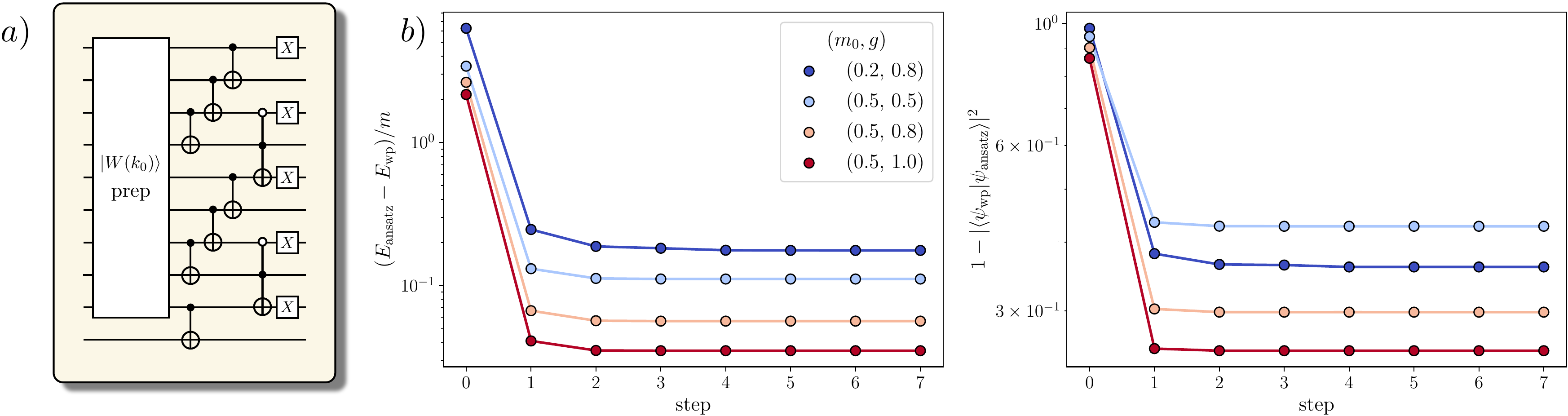}
    \caption{\textit{Wavepacket preparation in the Schwinger model.} a) A circuit that prepares the initial wavepacket of hadrons at strong coupling $|W(k_0)\rangle^{(\text{SC})}$ on $2L=10$ staggered sites.
    b) The deviation of the energy (left) and infidelity (right) of a wavepacket prepared with up to 7 steps of ADAPT-VQE.
    Results for $\sigma_{}=0.25,\,k_0=0.36\pi$ and $L=14$ (28 staggered sites) are shown.}
    \label{fig:SchwingerWP}
\end{figure}

An initial wavepacket of hadrons at SC is written as,\footnote{The SC hadron that crosses sites $2L-1$ and 0 has been omitted since, in practice, wavepackets will be localized to a spatial region away from site 0.}
\begin{align}
| W(k_0)\rangle^{(\text{SC})} \ &= \ {\cal N}\sum_{k}e^{-i k x_0}\, e^{-(k_0 - k)^2/(4\sigma_{}^2)}  |k\rangle^{(\text{SC})} \ \nonumber \\
&\equiv \ \sum_{n=0}^{L-1} c_{2n}e^{i\phi_{2n}}  |01\rangle^{\otimes L-n-1}|10\rangle |01\rangle^{\otimes n} \ + \ \sum_{n=0}^{L-2} c_{2n+1}e^{i\phi_{2n+1}} |01\rangle^{\otimes L-n-2}|00\rangle|11\rangle|01\rangle^{\otimes n} \  .
\label{eq:psiWPSC}
\end{align}
Hadron wavepackets at SC have previously been used in MPS simulations of scattering~\cite{Papaefstathiou:2024zsu,Barata:2025hgx}.
An example of a circuit that prepares $| W(k_0)\rangle^{(\text{SC})}$ over ten staggered sites is given in Fig.~\ref{fig:SchwingerWP}a).
First, $|W(k_0)\rangle$ in Eq.~\eqref{eq:psiWP02} is prepared over all but the last staggered site using one of the $|W(k_0)\rangle$ preparation circuits previously developed.
The CNOT and CCNOT sequence that follows turns each ``$1$" in $|W(k_0)\rangle$ into ``$11$''.
The desired wavepacket of SC hadrons is produced by applying $\hat{X}_n$ to all even-numbered staggered sites.
The circuits that prepare $|W(k_0)\rangle^{(\text{SC})}$ over an arbitrary number of spatial sites are determined by a straightforward generalization.

The next step in wavepacket preparation is to perform ADAPT-VQE using a pool of operators that are translationally-invariant, real, charge conjugation and parity symmetric, and conserve electric charge.
We choose an operator pool generated by taking all nested commutators of the mass and kinetic terms (the Lie algebra of the free Hamiltonian),
\begin{align}
\{\hat{O} \}_{\text{Schwinger}} \ &= \ \{\hat{O}_{mh}(s)\} \ , \nonumber \\[4pt]
\hat{O}_{mh}(s) \ &=\ i \left [\, \sum_{n=0}^{2L-1} (-1)^n \hat{Z}_n   \ , \ \sum_{n=0}^{2L-1} v(n,s)(1-\frac{1}{2}\delta_{s,L})\left (\hat{\sigma}^+_{n} \hat{Z}^{s-1}\hat{\sigma}^-_{n+s} + {\rm h.c.}\right ) \right ]
\label{eq:QCDpool}
\end{align}
where $v(n,s) = (\pm 1)^{L+1}$ with $(+)$ if $n+s \leq 2L-1$ and $(-)$ if $n+s > 2L-1$. 
This sign is needed to account for the minus sign in the kinetic term for even $L$.
The range of $s$ is $s\in \{1,3,\ldots,L\}$, and only odd $s$ is generated as a consequence of charge conjugation symmetry.
This operator pool is the PBC version of the pool used in Refs.~\cite{Farrell:2023fgd,Farrell:2024fit,Farrell:2024mgu} to prepare the Schwinger model vacuum.
Quantum circuits corresponding to the unitary evolution of these operators are a straightforward extension of those in Ref.~\cite{Farrell:2023fgd}.
For convenience, the {\tt PauliEvolutionGate} method in {\tt qiskit} is used to generate these circuits.

The results from using ADAPT-VQE to prepare wavepackets are shown in Fig.~\ref{fig:SchwingerWP}b) for $L=14$ and wavepacket parameters $k_0=0.36\pi$ and $\sigma_{}=0.25$.
Four sets of couplings were chosen
$\{(m_0,g,m)\}=\{(0.5,1.0,1.61),(0.5,0.8,1.44),(0.5,0.5,1.23),(0.2,0.8,0.82)\}$ that approach the continuum limit of vanishing mass gap $m\to0$. 
Convergence to the target wavepacket mostly plateaus after the second step of ADAPT-VQE.
The convergence is better with larger $m_0$ and $g$ likely because the mass gap is larger and/or it is closer to the SC limit.
These fidelities will likely be sufficient for near-term demonstrations of scattering.
However, higher-quality wavepackets will eventually be needed for simulations that approach the continuum limit.
Preliminary attempts to expand the operator pool by including commutators between the kinetic term and $\hat{Q}_n\hat{Q}_{n+s}$ did not improve convergence.
It is possible that an initial state with hadrons of different lengths, i.e., not only $e^-e^+$ on neighboring staggered sites, would perform better.
This would allow for the valence fermion and anti-fermion to have different relative momentum, like in Refs.~\cite{Davoudi:2024wyv,Rigobello:2021fxw}.

\subsubsection{Two-dimensional Ising field theory}
\label{sec:2DIsing}
\noindent
The wavepacket preparation method described in Sec.~\ref{sec:WPsummary} readily generalizes to lattice QFTs beyond one dimension.
Consider the two-dimensional tilted-field Ising model on a PBC square lattice defined by the Hamiltonian,
\begin{align}
\hat{H} \ = \ -\sum_{\langle ij\rangle}\hat{Z}_i \hat{Z}_j \ - \sum_i \left (g_x \hat{X}_i \ + \ g_z \hat{Z}_i \right ) \ .
\label{eq:2DIsingH}
\end{align}
The first sum is over nearest neighbors $\langle i,j\rangle$ and the second sum is over every lattice site.
This Hamiltonian has a translational symmetry in both the $x$- and $y$-direction, as well as a symmetry under $\pi/2$ rotations.\footnote{The recovery of the $SO(2)$ rotational symmetry in the continuum limit is subtle, and may require lattice and operator smearing.
See, e.g., Refs.~\cite{HadronSpectrum:2009krc,Davoudi:2012ya}.}
On an infinite lattice and at the critical point, $g_c=3.04438(2)$~\cite{PhysRevE.66.066110} and $g_z = 0$, this system is described by the 3D Ising CFT.
A massive interacting QFT is obtained by taking $g_x \to g_c$ and $g_z\to0$ but keeping fixed
the scaling invariant ratio,
\begin{align}
\eta_{\text{latt}}^{2d} \ = \ 
\frac{g_x-g_c}{|g_z|^{\frac{D-\Delta_{\epsilon}}{D-\Delta_{\sigma}}}} \ = \ 
\frac{g_x-g_c}{|g_z|^{0.63959303(29)}} \ .
\label{eq:eta2d}
\end{align}
The second equality has used $D=3$ spacetime dimensions and the scaling dimensions of the relevant deformations, $\Delta_{\epsilon} = 1.41262528(29)$ and $\Delta_{\sigma} = 0.518148806(24)$, determined using conformal bootstrap~\cite{Chang:2024whx}.
The universal physics of the 3D Ising CFT also describes the Wilson-Fisher fixed point in two-dimensional scalar field theory~\cite{PhysRevLett.28.240}.
As a result, the continuum limit for arbitrary $\eta_{\text{latt}}^{2d}$ corresponds to a massive interacting scalar field theory in two dimensions.

A wavepacket on a $L\times L$ lattice is,
\begin{equation}
\vert \psi_{\text{wp}} \rangle  \ = \ {\cal N}\sum_{\vec{k}} e^{-i \vec{k}\cdot \vec{x}_0}\, e^{-|\vec{k}_0 - \vec{k}|^2/(4\sigma_{}^2)} \vert \psi_{\vec{k}} \rangle \ ,
\label{eq:psiWPFull2D}
\end{equation}
where, e.g., $\vec{k} = (k_x,k_y)$ and $\vert \psi_{\vec{k}} \rangle$ are the lowest-energy single-particle states with momentum $\vec{k}$.\footnote{For simplicity the same spread in momentum $\sigma_{}$ is used for $k_x$ and $k_y$.}
The sum over $\vec{k}$ runs over $L^2$ terms where $k_x=2\pi n_x/L$ and $k_y=2\pi n_y/L$ with $n_x,n_y$ integers and $k_x,k_y \in (-\pi,\pi]$.
The Hamiltonian can be block diagonalized into $L^2$ different blocks labeled by $\vec{k}$.
A simple initial state with the correct amplitude and phase in each momentum block is
\begin{equation}
\vert W(k_0)\rangle  \ = \ {\cal N}\sum_{\vec{k}} e^{-i \vec{k}\cdot \vec{x}_0}\, e^{-|\vec{k}_0 - \vec{k}|^2/(4\sigma_{}^2)} \vert \vec{k} \rangle  \ = \  \sum_{n_x=0}^{L-1}\sum_{n_y=0}^{L-1} e^{i \phi_{n_x,n_y}}c_{n_x,n_y}|2^{L n_y + n_x}\rangle \ ,
\label{eq:psiWP02d}
\end{equation}
where
\begin{align}
|\vec k\rangle \ = \ \frac{1}L\sum_{n_x=0}^{L-1}\sum_{n_y=0}^{L-1} e^{i \vec{k}\cdot \vec{n}}|2^{L n_y + n_x}\rangle \ .
\end{align}
These are the two-dimensional generalizations of the states given in Eqs.~\eqref{eq:psik0} and~\eqref{eq:psiWP0}, and $\vert W(k_0)\rangle$ can be prepared with the circuit in the left panel of Fig.~\ref{fig:IsingWPCircs}.

\begin{figure}
    \centering
    \includegraphics[width=\linewidth]{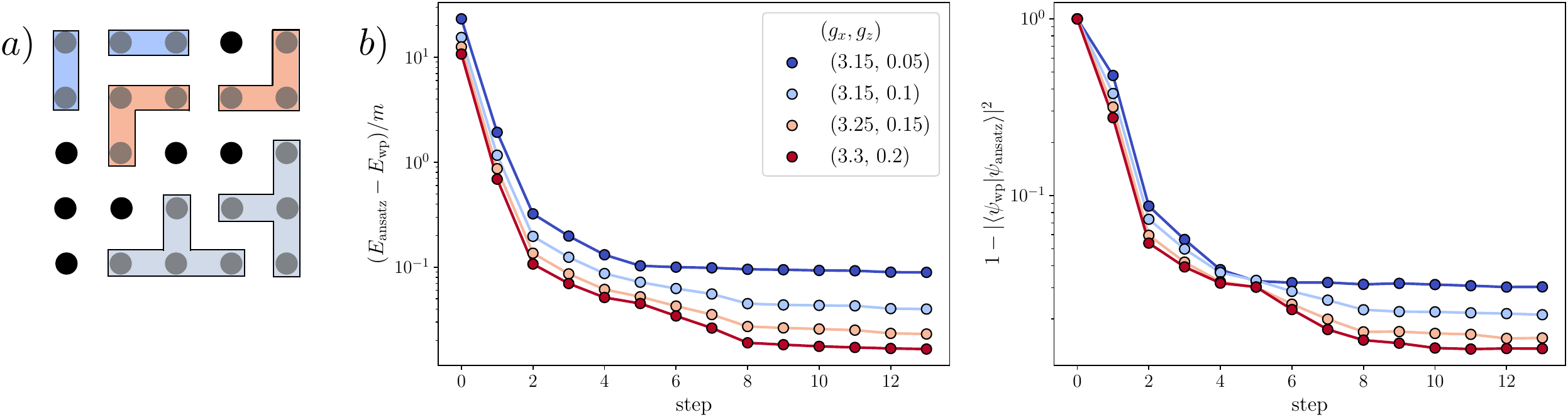}
    \caption{\textit{Wavepacket preparation in two-dimensional Ising field theory.} a) Operators in the ADAPT-VQE pool Eq.~\eqref{eq:opPool2d}, that are related by rotations.
    The blue operators span two sites, and the vertical strip is related to the horizontal strip by a $\pi/2$ rotation.
    The tan operators span three sites in an `L' shape and are related by a $\pi$ rotation.
    The gray operators span four sites in a `T' shape and are related by a $\pi/2$ rotation.
    b) The deviation of the energy (left) and infidelity (right) of a wavepacket prepared with up to 13 steps of ADAPT-VQE. 
    Results for a $5\times 5$ lattice with $\vec{k}_0=(0.4\pi,0.4\pi)$ and $\sigma_{}=0.5$ are shown.}
    \label{fig:2DWP}
\end{figure}
The next step in wavepacket preparation is to minimize the energy using translationally-invariant and real circuits starting from $\vert W(k_0)\rangle$.
This is done using ADAPT-VQE and an operator pool inspired by the Lie algebra of the Hamiltonian is found to be effective.
Keeping all unique operators up to third-order commutators, $i\sum_n [\hat{H}, \hat{X}_n ]$ and $i\sum_n [ \hat{H}, [ \hat{H}, [\hat{H}, \hat{X}_n ] ]  ]$, gives an operator pool
\begin{align}
&\{ {\hat O}\}_{2d \ \text{Ising}}\ = \ \sum_{n_x=0}^{L-1}\sum_{n_y=0}^{L-1} \bigg \{ \hat{Y}_{n_x,n_y} \,  
 , \nonumber \\
 &\left (\hat{Y}_{n_x,n_y}\hat{Z}_{n_x+1,n_y}+ \hat{Z}_{n_x,n_y}\hat{Y}_{n_x+1,n_y}+\hat{Y}_{n_x,n_y}\hat{Z}_{n_x,n_y+1}+ \hat{Z}_{n_x,n_y}\hat{Y}_{n_x,n_y+1}\right )\, , \nonumber \\
 &\left (\hat{Y}_{n_x,n_y}\hat{X}_{n_x+1,n_y}+ \hat{X}_{n_x,n_y}\hat{Y}_{n_x+1,n_y}+\hat{Y}_{n_x,n_y}\hat{X}_{n_x,n_y+1}+ \hat{X}_{n_x,n_y}\hat{Y}_{n_x,n_y+1}\right )\, , \nonumber \\
 &\left ( \hat{Z}_{n_x,n_y}\hat{Y}_{n_x+1,n_y}\hat{Z}_{n_x+2,n_y}+\hat{Z}_{n_x,n_y}\hat{Y}_{n_x,n_y+1}\hat{Z}_{n_x,n_y+2} \right ) \, , \nonumber \\
 &\left (\hat{Z}_{n_x,n_y}\hat{X}_{n_x+1,n_y}\hat{Y}_{n_x+2,n_y}+ \hat{Y}_{n_x,n_y}\hat{X}_{n_x+1,n_y}\hat{Z}_{n_x+2,n_y}+\hat{Z}_{n_x,n_y}\hat{X}_{n_x,n_y+1}\hat{Y}_{n_x,n_y+2}+ \hat{Y}_{n_x,n_y}\hat{X}_{n_x,n_y+1}\hat{Z}_{n_x,n_y+2}\right )\, , \nonumber \\
 &\left ( \hat{Z}_{n_x,n_y}\hat{Y}_{n_x+1,n_y}\hat{Z}_{n_x+1,n_y+1}+\hat{Z}_{n_x,n_y}\hat{Y}_{n_x,n_y+1}\hat{Z}_{n_x-1,n_y+1}+\hat{Z}_{n_x,n_y}\hat{Y}_{n_x-1,n_y}\hat{Z}_{n_x-1,n_y-1}+\hat{Z}_{n_x,n_y}\hat{Y}_{n_x,n_y-1}\hat{Z}_{n_x+1,n_y-1} \right ) \, , \nonumber \\
 &\Big ( \hat{Z}_{n_x,n_y}\hat{X}_{n_x+1,n_y}\hat{Y}_{n_x+1,n_y+1}+\hat{Z}_{n_x,n_y}\hat{X}_{n_x,n_y+1}\hat{Y}_{n_x-1,n_y+1}+\hat{Z}_{n_x,n_y}\hat{X}_{n_x-1,n_y}\hat{Y}_{n_x-1,n_y-1}+\hat{Z}_{n_x,n_y}\hat{X}_{n_x,n_y-1}\hat{Y}_{n_x+1,n_y-1} \nonumber \\
&+\hat{Y}_{n_x,n_y}\hat{X}_{n_x+1,n_y}\hat{Z}_{n_x+1,n_y+1}+\hat{Y}_{n_x,n_y}\hat{X}_{n_x,n_y+1}\hat{Z}_{n_x-1,n_y+1}+\hat{Y}_{n_x,n_y}\hat{X}_{n_x-1,n_y}\hat{Z}_{n_x-1,n_y-1}+\hat{Y}_{n_x,n_y}\hat{X}_{n_x,n_y-1}\hat{Z}_{n_x+1,n_y-1} \Big ) \ , 
\nonumber \\
&\Big ( \hat{Y}_{n_x,n_y}\hat{Z}_{n_x+1,n_y}\hat{Z}_{n_x-1,n_y}\hat{Z}_{n_x,n_y+1} + \hat{Y}_{n_x,n_y}\hat{Z}_{n_x+1,n_y}\hat{Z}_{n_x-1,n_y}\hat{Z}_{n_x,n_y-1}+ \hat{Y}_{n_x,n_y}\hat{Z}_{n_x+1,n_y}\hat{Z}_{n_x,n_y+1}\hat{Z}_{n_x,n_y-1}\nonumber \\
&+ \hat{Y}_{n_x,n_y}\hat{Z}_{n_x-1,n_y}\hat{Z}_{n_x,n_y+1}\hat{Z}_{n_x,n_y-1} \Big )  \bigg \} \ .
\label{eq:opPool2d}
\end{align}
The rotational symmetry of the Hamiltonian on a square lattice groups together all operators related by $\pi/2$ rotations.
Lines 2-5 combine together horizontal and vertical strips of operators.
Operators spanning three sites can additionally be arranged in an `L' shape,
and lines 6-8 group together the four rotations of `L'. 
The four operators in lines 9 and 10 have a `T' shape and are related by rotations.
See Fig.~\ref{fig:2DWP}a) for an illustration.
The {\tt PauliEvolutionGate} method in {\tt qiskit} is used to convert the exponential of the operators in Eq.~\eqref{eq:opPool2d} into quantum circuits.

The results from preparing wavepackets  on a $5\times 5$ lattice using ADAPT-VQE are shown in Fig.~\ref{fig:2DWP}b).
Wavepacket parameters $\sigma_{}=0.5$ and $\vec{k}_0=(0.4\pi,0.4\pi)$, and four sets of couplings
$\{ (g_x,g_z,m) \} = \{ (3.3,0.2,3.27) , (3.25,0.15,2.77) , (3.15,0.1,2.17) , (3.15,0.05,1.48) \}$ with decreasing mass gap $m$ are chosen.
By minimizing the energy, the solution converges to the desired wavepacket, as seen by the decreasing infidelity with the exact wavepacket.
Like in all the other examples, convergence is slower for smaller mass gaps.
The initial operators chosen by ADAPT-VQE are ordered: one site, two sites nearest-neighbor, three sites in an `L' configuration and three sites in a line configuration.
This is a result of the hierarchy in correlations (short-distance correlations are more significant than long-distance ones) present in gapped systems.

\subsection{Quantum circuits for simulating scattering in one-dimensional Ising field theory}
\label{sec:qcirc_scatt}
\noindent
To reduce finite-size effects, the system size $L$ must be much larger than the spatial extent of the wavepackets $\sigma_{}^{-1}$.
Because of this, it is pragmatic to set amplitudes in the $|W(k_0)\rangle$ wavefunction below some threshold to zero (and then normalize the wavefunction).
This defines a wavepacket size $d$ that depends on $\sigma_{}$ but is independent of $L$.
The initial state is of the form
\begin{align}
|\psi_{\text{ansatz}}\rangle \ \sim \ |0\rangle^{\otimes (L-d)/2} \ \otimes \ |W(k_0)\rangle \ \otimes \ |0\rangle^{\otimes (L-d)/2} \ ,
\label{eq:psiinit}
\end{align}
and has errors that can be exponentially suppressed by increasing $d$. 
A wavepacket size of $d=22$ is shown to have negligible truncation errors for $\sigma_{}=0.13$ (see App.~\ref{app:systematics}), and will be used throughout this section. 
Roughly speaking, the ADAPT-VQE circuit $\hat{U}(\vec{\theta}_\star)$ that minimizes the energy acts as
\begin{align}
\hat{U}(\vec{\theta}_\star)\left (|0\rangle^{\otimes (L-d)/2}  \otimes |W(k_0)\rangle \otimes |0\rangle^{\otimes (L-d)/2}\right ) \ \rightarrow \ \left (|\psi_{\text{vac}}\rangle  \otimes |\psi_{\text{wp}}\rangle \otimes |\psi_{\text{vac}}\rangle\right ) \ .
\label{eq:psiinit2}
\end{align}
This expression is only schematic, as the wavefunction on the RHS does not factorize between vacuum and wavepacket components.
However, sufficiently far from the region where $|W(k_0)\rangle$ was constructed, local observables are indistinguishable from their vacuum expectation value.
This is due to the exponential decay of correlations inherent to one-dimensional gapped systems~\cite{Hastings:2005pr}. 
As a result, the vacuum can be approximately prepared from
\begin{align}
\hat{U}(\vec{\theta}_\star) |0\rangle^{\otimes L}  \ \approx \ |\psi_{\text{vac}}\rangle \ .
\label{eq:ADAPTWPvac}
\end{align}
In App.~\ref{app:57Adapt} it is shown that this approximately-prepared vacuum is of even higher quality than the corresponding wavepacket.
All results presented have the energy density of this approximate vacuum subtracted, as defined in Eq.~\eqref{eq:vacsubEn}. 
This includes the results from scattering simulations, which have the energy density of the time-evolved approximate vacuum subtracted. 

\begin{figure}[t]
    \centering
    \includegraphics[width=0.75\linewidth]{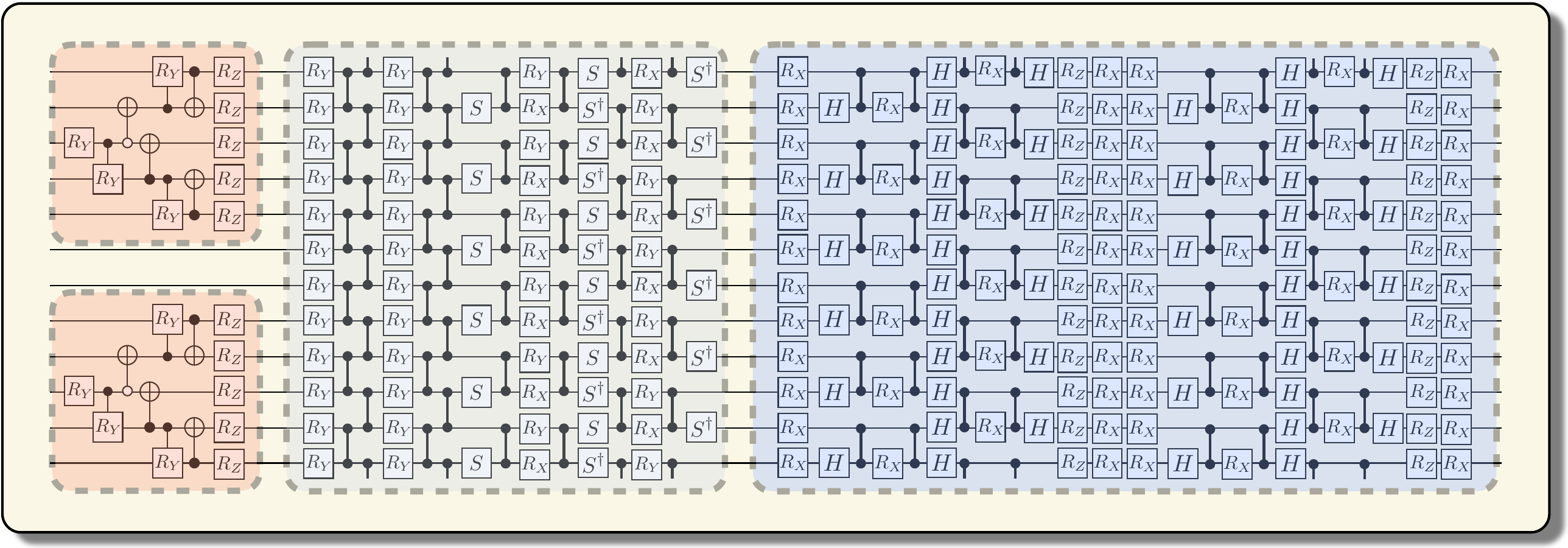}
    \caption{\textit{Circuits for simulating scattering in one-dimensional Ising field theory.}
    A circuit that simulates the scattering of two $d=5$ wavepackets on a $L=12$ lattice.
    The pieces of the circuit highlighted in light red prepare $|W(k_0)\rangle$ and $|W(-k_0)\rangle$, the layer in gray represents the symmetry-preserving energy minimization circuit $\hat{U}(\vec{\theta}_*)$ and the layer in blue is two steps of second-order Trotterized time evolution $\hat{U}_2(t)$.}
    \label{fig:scatcirc}
\end{figure}

The structure of the circuit that is used to simulate scattering in one-dimensional Ising field theory is shown in Fig.~\ref{fig:scatcirc}.
First, two wavepackets are prepared with opposite momenta.
This is done by initializing $|W(k_0)\rangle$ and $|W(-k_0)\rangle$ at the desired locations (light red), and then acting with the circuit that minimizes the energy, $\hat{U}(\vec{\theta}_\star)$ (gray).
Note that the $\hat{U}(\vec{\theta}_\star)$ for preparing  $|\psi_{\text{ansatz}}(k_0)\rangle$ works equally well for preparing $|\psi_{\text{ansatz}}(-k_0)\rangle$ since 
\begin{equation}
| \psi_{{\rm ansatz}}(-k_0)\rangle \ = \ \left (\hat{U}(\vec{\theta}_\star)|W(k_0)\rangle\right )^* \ = \ \hat{U}(\vec{\theta}_\star)|W(-k_0)\rangle \ . 
\end{equation}
The first equality follows from time-reversal symmetry, and the second equality follows from using real unitaries in ADAPT-VQE. Lastly, steps of second-order Trotterized time evolution are applied (blue).
%

\subsection{MPS simulations of scattering in one-dimensional Ising field theory}
\label{sec:csimscatt}
\noindent
In this section, the circuit elements in Fig.~\ref{fig:scatcirc} are used to simulate scattering in one-dimensional Ising field theory with MPS.
The simulation parameters are: $L=256$,\, $\sigma_{}=0.13$,\, $d=22$, and $g_x=1.25,\,g_z=0.15$, with corresponding mass gap $m=1.59$ determined from finite-size extrapolations (see App.~\ref{app:kinematics}). 
The circuits that minimize energy and prepare wavepackets are determined by implementing ADAPT-VQE on a MPS simulator.\footnote{It is important to set optimization bounds on the variational parameters so that the optimizer does not explore highly-entangled states that are not well represented by MPS.}

\begin{figure}[t]
    \centering
    \includegraphics[width=0.5\linewidth]{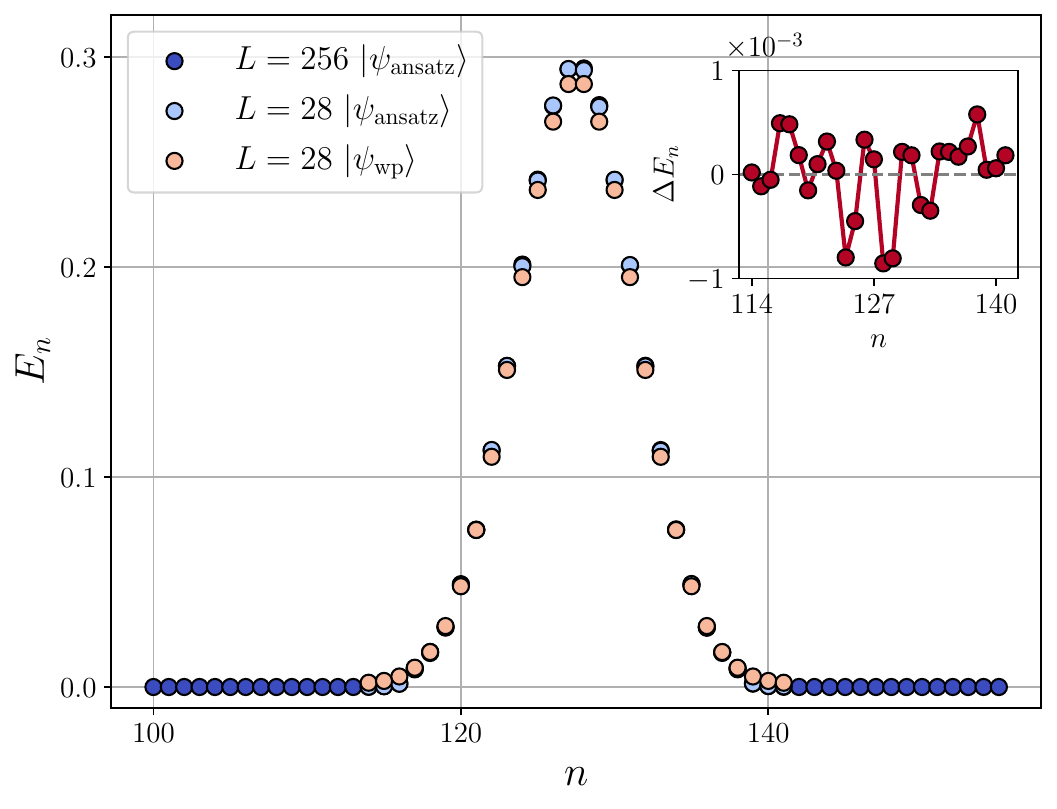}
    \caption{
    \textit{Comparison of wavepackets prepared on large and small lattices in one-dimensional Ising field theory.} 
    The vacuum-subtracted energy density $E_n$ of a wavepacket with $g_x=1.25$,\, $g_z=0.15$,\, $k_0=0.36\pi$,\, $\sigma_{}=0.13$ and $d=22$.
    The approximate wavepackets $|\psi_{\text{ansatz}}\rangle$ are prepared from 8 steps of ADAPT-VQE on a $L=256$ lattice (dark blue) and a $L=28$ lattice (light blue).
    The energy density of the exact $L=28$ wavepacket is also shown (tan).
    The inset shows the difference in the energy density $\Delta E_n$ between the approximate $L=28$ and $L=256$ wavepackets.
    Only the energy density of the 56 sites in the center of the lattice is shown, and the $L=28$ wavepackets are translated to align with the $L=256$ wavepacket.}
    \label{fig:L108L28WP}
\end{figure}
In Methods~\ref{sec:IsingStatevector} the prepared wavepackets were benchmarked against the exact wavepacket for $L=28$.
Since comparison with the exact wavepacket on a $L=256$ lattice is not possible, the energy densities of wavepackets on $L=256$ and $L=28$ lattices are compared instead.
Differences in the energy density of the exact wavepackets are expected to be small due to the exponential convergence of the single-particle spectrum with increasing system size (see App.~\ref{app:kinematics}).
The energy density of wavepackets with $k_0=0.36\pi$ prepared from 8 steps of ADAPT-VQE, 
as well as the energy density of the exact $L=28$ wavepacket are shown in Fig.~\ref{fig:L108L28WP}.\footnote{Due to the separation of length scales, $L\gg d$, special care must be taken with the energy minimization in ADAPT-VQE.
When $L\gg d$, the majority of $|\psi_{\text{ansatz}}\rangle$ in Eq.~\eqref{eq:psiinit} is $|00 \ldots 0\rangle$, 
and ADAPT-VQE prioritizes circuits that locally prepare the highest-quality vacuum.
We address this problem by instead minimizing the energy density summed across an interval that is a few correlation lengths larger than $d$.
This instance of ADAPT-VQE minimizes the energy density over $28=d+6 \approx d+10m^{-1}$ sites, and slight variations in the size of the interval have negligible effects.}
The inset shows that the differences in the energy density between the two prepared wavepackets are $\Delta E_n<10^{-3}$.
This is strong evidence that the $L=256$ and $L=28$ wavepackets are of similar quality.
Additionally, the energy density of the approximate vacuum on $L=256$ sites, prepared as in  Eq.~\eqref{eq:ADAPTWPvac}, agrees to 0.01\% with the vacuum on $L=28$.
The prepared wavepackets at other momenta exhibit similar behavior.
Further verification of the ADAPT-VQE circuits obtained with MPS comes from comparing the variational parameters determined for $L=28$ and $L=256$, which agree to 4 decimal places.
The variational parameters and operators that prepare all of the wavepackets used in this section are provided in App.~\ref{app:ADAPTparam}.

A goal of our simulations is to observe a clear distinction between scattering at energies above and below inelastic threshold.
There are two stable particles for $g_x=1.25,\,g_z=0.15$: $|1\rangle$ with mass $m_1=m = 1.59$ and $|2\rangle$ with mass $m_2=2.98$.\footnote{Away from $E_8$ theory~\cite{Zamolodchikov:1989hfa,Zamolodchikov:1989fp}, a stable particle is defined to have a mass less than $2m_1$ so that decay is kinematically forbidden. 
The existence of two stable particles at $\eta_{\text{latt}} = \frac{(g_x-1)}{|g_z|^{8/15}}\approx0.69$ is consistent with Ref.~\cite{Jha:2024jan}.}
The lowest-energy inelastic threshold corresponds to two $|1\rangle$ particles colliding and producing one $|1\rangle$ particle and one $|2\rangle$ particle, i.e., the process $11\to12$.
This threshold is at a center-of-mass energy of $E_{\text{thr}} = m_1 + m_2$.
The corresponding threshold momentum can be found from inverting the single-particle dispersion relation and solving $E(k_{\text{thr}}) = E_{\text{thr}}/2$.
In App.~\ref{app:kinematics}, the dispersion relation is computed exactly for $L=\{16,17,\ldots,28\}$ and shown to be exponentially converged in system size.
The predicted threshold momentum is $|k_{\text{thr}}|\approx0.24\pi$.
However, since our wavepackets are obtained via variational energy minimization, their energy is strictly higher than that of the exact wavepacket.
Because of this, the actual momentum threshold is a bit lower, closer to $|k_{\text{thr}}|\approx0.22\pi$.
Additionally, the inelastic threshold is not sharp due to the spread in momentum (or equivalently in energy) coming from a finite $\sigma_{}$.

\begin{figure}
    \centering
    \includegraphics[width=\linewidth]{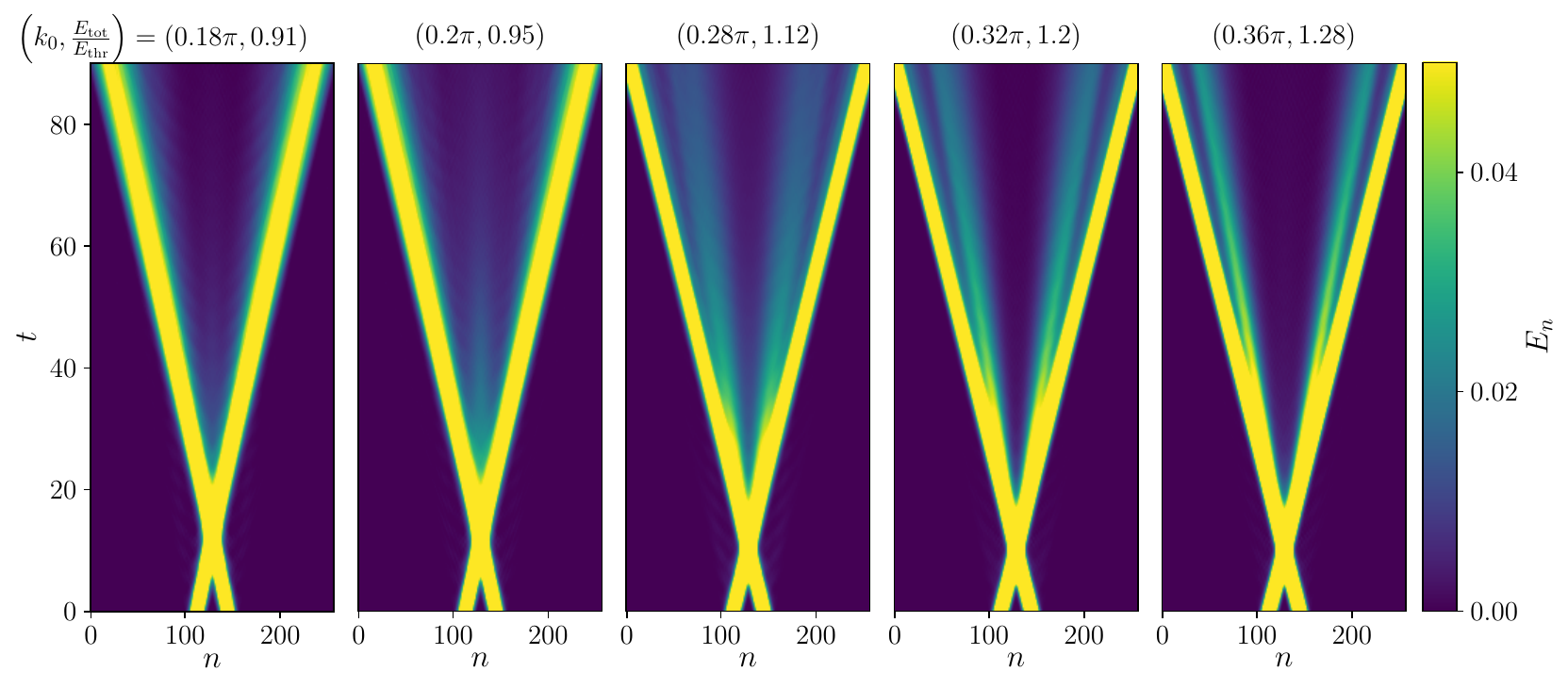}
    \caption{\textit{MPS circuit simulations of scattering in one-dimensional Ising field theory.}
    The vacuum-subtracted energy density $E_n$ throughout the scattering process for a selection of center-of-mass energies $E_{\text{tot}}$.
    The initial wavepackets are separated by 10 sites and evolved with a Trotter time step of $\delta t = 1/16$.
    The wavepackets are constructed from 8 steps of ADAPT-VQE except for $k_0=0.2\pi$ which is constructed from 10 steps.
    Results for $L=256$ and $\sigma_{}=0.13,\, d=22$,\, $g_x=1.25$,  $g_z=0.15$ and max bond dimension 350 are shown.}
    \label{fig:2WP_scattering_MPS}
\end{figure}
Results from MPS circuit simulations of scattering are shown in 
Fig.~\ref{fig:2WP_scattering_MPS}. 
These simulations compute the vacuum-subtracted energy density $E_n$ in Eq.~\eqref{eq:vacsubEn} as a function of lattice position and simulation time.
Each plot is labeled by the total energy $E_{\text{tot}}$ and momentum $k_0$ of the wavepackets.
The elastic process $11\to11$ is identified from outgoing particles with the same velocity as the incoming particles.
As shown schematically in Fig.~\ref{fig:elastic_vs_inelastic}, the inelastic process $11\to12$ converts some of the kinetic energy to mass to produce the heavier $|2\rangle$ particle.
As a result, the outgoing particles travel slower than the incoming ones.
The lowest momentum, $k_0=0.18\pi$, is well below inelastic threshold, and only elastic scattering is observed. 
Scattering at $k_0=0.2\pi$ reveals a very faint track in the middle of the lattice that does not propagate.
This corresponds to $11\to12$ where the post-collision particles are at rest.
Increasing the momentum further causes the left- and right-moving particle $|2\rangle$ tracks to be more identifiable, as shown in the plots for $k_0\geq0.28\pi$.
The kinematics are such that the particle $|1\rangle$ tracks in $11\to12$ are hidden behind the elastic scattering trajectories.
It is important to remember that expectations are averaged over the whole scattering wavefunction, which, in the case of $k_0\leq0.36\pi$, is a superposition of components with at most two particles.
This discussion has ignored three-particle production that can occur for $E_{\text{tot}}>3m_1=1.04 E_{\text{thr}}$, as it has a much smaller branching ratio compared to $11\to12$~\cite{Jha:2024jan}. See App.~\ref{app:kinematics} for a discussion of the other inelastic processes.
The MPS simulations used a maximum bond dimension of ${\tt max\_bond}=350$ and the energy density was converged to $.01\%$.

\subsection{Details on the simulations performed using IBM's quantum computers}
\label{sec:qsimDetails}
\noindent 
The simulations performed on IBM's quantum computers in Sec.~\ref{sec:qsim} use parameters that are chosen to maximize the clarity of the inelastic effects in the results. 
The interplay between the parameters is nontrivial, and our simulation choices were informed by the following considerations:
\begin{itemize}
    \item OBCs: OBCs are chosen instead of PBCs due to increased flexibility in the lattice-to-qubit mapping. See next bullet point.
    \item $L=104$: this was the largest lattice that could be mapped to {\tt ibm\_marrakesh} while  maintaining a maximum $CZ$ gate error of $ 7.5\times10^{-3}$.
    \item $\delta t=0.55$: this was the largest Trotter step size that did not have significant Trotter errors. See App.~\ref{app:systematics} for a quantification of the Trotter errors. 
    \item $t_{\text{max}}=24.75$ ($n_T=45$ Trotter steps): this was the maximum simulation time that could be reached before the results were overwhelmed by device errors.
    \item $k_0=0.32\pi$: informed by the results of MPS scattering simulations shown in Fig.~\ref{fig:2WP_scattering_MPS}, we determined that a wavepacket momentum of $k_0=0.32\pi$ has the strongest signal of inelastic scattering at $t_{\text{max}}=24.75$.
    \item $\sigma_{}=0.13$: the wavepacket extent in momentum space
    was chosen to minimize the wavepacket's spatial size and its spreading under time evolution. 
    Wavepacket spreading is most significant for low-energy, elastic scattering, e.g., $k_0=0.18\pi$ in Fig.~\ref{fig:2WP_scattering_MPS}.
    Clearly resolving elastic and inelastic scattering, i.e., distinguishing wavepacket delocalization and particle production,
    requires $\sigma_{}\leq0.13$.
    This is shown in App.~\ref{app:systematics}.
    \item $d=21$: this is the smallest spatial extent of $|W(k_0)\rangle$ that could be chosen for $\sigma_{}=0.13$ while keeping truncation effects small, see App.~\ref{app:systematics}. 
    \item No MCM-FF: the unitary $|W(k_0)\rangle$ preparation circuit in Fig.~\ref{fig:IsingWPCircs} was significantly less noisy when run on {\tt ibm\_marrakesh} than the constant-depth circuit in Fig.~\ref{fig:ConstantDepth} that utilizes MCM-FF.
    \item 7 steps of ADAPT-VQE: The quality of the prepared wavepacket significantly improves at 7 steps.
    \item 2 site wavepacket separation:
    this is the smallest separation that keeps interactions between the initial wavepackets negligible and preserves the $E_n = E_{L-1-n}$ reflection symmetry.

\end{itemize}

\begin{figure}
    \centering
    \includegraphics[width=0.8\linewidth]{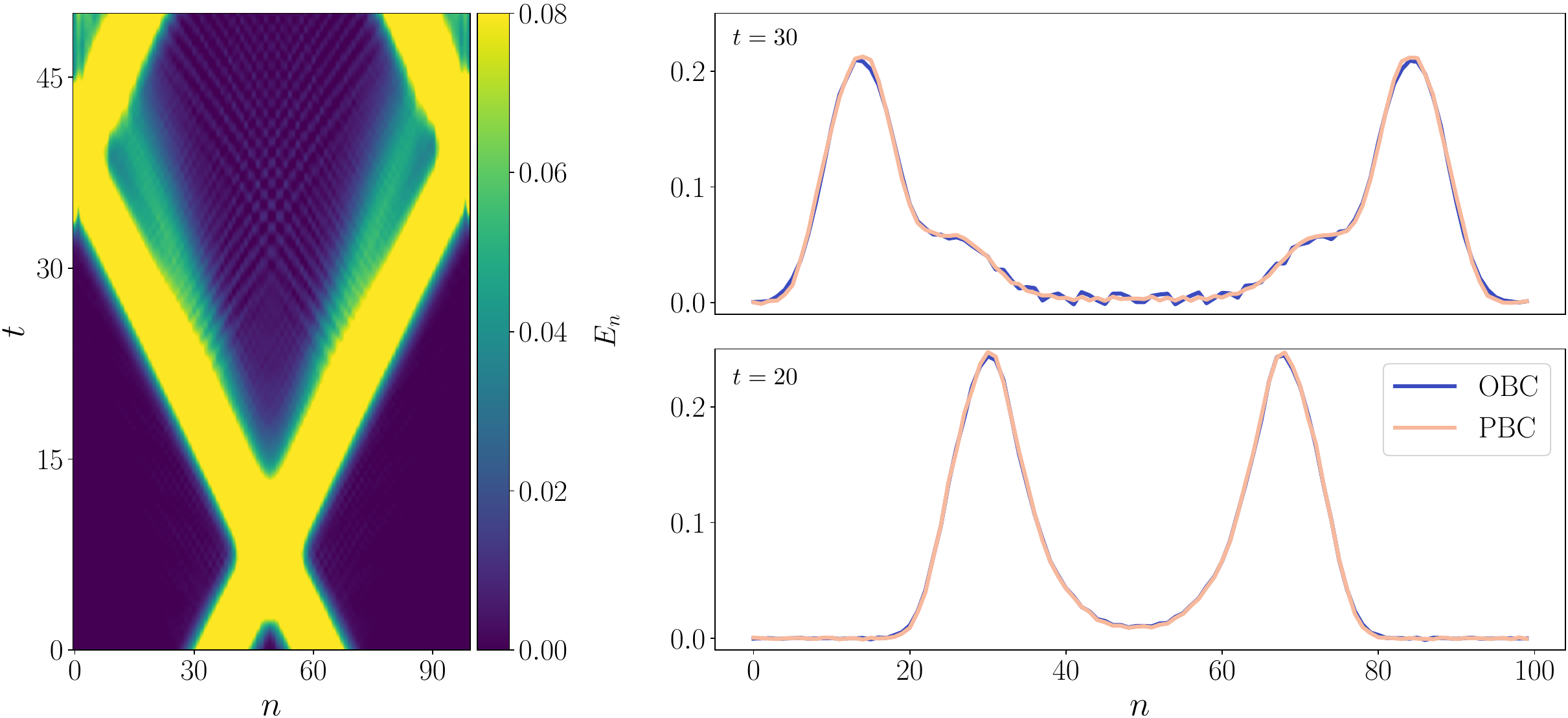}
    \caption{\textit{Comparison of OBCs and PBCs in one-dimensional Ising field theory.} Left: the vacuum-subtracted energy density $E_n$ throughout a MPS simulation of scattering on a $L=100$ lattice with OBCs.
    Wavepacket parameters $\sigma_{}=0.13,\,k_0=0.36\pi$ and a time step of $\delta t=1/16$ are used. 
    Right: the energy density at $t=20$ and $t=30$ for OBCs (blue) and PBCs (tan).}
    \label{fig:PBCOBC}
\end{figure}
The wavepacket preparation algorithm presented in this work relies on having a Hamiltonian that is block-diagonal in momentum space.
This is a consequence of translational symmetry, which is broken in a system with OBCs.
However, with OBCs, there is still an approximate translational symmetry in the bulk. 
For this reason, the quantum circuits that prepare PBC wavepackets can also be used to prepare wavepackets in a system with OBCs, if they act sufficiently far from the boundaries.
The initial state of the quantum simulations in Sec.~\ref{sec:qsim} has wavepackets localized $\gtrsim 30$ sites from boundary. 
This is much larger than the correlation length and therefore boundary effects may be ignored.
The circuits that minimize the energy and prepare wavepackets are determined in a system with PBCs to preserve the momentum content established by $|W(k_0)\rangle$.
The wavepacket with OBCs is prepared using the same circuit as for PBCs, but with all elements corresponding to Pauli strings that couple $q_0, q_{L-1}$ removed.
This causes the local vacuum near the boundaries to be of slightly lower quality than the rest of the state.
Time evolution is then implemented with the OBC Ising field theory Hamiltonian,  Eq.~\eqref{eq:HIFT} without the $\hat{Z}_0 \hat{Z}_{L-1}$ term.

The effects of OBCs on the quantum simulations in Sec.~\ref{sec:qsim} are illustrated in Fig.~\ref{fig:PBCOBC}.
The left plot shows a MPS simulation of the energy density throughout the inelastic scattering process in a $L=100$ system with OBCs.
The quality of the vacuum near the boundaries is lower than in the bulk, causing small perturbations to propagate inwards.
At $t\approx 20$, these perturbations interact with the post-collision state and generate small ripples in the energy density between the outgoing particles.
This is shown in more detail in the right plot of Fig.~\ref{fig:PBCOBC} which compares the energy density in a OBC and PBC simulation.
At $t=20$ the energy density is nearly identical and there are almost no boundary effects.
At $t=30$, fluctuations coming from the boundaries can be identified in the energy density between the particles.
These small fluctuations could be mitigated by adding boundary operators to the ADAPT-VQE operator pool to improve the vacuum near the edges~\cite{Farrell:2023fgd}.

The structure of the circuit that simulates scattering in Ising field theory is shown in Fig.~\ref{fig:scatcirc}. 
The initial state is prepared using the unitary circuit whose structure is shown in the left panel of Fig.~\ref{fig:IsingWPCircs}.
The symmetry-preserving circuits that minimize the energy and prepare wavepackets are determined by implementing ADAPT-VQE with a MPS circuit simulator.
This is described in Methods~\ref{sec:csimscatt}, and the operator ordering and variational parameters that minimize the energy are given in App.~\ref{app:ADAPTparam}.
The corresponding circuits used in ADAPT-VQE are shown in the right panel of Fig.~\ref{fig:IsingWPCircs}.
Time evolution is implemented using $n_T$ second-order Trotter steps with ordering $\{R_X, R_{Z}, R_{ZZ}, R_X\}$.
The total circuit depth for the  simulations is
\begin{equation}
\text{two-qubit gate depth: }\ 22 \ + \ 18 \ + \ 2n_T \ ,
\label{eq:circuitDepth}
\end{equation}
with the terms corresponding to the preparation of $|W(k_0)\rangle$ with $d=21$, 7 steps of ADAPT-VQE and $n_T$ steps of Trotterized time evolution. The total circuit depth and number of two-qubit gates used in the quantum simulations are provided in Table~\ref{tab:device_run_params}.

IBM's {\tt heron} quantum computers support fractional gates \cite{IBM_Quantum_Computing_Blog_2024}, which include a native $R_{ZZ}(\theta)=e^{-i\theta\hat{Z}\hat{Z}/2}$ for $\theta \in [0,\pi/2)$.
Trotterized time evolution requires $e^{i \delta t \hat{Z}\hat{Z}}$ with the opposite sign of $\theta$.
This could be overcome by sandwiching the $R_{ZZ}$ gates with $\hat{X}$ to flip the sign of $\theta$ at the cost of two additional single-qubit gates per $R_{ZZ}$. 
To eliminate the single-qubit gate overhead, we instead swap the positions of the wavepackets and evolve backwards in time.
Time reversal symmetry ensures that this simulates exactly the same scattering process.
By using $R_{ZZ}$ instead of $CZ$ gates, the two-qubit gate depth of each Trotter step is reduced from 4 to 2.
At the beginning of each run, we execute a suite of $R_{ZZ}$ calibration circuits to tune the rotation angles and reduce coherent errors. 

At the end of the simulations, the energy density in Eq.~\eqref{eq:vacsubEn} is measured and compared to the expected, noise-free, value obtained by simulating the same circuits with the Quimb MPS simulator.\footnote{With OBCs, the energy density $\hat{H}_n$ in Eq.~\eqref{eq:HIFT} on sites 0 and $L-1$ contains only a single $\hat{Z}\hat{Z}$ term without the $1/2$ prefactor.}
A relatively large $\delta t=0.55$ causes the bond dimension to become quite large to maintain accuracy.
Our MPS simulations required a bond dimension up to ${\tt max\_bond}=2250$ to be well converged at $t=40$.\footnote{Increasing the maximum bond dimension from 2000 to 2250 only changes the energy density by $8\times 10^{-4}$.}
In comparison, ${\tt max\_bond}=350$ was sufficient for the MPS scattering simulations shown in Fig.~\ref{fig:2WP_scattering_MPS} because they used a much smaller Trotter step size of $\delta t=1/16$. 

\subsubsection{Error mitigation}
\begin{figure}
  \begin{minipage}[c]{.5\linewidth}
    \includegraphics[width=0.4\linewidth]{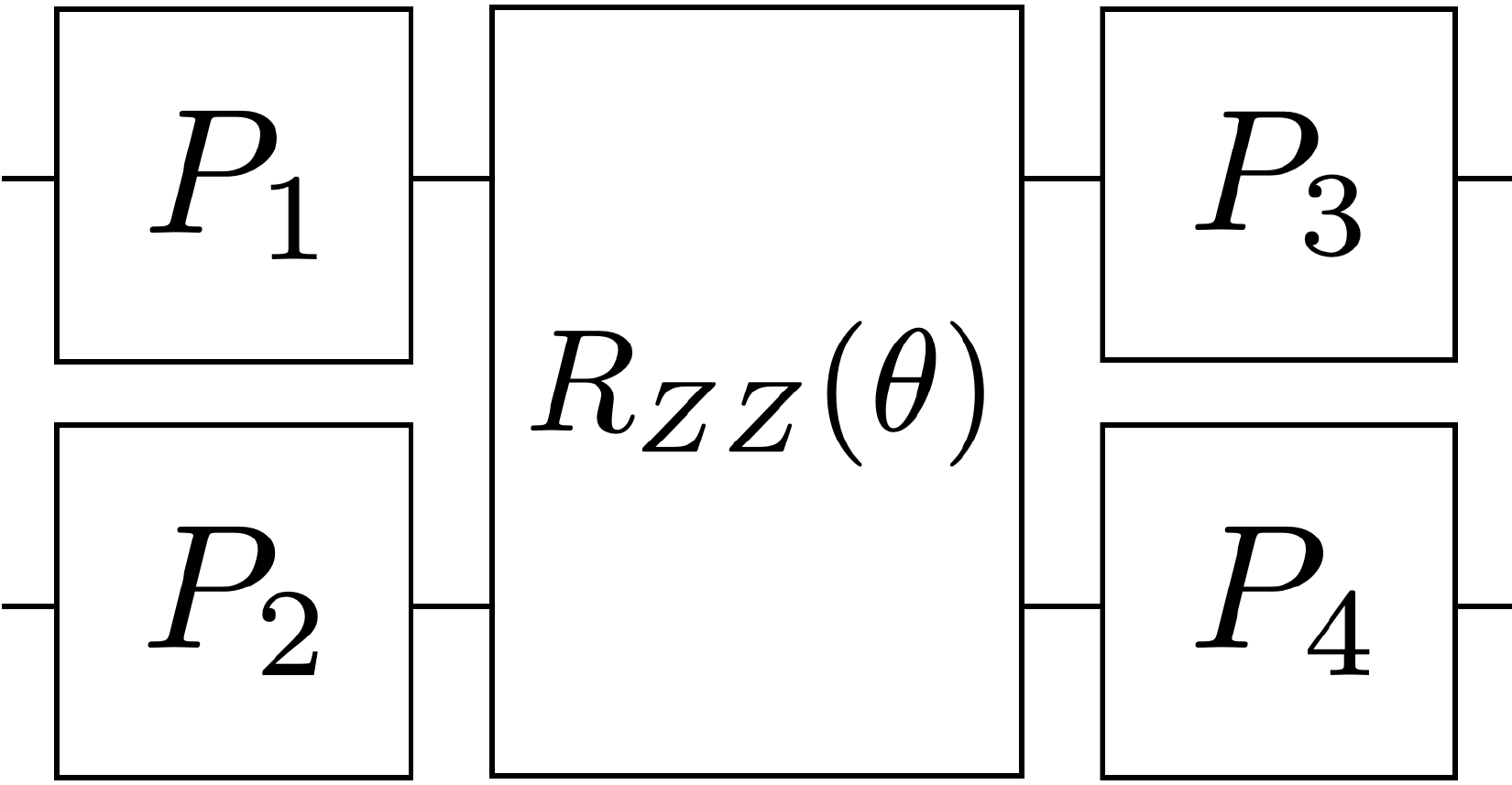}
  \end{minipage}\hfill
  \begin{minipage}[c]{.5\linewidth}
    \centering
    \renewcommand{\arraystretch}{1.4}
    \begin{tabularx}{\linewidth}{|c||Y|Y|Y|Y|Y|Y|Y|Y|}
    \hline
    $P_1$ & $I$ & $X$ & $Y$ & $Z$ & $X$ & $Y$ & $Z$ & $I$ \\\hline
    $P_2$ & $I$ & $X$ & $Y$ & $Z$ & $Y$ & $X$ & $I$ & $Z$ \\\hline
    $P_3$ & $I$ & $X$ & $Y$ & $Z$ & $X$ & $Y$ & $Z$ & $I$ \\\hline
    $P_4$ & $I$ & $X$ & $Y$ & $Z$ & $Y$ & $X$ & $I$ & $Z$ \\\hline
    \end{tabularx}
    \renewcommand{\arraystretch}{1}
  \end{minipage}
  \caption{\textit{Pauli Twirling of the $R_{ZZ}(\theta)$ gate.} The Pauli gates $P_i$ are appended before and after $R_{ZZ}(\theta)$ as shown on the left. The table on the right shows the sets of $P_i$ that leave $R_{ZZ}(\theta)$ invariant for all $\theta$. }
  \label{fig:rzz_twirls}
\end{figure}

\noindent
A suite of error mitigation methods allows accurate predictions of observables to be made from noisy device data.  
Dynamical decoupling (DD) \cite{Viola:1998jx,Ezzell_2023} removes the effects of idle noise and crosstalk between qubits. 
Pauli Twirling (PT) \cite{Wallman:2015uzh} is applied to every two-qubit gate to convert coherent errors into stochastic noise. 
The 16 combinations of Paulis are used to twirl CZ gates, while the 8 combinations given in Fig.~\ref{fig:rzz_twirls} are used for $R_{ZZ}$ (a larger set compared to the one used in Ref.~\cite{Kim:2021gvc}).  
The complete Pauli basis for twirling $R_{ZZ}(\theta)$ is not available because it is a non-Clifford operation for arbitrary $\theta$ .
Twirled Readout Error eXtinction (TREX) \cite{Berg:2020ibi} is used to mitigate measurement errors.
In the limit of infinite PT and TREX, the noise is characterized by a Pauli channel~\cite{Wallman:2015uzh}, 
\begin{align}
    \rho \ \rightarrow \ \sum_i p_i \hat{P}_i \rho \hat{P}_i \ ,
    \label{eq:pauli_channel}
\end{align} 
where the sum runs over all $4^L$ Pauli operators $\hat{P}_i$ and $\sum_i p_i = 1$ with $p_i\geq0$.
Under this channel, expectation values of observables $\hat{O}$ are given by 
\begin{align}
    \langle \hat{O}\rangle_\text{meas} \ = \ \sum_i p_i \text{Tr}(\hat{P}_i\rho\hat{P}_i\hat{O}) \ ,
    \label{eq:pauli_channel_expectation_value}
\end{align}
For Pauli observables, $\hat{P}_i\hat{O}\hat{P}_i=\pm\hat{O}$, and this channel scales the measured values $\langle \hat{O}\rangle_\text{meas}$ relative to their predicted (noise-free) counterparts $\langle \hat{O}\rangle_\text{pred}$: $\langle \hat{O}\rangle_\text{meas} = p_{\hat{O}}\langle \hat{O}\rangle_\text{pred}$.
The signal strength $|p_{\hat{O}}|\leq1$ characterizes how much an observable is impacted by the Pauli noise channel.

Expectation values of local observables are then estimated using Operator Decoherence Renormalization (ODR) \cite{Farrell:2023fgd,Farrell:2024fit,Urbanek:2021oej,ARahman:2022tkr}.
For each observable $\hat{O}$, the signal strength $p_{\hat{O}}$ is determined using a ``mitigation" circuit whose noise profile is similar to the original ``physics" circuit, but whose output can be efficiently computed with classical computers. 
In this work, the Trotterized time evolution of the vacuum state, $\hat{U}_2(t)|\psi_\text{vac}\rangle$
is used for the mitigation circuit.
With exact state preparation and time evolution, $e^{-i\hat{H} t}|\psi_\text{vac}\rangle$ is trivial since $|\psi_\text{vac}\rangle$ is an eigenstate of $\hat{H}$.
In our simulations, errors coming from the approximate preparation of the vacuum in Eq.~\eqref{eq:ADAPTWPvac} are negligible.
However, Trotter errors are large and cause the energy density to fluctuate up to $15\%$ during time evolution (see App.~\ref{app:systematics}).
These fluctuations can be mitigated by enforcing energy conservation in post-processing, see App.~\ref{app:energy_rescale} for more details.

The observables evaluated in the time-evolved (approximate) vacuum are calculated via MPS.
For early times, this is efficient in one dimension due to the correspondence between ground states of gapped systems and area law entanglement~\cite{Hastings:2007iok}.
The signal strength is computed from  the mitigation circuit results, 
\begin{align}
    p_{\hat{O}} \ = \ \frac{\langle \hat{O}\rangle_\text{meas}}{\langle \hat{O}\rangle_\text{pred}} \ = \ \frac{\langle \psi_\text{vac}|\hat{U}^{\dagger}_2(t) \hat{O} \hat{U}_2(t)|\psi_\text{vac}\rangle_\text{meas}}{\langle \psi_\text{vac}|\hat{U}^{\dagger}_2(t) \hat{O} \hat{U}_2(t)|\psi_\text{vac}\rangle_\text{pred}} \ .
    \label{eq:odr_p_j}
\end{align}
This $p_{\hat{O}}$ is then used to rescale $\langle \hat{O}\rangle_\text{meas}$ in the physics circuit and estimate its noise-free value. 
The $p_{\hat{O}}$ are also able to identify device runs with anomalously large amounts of noise. 
We filter out observables with $p_{\hat{O}}<0.01$~\cite{Farrell:2024fit}, and observables that involve a qubit with readout error $\epsilon_{\text{readout}}>0.035$.
Additionally, there is a parity symmetry that equates the energy density of sites that are related by reflection about the collision point.
For our simulations this implies $ E_n = E_{L-1-n}$.
By combining the uncorrelated measurements of observables related by this symmetry, our number of shots is effectively doubled.

\begin{table}[t]
\centering
\renewcommand{\arraystretch}{1.4}
\begin{tabularx}{0.9\linewidth}{|c||c|Y|Y|Y|Y|} \hline
\makecell{$t$} & $n_T$ & \makecell{\# of two-qubit gates} & \makecell{Two-qubit gate depth} & \makecell{\# of PTs}  &\makecell{\# of shots} \\\hline\hline
0 & 0 & 954 & 40 & 40 &$2.56\times10^6$ \\\hline
8.25 & 15 & 2,499 & 70 & 40 &$1.28\times10^6$ \\\hline
16.5 & 30 & 4,044 (4,022) & 100 & 40 (80) & $1.28\times10^6$ ($2.56\times10^6$)\\\hline
24.75 & 45 & 5,589 (5,567) & 130 & 80 (160) & $2.56\times10^6$ ($5.12\times10^6$)\\\hline
\end{tabularx}
\renewcommand{\arraystretch}{1}
\caption{\textit{Resources used in the quantum simulations performed on 104 qubits of {\tt ibm\_marrakesh}.} For a given simulation time $t$ (first column), the number of Trotter steps $n_T$ is given in the second column. 
The total number of two-qubit gates and corresponding two-qubit gate depth is given in columns three and four, respectively.
The total number of Pauli twirls and shots per simulation time, including all error mitigation overhead, are given in column five and six, respectively. The quantities in parenthesis correspond to the single wavepacket simulations.}
\label{tab:device_run_params}
\end{table}
A total of 16 TREX twirls, each with 500 shots, is used to mitigate the measurement errors for every PT, and the number of PTs run for each of the physics and mitigation circuits are given in Table~\ref{tab:device_run_params}. 
Additionally, measurements must be performed in both the $X$- and $Z$-bases to determine the energy density.
For the $t_3=24.75$, this gives a total of $500\times16\times80\times2\times2=2.56\times10^6$ shots. 
It is important that the PT and TREX twirls are applied identically for each pair of physics and mitigation circuits, so that their noise profiles are as similar as possible.\footnote{The mitigation circuits also work for TREX calibration, and no additional circuits are needed to mitigate readout errors.}
The initial state used for mitigation, $|\psi_\text{vac}\rangle$, is prepared by acting the wavepacket ADAPT-VQE circuit on the $|000...\rangle$ state, see the discussion around Eq.~\eqref{eq:ADAPTWPvac}.
It is desirable to have maximal similarity between the structure and noise profiles of the physics and mitigation circuits.
This is accomplished by preparing $|000...\rangle$ with the $|W(k_0)\rangle$ circuit in Fig.~\ref{fig:scatcirc}, but with the angle of the first $R_Y$-gate set to zero and the second CNOT changed from a control on $|0\rangle$ to a control on $|1\rangle$.
An average over TREX twirls is taken to compute an expectation value for each observable for each PT. 
This is then used as input to the rest of the error mitigation pipeline.
The expectation values $\langle\hat{Z}_n\rangle$, $\langle\hat{X}_n\rangle$, and $\langle\hat{Z}_n\hat{Z}_{n+1}\rangle$ are computed separately then added together to form the energy density.

\begin{figure}
    \centering
    \includegraphics[width=\linewidth]{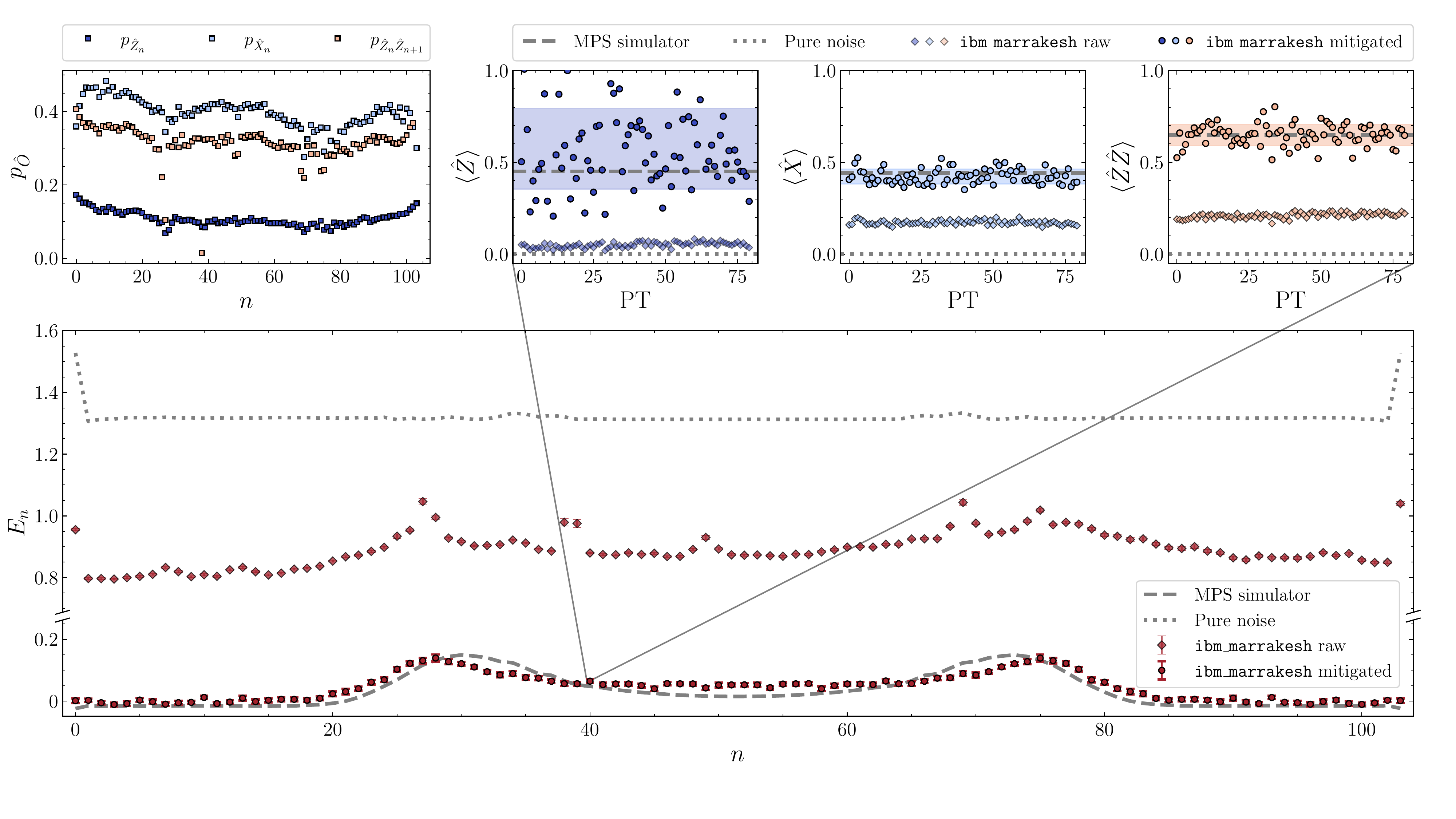}
    \caption{{\it The effect of error mitigation on the quantum simulation results from {\tt ibm\_marrakesh}.} 
    The data corresponds to the quantum simulation results in Fig.~\ref{fig:ibm_results} for $L=104$ and $t_3=24.75$ (130 two-qubit gate depth).
    Top left: the median signal strengths $p_{\hat{O}}$ defined in Eq.~\eqref{eq:odr_p_j} as a function of lattice position $n$ for $\hat{O}=\{ \hat{Z}_n,\,\hat{X}_n,\,\hat{Z}_n\hat{Z}_{n+1} \}$. 
    The signal strength is computed for each $n$ by averaging over the TREX twirls, and then computing the median over the set of PTs. 
    Top right: the effect of ODR on the expectation values of the same $\hat{O}$ for one lattice site, $n=40$. The data points are determined by averaging over TREX twirls for each PT. The gray dotted lines represent the expectation value of a completely decohered state, while the gray dashed lines are the MPS predictions. The colored bands represent $\pm1$ standard deviation of the mitigated results.  
    Bottom: the vacuum-subtracted energy density $E_n$ of the raw device data (diamonds) and after all error mitigation (circles). 
    The raw device data represents the median over PTs (averaged over TREX twirls). 
    The error bars for both the raw and the mitigated data are obtained via bootstrap resampling.}
    \label{fig:error_mitigation}
\end{figure}

The raw and error-mitigated results are compared in Fig.~\ref{fig:error_mitigation} for $t_3=24.75$. 
The raw data from the device is close to the limit of pure depolarizing noise, which is far from the expected values determined from MPS.
The results after error mitigation are significantly improved, and lie close to MPS expectations.
There exist points in the bottom panel of Fig.~\ref{fig:error_mitigation} where the device results disagree with the MPS expectations by several standard deviations, despite all error mitigation techniques.
This is particularly noticeable in the regions containing the outgoing particles, highlighting a weakness of the chosen mitigation circuits for ODR. 
The time evolution of the vacuum used as the reference state in ODR, $U_2(t)|\psi_\text{vac}\rangle$, does not capture effects of state-dependent noise on the wavepackets. 

The signal strengths for $\langle\hat{Z}_n\rangle$, $\langle\hat{X}_n\rangle$, and $\langle\hat{Z}_n\hat{Z}_{n+1}\rangle$ at $t_3=24.75$ are shown in the top left panel of Fig.~\ref{fig:error_mitigation}.
The signal strength for $\hat{Z}$ is consistently $\sim\!3\times$ lower than for the other observables.
This indicates that the weights $p_i$ in the Pauli noise channel Eq.~\eqref{eq:pauli_channel}, are larger for the Paulis that anticommute with $\hat{Z}_n$.
The top right panels of Fig.~\ref{fig:error_mitigation} show both the raw and ODR-predicted estimates for the different observables at a single site $n=40$.
The extra noise in $\langle\hat{Z}_n\rangle$ leads to significantly larger variations in the ODR-predicted $\langle\hat{Z}_n\rangle$ across Pauli twirls compared to the other observables.
These large variations are a major cause of uncertainty in the final results, shown in the bottom panel of Fig.~\ref{fig:error_mitigation}. 
We additionally computed $\langle\hat{X}_n\hat{X}_{n+1}\rangle$ and found that it is as noisy as $\langle\hat{Z}_n\rangle$. 
More extensive simulations are needed to identify the origin of this asymmetry.
One cause could be the incomplete twirling of the $R_{ZZ}$ gate (only 8 combinations of the 16 are possible).
Another could be the imbalance of the couplings $g_x=1.25$, $g_z=0.15$.

\subsection{Calculating skewness}
\label{sec:skew}
\noindent
The inelastic process $11\to12$ probed by the quantum simulations presented in Sec.~\ref{sec:qsim} produces a heavy $|2\rangle$ particle that travels slower than the lighter $|1\rangle$ particle.
At late times, the $|2\rangle$ particle can be identified as a distinct bump in the energy density that travels slower.
Before this, but after the collision, the emergence of the heavy $|2\rangle$ particle skews the outgoing energy profile toward the point of the collision.
This skewness was identified in the data presented in Sec.~\ref{sec:qsim} and is evidence for inelastic particle production.

\begin{figure}
    \centering
    \includegraphics[width=\linewidth]{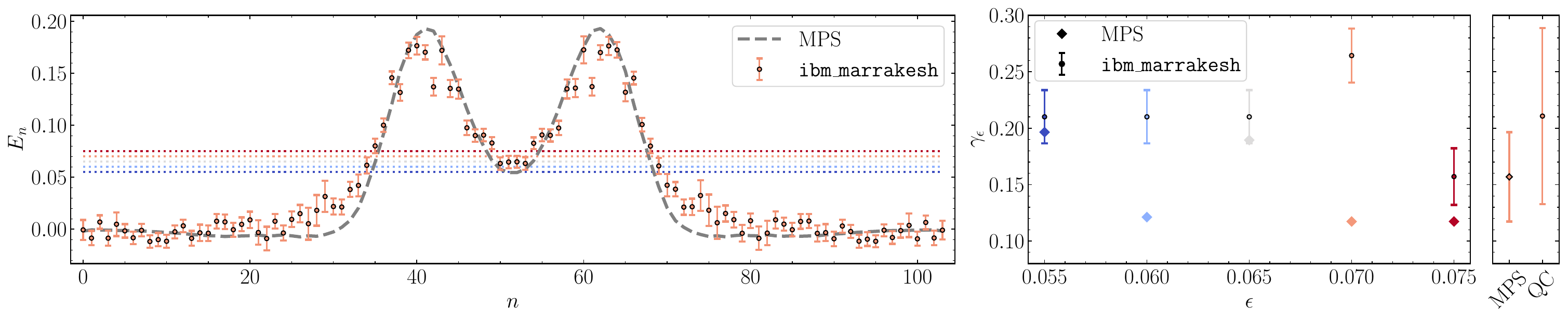}
    \caption{{\it Calculation of skewness for $t_2=16.5$.} 
    Left: the range of energy cutoffs $\epsilon$ (dotted lines) are overlaid on the energy density from Fig.~\ref{fig:ibm_results}. 
    Middle: the third moment of the energy density distribution  $\gamma_\epsilon$ is calculated for a contiguous interval of points where $E_n+\sigma_n>\epsilon$. 
    The error bars represent the standard deviation computed via bootstrap resampling. 
    Right: the reported value of $\gamma$ is computed via Eq.~\eqref{eq:gamma}, with error bars that cover the range of $\gamma_\epsilon$ coming from the variation of the energy cutoff and, for the quantum results, statistical errors from bootstrap resampling.}
    \label{fig:skewness_cutoffs}
\end{figure}

Our metric for the skewness is designed to account for the discrete nature of the data and incorporate statistical errors and device noise.
We compute the skewness of the energy density over an interval of sites $j\in [n_{\text{min}},n_{\text{max}}]$ that captures the region of positive energy density on one half of the lattice.\footnote{The energy density is symmetrized about the center of the collision as a part of error mitigation. See Methods~\ref{sec:qsimDetails} for details.} 
Given an energy cutoff $\epsilon$, the window $[n_\text{min}, n_\text{max}]$ is chosen by contiguously including sites where $E_n+\sigma_n \geq \epsilon$, where $E_n$ and its standard deviation $\sigma_n$ are computed via bootstrap resampling as detailed in Methods~\ref{sec:qsimDetails}.
The energy cutoffs are chosen to capture as much of the region of excitations as possible, while omitting outliers caused by noise. 
For each time, the energy cutoff is varied across a range given in Table~\ref{tab:skewness_cutoffs} to estimate systematic errors.
For each $\epsilon$, we compute the third moment $\gamma_\epsilon$ of the distribution of the energy density,
\begin{align}
    \gamma_\epsilon \ = \ \frac{\sum_j (j -\mu)^3  \tilde{E}_j}{\sigma_{\epsilon}^3} \ ,
\end{align}
where the normalized energy density $\tilde{E}$, mean $\mu$ and standard deviation $\sigma_{\epsilon}$ are defined as
\begin{align}
\tilde{E}_n \ = \ \frac{E_{n}}{\sum_j E_j} \ \ , \ \ \mu \ = \ \sum_{j} j\, \tilde{E}_j  \ \ , \  \ \sigma_{\epsilon}^2 \ = \ \sum_j (j -\mu)^2  \tilde{E}_j \ .
\end{align}
The sums run over all sites in the window $[n_{\text{min}},n_{\text{max}}]$.
The skewness metric $\gamma$ reported in Sec.~\ref{sec:qsim} is calculated from $\{\gamma_\epsilon\}$
\begin{align}
\gamma \ = \ \frac{\max \gamma_\epsilon + \min \gamma_\epsilon}{2}.
\label{eq:gamma}
\end{align}
A positive value of $\gamma$ corresponds to a right-skewed distribution, while a negative $\gamma$ corresponds to a left-skewed one.
Considerable variations are observed in the $\gamma_\epsilon$ values computed from different $\epsilon$. 
Reflecting this, the error bars for $\gamma$ include both the statistical errors from bootstrap resampling $\gamma_\epsilon$ and the variation of $\gamma_\epsilon$ with the cutoff $\epsilon$.
Figure~\ref{fig:skewness_cutoffs} shows an example of the choice of cutoffs for $t_2=16.5$, as well as the resulting values of $\gamma_\epsilon$ for each $\epsilon$, and the reported values of $\gamma$ along with the error bars. 

\begin{table}[t]
\centering
\renewcommand{\arraystretch}{1.4}
\begin{tabularx}{0.7\linewidth}{|c|c||Y|Y||Y|Y|} \hline
Initial state & $t$ & $\epsilon_\text{MPS}$ & $\epsilon_{\tt ibm\_marrakesh}$ & $\gamma_{\text{MPS}}$ & $\gamma_{{\tt ibm\_marrakesh}}$ \\\hline\hline
$|\psi_\text{2wp}\rangle$ & $24.75$ & $0.02-0.04$ & $0.065-0.085$ & 0.37(13) & 0.27(13) \\\hline
$|\psi_\text{1wp}\rangle$ & $24.75$ & $0.03-0.05$ & $0.05-0.07$ & 0.13(4) & 0.12(6) \\\hline
$|\psi_\text{2wp}\rangle$ & $16.5$ & $0.055-0.075$ & $0.055-0.075$ & 0.16(4) & 0.21(8)\\\hline
$|\psi_\text{1wp}\rangle$ & $16.5$ & $0.03-0.05$ & $0.05-0.07$ & 0.09(7) & 0.14(9) \\\hline
$|\psi_\text{2wp}\rangle$ & $0.0$ & $0.02-0.04$ & $0.02-0.04$ & 0.0(0)& -0.09(7)\\\hline
\end{tabularx}
\renewcommand{\arraystretch}{1}
\caption{The ranges of energy cutoffs $\epsilon$ used for calculating $\gamma$ from the energy density obtained from MPS simulations and {\tt ibm\_marrakesh}. Cutoffs in increments of $0.005$ are used.}
\label{tab:skewness_cutoffs}
\end{table}

\begin{figure}
    \centering
    \includegraphics[width=\linewidth]{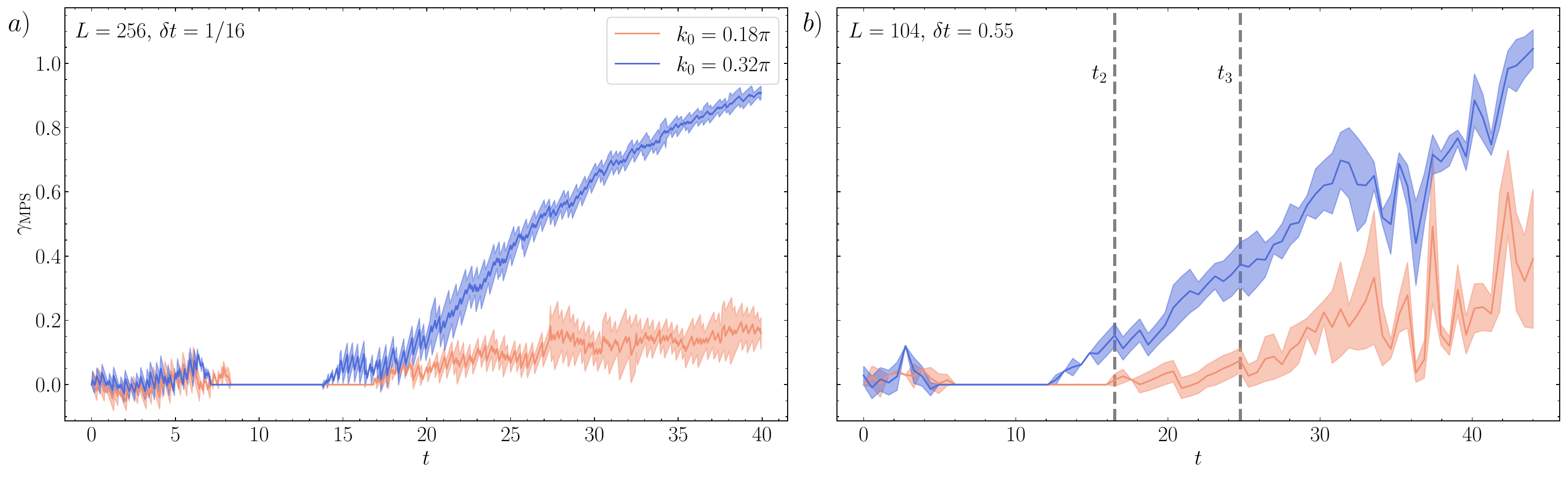}
    \caption{{\it Skewness as an indicator of inelasticity.} a) The skewness $\gamma_\text{MPS}$ calculated for MPS simulations of elastic $(k_0=0.18\pi)$ and inelastic $(k_0=0.32\pi)$ scattering with $L=256,\,\delta t=1/16,\,d=27,\,\sigma_{}=0.1$, PBCs, and max bond dimension 350.
    The energy cutoffs $\epsilon$ are chosen to be [$E_{L/2}+0.01,E_{L/2}+0.03$] in increments of $0.005$, and $\gamma_\text{MPS}$ is set to 0 if the maximum value of $E_n$ occurs at $L/2$ or if there are no points with $E_n>\epsilon$. 
    The shading represents the range of $\gamma_\text{MPS}$ coming from the variation of $\epsilon$. 
    b) The same but for $L=104,\,\delta t=0.55,\,d=21,\,\sigma_{}=0.13$, OBCs, and a max bond dimension 1500. 
    The vertical lines show the times $t_2=16.5$ and $t_3=24.75$ at which the skewness in Fig.~\ref{fig:1wp_vs_2wp} is calculated.}
    \label{fig:skewness_mps}
\end{figure}

Figure~\ref{fig:skewness_mps} compares the skewness in MPS simulations of elastic $(k_0=0.18\pi)$ and inelastic $(k_0=0.32\pi)$ scattering. 
Despite small fluctuations coming from the discrete nature of the data, a clear trend is identified distinguishing the elastic and inelastic channels.
The skewness is consistent with 0 before the collision for both momenta. After the collision $(t \approx 12-15)$, the skewness increases for inelastic scattering and stays close to zero for elastic scattering. 
This reinforces the use of skewness of the energy density as a quantitative metric to detect inelastic effects shortly after the collision.
Fig.~\ref{fig:skewness_mps}a) shows the skewness in a large system with PBCs and larger wavepackets, while b) replicates the simulations that were ran on {\tt ibm\_marrakesh} for Figs.~\ref{fig:ibm_results} and \ref{fig:1wp_vs_2wp}.
Differences between Fig.~\ref{fig:skewness_mps}a) and b) are primarily due to boundary effects and Trotter errors. 

\clearpage

\section*{Appendices}
\label{sec:appendix}
\setcounter{subsection}{0}
\renewcommand\thesubsection{\Alph{subsection}}

\subsection{A lower bound on the success probability of initializing \texorpdfstring{$|W(k_0)\rangle$}{}}
\label{app:psuccess}
\noindent
The probability that the MCM-FF protocol in Methods~\ref{sec:WKprep} succeeds is the probability of measuring odd parity in the state given in Eq.~\eqref{eq:psi0}.
This is,
\begin{align}
p_{\text{success}} \ &= \ \frac{1}{2}\left (1 - \langle(-1)^{\sum_{n=0}^{\frac{d}{2}-1} \hat{Z}_{2n+1}} \rangle\right ) \ = \ \frac{1}{2}\left (1 - \prod_{n=0}^{\frac{d}{2}-1}\langle(-1)^{ \hat{Z}_{2n+1}} \rangle\right ) \ = \ \frac{1}{2} - \frac{1}{2}\prod_{n=0}^{\frac{d}{2}-1}\left [ 1-2\delta(c_{2n}^2 + c_{2n+1}^2) \right ] \ .
\label{eq:psuccess2}
\end{align}
The second equality follows from the state being a tensor product.
It is convenient to define $p_n = \delta(c_{2n}^2+c_{2n+1}^2)$ with $\sum_{n=0}^{\frac{d}{2}-1}p_n = \delta$.
We will assume $\delta\leq 1$.
Lower bounding $p_{\text{success}}$ is equivalent to obtaining an upper bound for $\prod_{n=0}^{d/2 - 1}\left (1-2p_n\right )$.
Consider instead bounding the log,
\begin{equation}
\log\left [\prod_{n=0}^{d/2 - 1}\left (1-2p_n\right ) \right ] \ = \ 
\sum_{n=0}^{d/2 - 1}\log\left (1-2p_n\right ) .
\end{equation}
An upper bound can be obtained by noticing that that $\log(1-2p)$ is concave and therefore Jensen's inequality can be applied.
Jensen's inequality states that for concave functions $f(p)$,
\begin{equation}
\sum_n w_nf(p_n)\ \leq \  f\left [\sum_n w_n p_n\right ] \ \ , \ \ \sum_n w_n = 1 \ .
\end{equation}
Identifying $w_n = 2/d$ and $f(p) = \log(1-2p)$ gives,
\begin{equation}
\frac{2}{d}\sum_{n=0}^{d/2 - 1}\log\left (1-2p_n\right ) \  \leq \ \log\left (1-\frac{4}{d}\sum_{n=0}^{d/2 - 1} p_n\right ) \ = \  \log\left (1-\frac{4 \delta }{d}\right ) \ .
\end{equation}
Exponentiating both sides gives,
\begin{equation}
\prod_{n=0}^{d/2 - 1}\left (1-2p_n\right ) \  \leq \   \left (1-\frac{4 \delta }{d}\right )^{d/2} \ .
\end{equation}
Inserting this into Eq.~\eqref{eq:psuccess2} it is easy to show that 
\begin{equation}
p_{\text{success}}  \ \geq  \ \frac{1}{2}\left (1-e^{-2\delta} \right )  \ \geq \ 0.43 \delta
\end{equation}
for $\delta \leq 1$.

\subsection{Single-particle spectra, inelastic thresholds and wavepacket spreading}
\label{app:kinematics}
\begin{figure}[t!]
    \centering
    \includegraphics[width=0.75\linewidth]{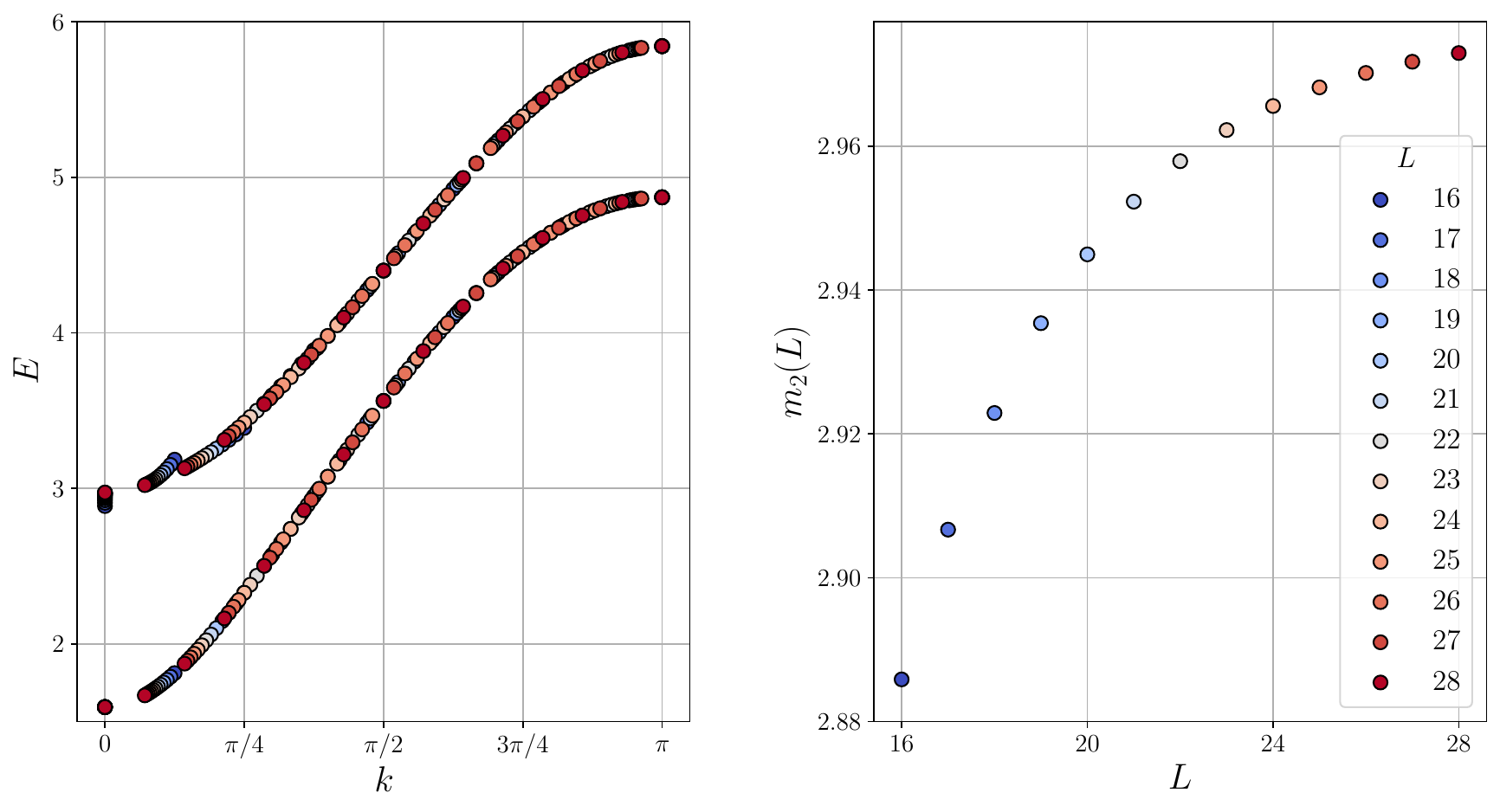}
    \caption{Left: the dispersion relation for particle $|1\rangle$ (lower) and $|2\rangle$ (upper) in one-dimensional Ising field theory determined from exact diagonalization for a range of system sizes.
    Right: the system size dependence of $m_2$.
    All results are for $g_x=1.25$ and $g_z=0.15$.}
    \label{fig:finite size}
\end{figure}

\noindent
For systems with a mass gap, the single-particle spectrum converges exponentially with increasing $L$ for sufficiently large $L$.
This is verified in the left plot of Fig.~\ref{fig:finite size} which shows the dispersion relation of particles $|1\rangle$ and $|2\rangle$ in one-dimensional Ising field theory for $g_x=1.25,\,g_z=0.15$ across a range of system sizes.
The smooth behavior of $E(k)$ for $L=\{16,17,\ldots,28\}$ provides strong evidence that $E(k)$ is well converged by $L=28$.
The finite-size effects, while small, are the most pronounced for particle $|2\rangle$ at low momentum.
The right plot of Fig.~\ref{fig:finite size} shows $m_2(L)$.
Both $m_{1,2}$ are fit well to an exponential of the form 
\begin{equation}
m_{1,2}(L) \ = \  m_{1,2}(\infty)\ + \ ae^{-b L} \ ,
\end{equation}
with $m_1(\infty) = 1.59377803(8)$ and $m_2(\infty) = 2.97682(4)$.  
The value of $m_1(\infty)$ has been confirmed by DMRG.

\begin{figure}[tb!]
    \centering
    \includegraphics[width=\linewidth]{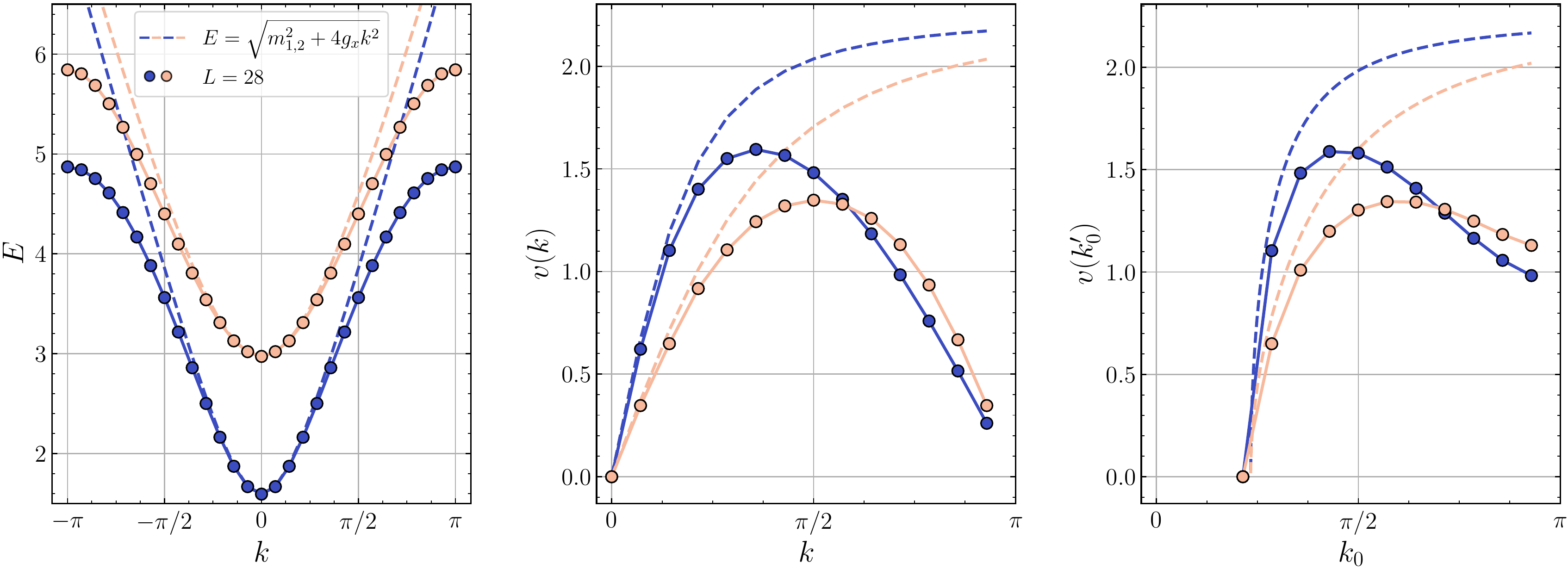}
    \caption{Left: the dispersion relation for particle $|1\rangle$ (blue) and $|2\rangle$ (tan) in one-dimensional Ising field theory with $g_x=1.25,g_z=0.15$. 
    The circles are obtained from exact diagonalization with $L=28$, and the dashed lines are from the approximation $E=\sqrt{m_{1,2}^2+4g_xk^2}$.
    Center: the corresponding group velocity $v(k)=dE/dk$. Right: the group velocity of the outgoing particles $v(k_0')$ as a function of the incoming momentum $k_0$ for the inelastic process $11\rightarrow12$.}
    \label{fig:dispersion}
\end{figure}

For arbitrary $g_x$ and $g_z$, the exact dispersion relation is unknown.
However, for $g_z=0$, the transverse field Ising model maps to a free massive Majorana fermion, and the exact dispersion relation is given by
\begin{align}
E(k) \ = \ 2\sqrt{1+g_x^2 - 2g_x\cos{k}} \ = \ \sqrt{4(g_x-1)^2 + 4g_xk^2} \ + \ {\cal O}(k^4)  \ \equiv\ \sqrt{m^2 + 4g_xk^2}\ + \ {\cal O}(k^4) \ .
\label{eq:Isingdispersion}
\end{align} 
In the second equality, the low-energy field theory limit of $k\to 0$ has been taken.
A nonzero $g_z$ will alter the dispersion relation.
An approximation that works quite well for $|k|\lesssim\pi/4$ is the RHS of Eq.~\eqref{eq:Isingdispersion}  with $m=m_{1,2}$ determined from exact diagonalization.
This is compared to the exact dispersion relation for $L=28$ in the left plot of Fig.~\ref{fig:dispersion}.
This approximation predicts a group velocity of
\begin{align}
v(k) \ = \ \frac{dE}{dk} \ = \  \frac{4g_x k}{\sqrt{m^2 + 4g_xk^2}}\ ,
\end{align}
and is compared to the lattice group velocity in the center plot of Fig.~\ref{fig:dispersion}.\footnote{The lattice derivative is computed via the discrete symmetric finite difference $f'(x) \approx \frac{f(x+\epsilon) - f(x-\epsilon)}{2\epsilon}$.}
Low momentum corresponds to ``non-relativistic" propagation with $v(k)$ increasing linearly with momentum.
At larger momentum there is a plateau in the group velocity corresponding to relativistic propagation with a ``speed of light" of $c\approx1.6$.
Further increasing the momentum leads to a decrease in the group velocity due to lattice artifacts.

Energy and total momentum are conserved throughout scattering, and schematically the inelastic process $11 \to 12$ is
\begin{align}
|1(k_0)\,\,1(-k_0)\rangle \ \to \ |1(k_0')\,\,2(-k_0')\rangle \ + \ 1\leftrightarrow2 
\end{align}
with $|k_0'| < |k_0|$ and 
\begin{align}
2E_1(k_0) \ = \ E_1(k_0') + E_2(k_0')
\label{eq:energy_condition_inelastic}
\end{align}
where $E_{1,2}(k)$ is the dispersion relation for particles $|1\rangle$, $|2\rangle$.
Energy and momentum conservation uniquely fix the velocity of the outgoing particles in the $11\to12$ process.
The outgoing velocities as functions of the incoming momenta $k_0$ are shown in the right plot of Fig.~\ref{fig:dispersion}.
Near $k_\text{thr}$, there is a large difference in the group velocities of the outgoing $|1\rangle$ and $|2\rangle$ particles.
Selecting an incoming momentum $k_0$ in this region ensures that the trajectories of the $|1\rangle$ and $|2\rangle$ particles do not overlap and can be clearly distinguished.
This guides the selection of $k_0$ that has the largest signal of inelastic effects.
The predicted velocities agree well with the MPS simulations of inelastic scattering shown in Fig.~\ref{fig:2WP_scattering_MPS}.

\begin{figure}
    \centering
    \includegraphics[width=0.8\linewidth]{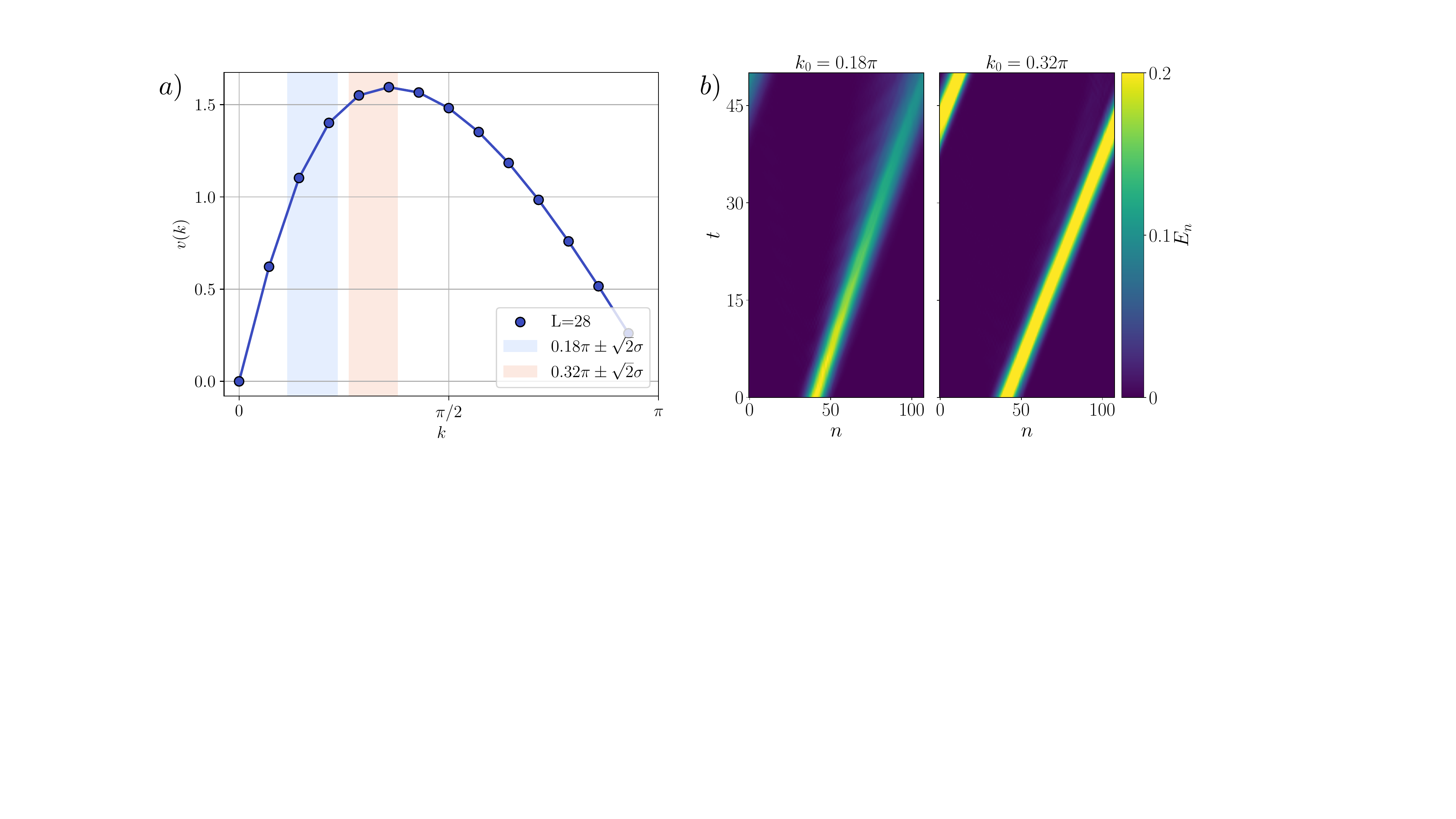}
    \caption{a) The group velocity $v(k)$ of particle $|1\rangle$ for $g_x=1.25,\, g_z=0.15$ and $L=28$ determined from exact diagonalization. 
    The shaded light blue (tan) regions correspond to the support ($\pm$ one standard deviation of the probability) of the wavepackets used to simulate elastic (inelastic) scattering.
    b) The time evolution of the vacuum-subtracted energy density $E_n$ for the propagation of a single wavepacket with $g_x=1.25$,\, $g_z=0.15$ and $L=108$.
    The wavepackets have $\sigma_{}=0.13,\, d=22$ and are constructed from 8 steps of ADAPT-VQE.}
    \label{fig:PBCWP}
\end{figure}

In Methods~\ref{sec:csimscatt}, elastic and inelastic scattering in Ising field theory were simulated at momentum $k_0 = 0.18\pi$ and $k_0=0.32\pi$ respectively.
The group velocity of the momentum modes making up the two wavepackets is shown in Fig.~\ref{fig:PBCWP}a).
The shaded regions represent the support of 68\% of the wavefunction probability ($\pm 1$ standard deviation of the gaussian wavepacket).
In these regions, the group velocity varies between $0.9\lesssim v\lesssim1.45$ and $1.5\lesssim v\lesssim1.6$ for $k_0=0.18\pi$ and $k_0=0.32\pi$ respectively.
The large variance in velocity at small momentum causes the wavepacket to spread out as it propagates.
This is illustrated in the left and right plots of Fig.~\ref{fig:PBCWP}b) which show the propagation of a single wavepacket.
The energy density of the $k_0=0.18\pi$ wavepacket delocalizes under time evolution, whereas the energy density remains focused for $k_0=0.32\pi$.
The wavepacket velocities can also be determined from these plots, and are consistent with dispersion relation estimates of $v(0.18\pi)\approx1.25$ and $v(0.32\pi)\approx1.6$.


The wavepacket parameters $k_0=0.32\pi$ and $\sigma_{}=0.13$ were used in Sec.~\ref{sec:qsim} to simulate inelastic scattering on IBM's quantum computers.
As seen in Fig.~\ref{fig:PBCWP}a), one standard deviation in momentum space includes up to $k_0=0.38\pi$.
Scattering at $k_0=0.32\pi$ and $\sigma_{}=0.13$ accesses additional inelastic processes beyond $11\to 12$.
All inelastic processes $11\to X$ with $k_{\text{thr}}\leq0.42\pi$ are given in Table~\ref{tab:thresholds}.
The thresholds have been estimated from the dispersion relation shown in Fig.~\ref{fig:dispersion}.
All of these processes occur in superposition for simulations at $k_0=0.32\pi$.
However, the higher-energy processes are suppressed due to the kinematics of the initial state (see right column of Table~\ref{tab:thresholds}).
Additionally, the $11\to111$ process has a much smaller branching ratio than $11\to12$~\cite{Jha:2024jan}. 
For these reasons, $11\to12$ is the dominant inelastic process for the simulations in this work.

\begin{table}[t]
\centering
\renewcommand{\arraystretch}{1.4}
\begin{tabularx}{0.5\linewidth}{|c||Y|Y|Y|} \hline
Process & $E_\text{thr}/m_1$ & $k_\text{thr}$ & $P(E_\text{tot}>E_\text{thr})$ \\\hline\hline
$\phantom{2}11\to12\phantom{121}$ & 2.87 & $0.24\pi$ & 0.9504 \\\hline
$\phantom{2}11\to111\phantom{21}$ & 3 & $0.26\pi$ & 0.8645 \\\hline
$\phantom{2}11\to22\phantom{121}$ & 3.75 & $0.38\pi$ & $5.92\times10^{-3}$ \\\hline
$\phantom{2}11\to112\phantom{21}$ & 3.87 & $0.4\pi$ & $7.90\times 10^{-4}$ \\\hline
$\phantom{2}11\to1111\phantom{1}$ & 4 & $0.42\pi$ & $6.99\times 10^{-5}$ \\\hline
\end{tabularx}
\renewcommand{\arraystretch}{1}
\caption{\textit{Kinematic thresholds of inelastic processes.} For each inelastic process (first column), the second column gives the threshold energy $E_\text{thr}$ in units of $m_1$, above which the process is kinematically allowed. 
The third column gives the corresponding momentum threshold $k_\text{thr}$ for a collision of two $|1\rangle$ particles. 
The last column shows the probability that the process is accessed in a collision of two $|1\rangle$ particles with $k_0=0.32\pi$ and $\sigma_{}=0.13$.}
\label{tab:thresholds}
\end{table}
%

\subsection{Additional circuits for preparing \texorpdfstring{$|W(k_0)\rangle$}{}}
\label{app:WP0prep}
\noindent
This appendix presents three additional ways of preparing the $\vert W(k_0) \rangle$ state in Eq.~\eqref{eq:psiWP02}.
These methods make use of beyond-linear connectivity and/or MCM-FF to reduce the CNOT depth relative to the circuit in the left panel of Fig.~\ref{fig:IsingWPCircs}.
Each circuit will focus on getting the amplitudes, $c_n$ in Eq.~\eqref{eq:psiWP02}, correct.
Afterwards, the phases are added with single qubit $R_Z$ rotations $\prod_n e^{-i \phi_n \hat{Z}_n /2}$.

\begin{figure}
    \centering
    \includegraphics[width=\linewidth]{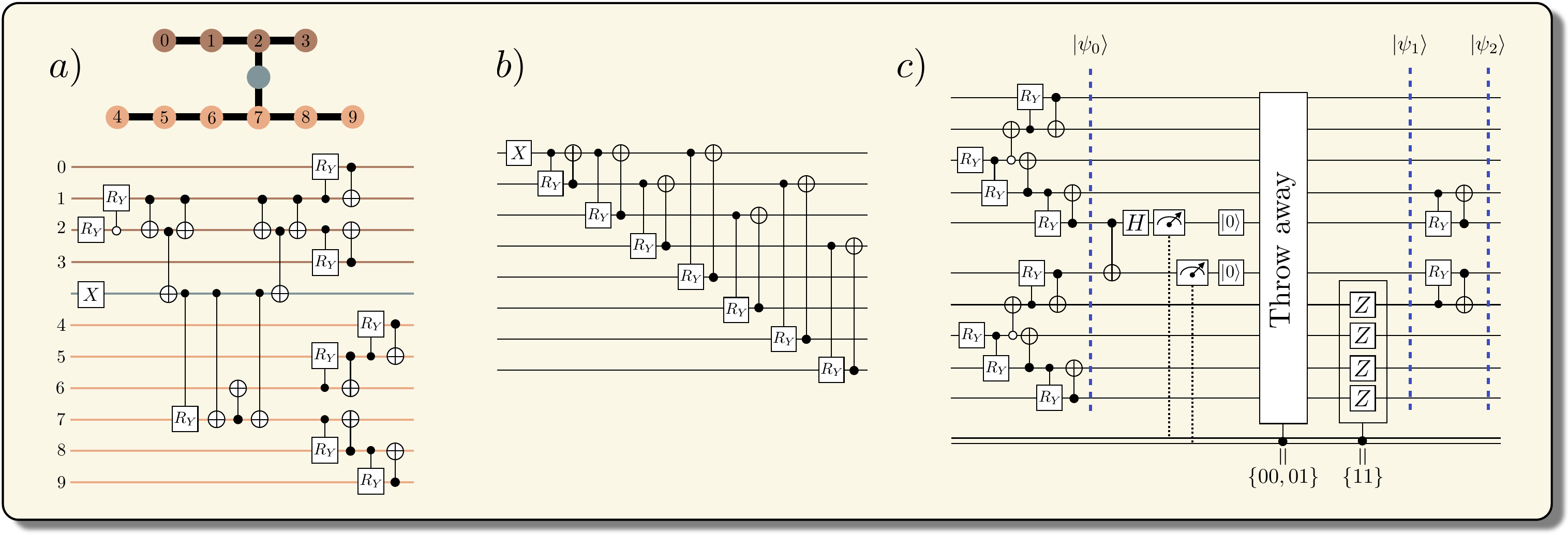}
    \caption{{\it Quantum circuits that prepare $|W(k_0)\rangle$}
    a) A circuit that uses an ancilla (gray) that is available with heavy-hex connectivity to simultaneously build $|W(k_0)\rangle$ on the dark and light brown-colored qubits.
    The qubit mapping is shown on top.
    b) A log-depth circuit that makes use of all-to-all connectivity.
    c) A circuit that makes use of MCM-FF to fuse together two smaller $|W(k_0)\rangle$ states.}
    \label{fig:WP0Prep}
\end{figure}

IBM's quantum computers natively support a heavy-hex connectivity.
A useful way to view heavy-hex is as a linear chain plus some connections between segments of the chain.
These connections can be used as ancillas to reduce the circuit depth.
A circuit illustrating this is shown in Fig.~\ref{fig:WP0Prep}a).
The idea is to split $|W(k_0)\rangle$ into two segments above and below the ancilla,
and then simultaneously build $|W(k_0)\rangle$ in both segments using the circuit in the left panel of Fig.~\ref{fig:IsingWPCircs}.
This strategy is the most efficient for $d = 2 + 4n$, with $n$ an integer. The angles for the $R_Y$ rotations are found by recursively solving the following equations,
\begin{align}
\left [\sin{\left ( \frac{\theta_{\eta}}{2}\right )} \right ]^2 \ &= \ \sum_{i=0}^{\eta-1}c^2_{\eta+i} \ , \nonumber \\
\left [\cos{\left ( \frac{\theta_{\eta}}{2}\right )}\sin{\left ( \frac{\theta_{\eta-1}}{2}\right )} \right ]^2 \ &= \ \sum_{i=0}^{\eta-1}c^2_{i} \ , \nonumber \\
\left [\sin{\left ( \frac{\theta_{3\eta+1}}{2}\right )}\cos{\left ( \frac{\theta_{\eta}}{2}\right )}\cos{\left ( \frac{\theta_{\eta-1}}{2}\right )} \right ]^2 \ &= \ \sum_{i=0}^{\eta}c^2_{3\eta+1+i} \ , \nonumber \\
\sin{\left ( \frac{\theta_{\eta}}{2}\right )}\cos{\left ( \frac{\theta_{\eta+j+1}}{2}\right )}\prod_{i=1}^j \sin{\left ( \frac{\theta_{\eta+i}}{2}\right )} \ &= \ c_{\eta+j} \ \ , \ \ j\in[0,1,\ldots,\eta-2] \ , \nonumber \\
\cos{\left ( \frac{\theta_{\eta}}{2}\right )}\sin{\left ( \frac{\theta_{\eta-1}}{2}\right )}\cos{\left ( \frac{\theta_{\eta-j-1}}{2}\right )}\prod_{i=2}^j \sin{\left ( \frac{\theta_{\eta-i}}{2}\right )} \ &= \ c_{\eta-j} \ \ , \ \ j\in[1,2,\ldots,\eta-1] \ , \nonumber \\
\cos{\left ( \frac{\theta_{3\eta+1}}{2}\right )}\cos{\left ( \frac{\theta_{\eta}}{2}\right )}\cos{\left ( \frac{\theta_{\eta-1}}{2}\right )}\cos{\left ( \frac{\theta_{3\eta-j-1}}{2}\right )}\prod_{i=1}^j \sin{\left ( \frac{\theta_{3\eta-i}}{2}\right )} \ &= \ c_{3\eta-j} \ \ , \ \ j\in[0,1,\ldots,\eta-1] \ , \nonumber \\
\sin{\left ( \frac{\theta_{3\eta+1}}{2}\right )}\cos{\left ( \frac{\theta_{\eta}}{2}\right )}\cos{\left ( \frac{\theta_{\eta-1}}{2}\right )}\cos{\left ( \frac{\theta_{3\eta+j+1}}{2}\right )}\prod_{i=2}^j \sin{\left ( \frac{\theta_{3\eta+i}}{2}\right )} \ &= \ c_{3\eta+j} \ \ , \ \ j\in[1,2,\ldots,\eta] \ ,
\end{align}
with $\eta \equiv (d-2)/4$.
The circuit depth is 
\begin{equation}
\text{CNOT depth: }\ 7+2\left \lceil \frac{d-2}{4}\right \rceil \ ,
\end{equation}
which is asymptotically half of the depth of the circuit in the left panel of Fig.~\ref{fig:IsingWPCircs}.

The state $|W(k_0)\rangle$ can be prepared with a log-depth circuit if the target connectivity is all-to-all.
This circuit is shown in Fig.~\ref{fig:WP0Prep}b) and generalizes the construction in Ref.~\cite{Cruz_2019}.
The set of equations relating the rotation angles $\{\theta\}$ of the controlled-$R_Y$ gates and the amplitudes $c_n$ can be determined from a $d\times d$ matrix $\Theta_{i,j}$.
The following pseudocode constructs this matrix:
\begin{lstlisting}[language=Python,mathescape=true]
$\Theta_{i,j}=1 \quad \forall \, i,j$
for $n$ in $[0,\ldots,d-1]$:
    for $j$ in $[0,\ldots,\min(2^n,d-2^n)-1]$:
        $\Theta_{j,2^n+j} = \cos{ \left( \frac{\theta_{2^n+j}}{2}\right) }$
        $\Theta_{2^n+j,2^n+j} = \sin{ \left( \frac{\theta_{2^n+j}}{2}\right) }$
for $n'$ in $[1,\ldots,d-1]$:        
    for $n$ in $[n',\ldots,d-1]$:
        for $j$ in $[0,\ldots,\min(2^n,d-2^n)-1]$:
            for $j'$ in $[j,j-2,\ldots,0]$:
                $\Theta_{2^n+j,j'} = \Theta_{2^{n-n'}+j-1,2^{n-n'}+j'-1}$
                $\Theta_{2^n+j,2^{n-n'}+j'} = \Theta_{j,2^{n-n'}+j'}$
\end{lstlisting}
The set of equations relating the $c_n$ and angles are found by multiplying each row, $c_i=\prod_{j=0}^{d-1}\Theta_{i,j}$.
This circuit has a circuit depth of,
\begin{equation}
\text{CNOT depth: }2\lceil \log_2(d)\rceil \ .
\end{equation}

Two mid-circuit measurements and one feedforward operation can be used to reduce the CNOT depth required to prepare $|W(k_0)\rangle$.
Inspired by the GHZ state preparation circuit in Ref.~\cite{Smith:2022nbd},
the idea is to split $|W(k_0)\rangle$ into two halves (assuming $d$ is even)
\begin{equation}
\sum_{n=0}^{d-1}  c_n |2^n\rangle  \ \rightarrow \ \left (\sum_{n=0}^{\frac{d}{2}-1}  a_n |2^n\rangle \right )\otimes\left (\sum_{n=0}^{\frac{d}{2}-1} b_n |2^n\rangle \right ) \ ,
\end{equation}
and fuse the two halves together with a Bell measurement followed by a feedforward operation.
One of the unitary constructions presented above can be used to prepare each half, and it will be shown that this method succeeds with probability $p_{\text{success}}=1/2$.

As an illustrative example, consider preparing a $d=6$ site wavepacket from fusing together two 3 site halves.
A circuit with the same structure extending this to the fusion of 5-site halves is shown in Fig.~\ref{fig:WP0Prep}c).
The input state $|\psi_0\rangle$ is
\begin{align}
\vert \psi_0\rangle\ = \  \left ( a_0|001\rangle + a_1|010\rangle + a_2|100\rangle \right )\otimes \left (b_0|001\rangle + b_1|010\rangle + b_2|100\rangle \right ) \ ,
\label{eq:mcminit}
\end{align}
with real coefficients that satisfy $a_0^2 + a_1^2 + a_2^2 = b_0^2 + b_1^2 + b_2^2 =1$.
Next, a Bell measurement is performed on qubits $q_3q_2$.
The states obtained conditioned on each possible Bell measurement outcome are given in Table~\ref{tab:Wmcm}.
\begin{table}[t]
\renewcommand{\arraystretch}{1.4}
\begin{tabular}{|c || c |c |}
\hline
     Measurement & State  & Probability \\
     \hline
     \hline
     $00$ & $a_0 b_2 |000\rangle |000\rangle  + \left ( a_1|010\rangle + a_2|100\rangle \right) \left (b_0|001\rangle + b_1|010\rangle \right )$ & $\frac{1}{2}\left (1-a_0^2-b_2^2+2a_0^2 b_2^2\right )$\\ 
     \hline 
     $01$ & $-a_0 b_2 |000\rangle |000\rangle  + \left ( a_1|010\rangle + a_2|100\rangle \right) \left (b_0|001\rangle + b_1|010\rangle \right )$ & $\frac{1}{2}\left (1-a_0^2-b_2^2+2a_0^2 b_2^2\right )$\\
     \hline 
     $10$ & $ a_0 |000\rangle \left (b_0|001\rangle + b_1 |010\rangle  \right ) + b_2\left (a_1|010\rangle + a_2 |100 \rangle \right ) |000\rangle$ & $\frac{1}{2}\left (a_0^2 + b_2^2 - 2a_0^2 b_2^2\right )$\\ 
     \hline 
     $11$ & $a_0 |000\rangle \left (b_0|001\rangle + b_1 |010\rangle  \right ) - b_2\left (a_1|010\rangle + a_2|100\rangle \right ) |000\rangle $ & $\frac{1}{2}\left (a_0^2 + b_2^2 - 2a_0^2 b_2^2\right )$\\ 
     \hline
\end{tabular}
\caption{The (unnormalized) state (middle column) obtained after a particular measurement outcome of qubits $q_2$ and $q_3$ (left column).
The state being measured is given in Eq.~\eqref{eq:mcminit} and
qubits $q_2$ and $q_3$ are reset after measurement.
The probability of each measurement outcome is given in the right column.}
\label{tab:Wmcm}
\end{table}
After the measurement, the qubits $q_3q_2$ are reset to $|00\rangle$.
The protocol has succeeded for measurement outcomes $ q_3 q_2  = \{10,11 \}$ and failed for $ q_3 q_2  = \{00, 01 \}$.
A feedforward of $\hat{Z}_4 \hat{Z}_5$ is applied if $ q_3 q_2 = \{11 \}$, and the resulting state $|\psi_1\rangle$ has the structure of $|W(k_0)\rangle$ on the unmeasured qubits,
\begin{align}
|\psi_1\rangle \ = \  \frac{1}{\sqrt{a_0^2 + b_2^2 -2a_0^2 b_2^2}}\left (a_0b_0|000001\rangle + a_0b_1 |000010\rangle   + b_2a_1|010000\rangle + b_2a_2 |100000 \rangle  \right ) \
 .
\label{eq:psi1}
\end{align}
From Table~\ref{tab:Wmcm}, the probability of success is $p_{\text{success}} = p_{10} + p_{11} = a_0^2 + b_2^2 -2a_0^2 b_2^2$.
This has a maximum of $p_{\text{success}} =1/2$ when $a_0^2 = b_2^2 = 1/2$ which can be shown to correspond to a wavepacket where $\sum_{n=0}^{\frac{d}{2}-1} c_n^2 \ = \ \sum_{n=0}^{\frac{d}{2}-1} c_{\frac{d}{2}+n}^2 $.
This condition is always satisfied for even $d$ since the magnitudes of Gaussian wavepackets are symmetric about their midpoints.
Because of this, the remaining expressions will assume $a_0^2 = b_2^2 = 1/2$.
The last step is to perform a controlled-$R_Y$ CNOT sequence between qubits $q_1 q_2$ and $q_4 q_3$, giving
\begin{align}
|\psi_2\rangle \ =& \  b_0|000001\rangle + b_1\cos{\left (\frac{\theta_2}{2}\right )} |000010\rangle + b_1\sin{\left (\frac{\theta_2}{2}\right )}|000100\rangle  \nonumber \\
&+ a_1\sin{\left (\frac{\theta_3}{2}\right )}|001000\rangle + a_1\cos{\left (\frac{\theta_3}{2}\right )}|010000\rangle + a_2 |100000 \rangle  \ ,
\label{eq:psi2}
\end{align}
where $\theta_2$ and $\theta_3$ are the value of the $R_Y$ rotation angles.
$|W(k_0)\rangle$ is prepared by identifying
\begin{align}
&b_0=c_0 \ , \quad b_1^2 = c_1^2 + c_2^2 \ , \quad  \tan{\left ( \frac{\theta_2}{2}\right )} = \frac{c_2}{c_1} \ , \nonumber \\
&a_1^2 = c_3^2 + c_4^2 \ , \quad   \tan{\left ( \frac{\theta_3}{2}\right )} = \frac{c_3}{c_4} \ , \quad  a_2 = c_5 \ .
\end{align}

This method generalizes to any even $d$ and succeeds with a probability of success $p_{\text{success}}=1/2$.
The procedure is:
\begin{itemize}
    \item[1.] Prepare the initial state $ \left (\sum_{n=0}^{\frac{d}{2}-1}  a_n |2^n\rangle  \right )\otimes \left (\sum_{n=0}^{\frac{d}{2}-1} b_n |2^n\rangle \right )$ using one of the unitary $|W(k_0)\rangle$ preparation circuits presented previously.
    The coefficients are given by
    \begin{align}
        b_n \ &= \ c_n \ , \ \  n= \{0,1,\ldots,\frac{d}{2}-3\} \ , \nonumber \\ 
        a_n \ &= \ c_{\frac{d}{2}+n} \ , \ \  n= \{2,3,\ldots,\frac{d}{2}-1\} \ , \nonumber \\
        a_{0} \ &= \ b_{\frac{d}{2}-1} = \frac{1}{\sqrt{2}} \ , \nonumber \\
        a_1^2 \ &= \ c^2_{\frac{d}{2}} + c^2_{\frac{d}{2} + 1 } \ ,  \nonumber \\
        b_{\frac{d}{2}-2}^2\  &= \ c^2_{\frac{d}{2}-1} + c^2_{\frac{d}{2} -2 } \ .
    \end{align}
    \item[2.] Measure the qubits $q_{d/2} q_{d/2 - 1}$ in the Bell basis. 
    If the measurement outcome is $q_{d/2 }q_{d/2 - 1} =\{00,01 \}$, restart the procedure.
    If the measurement outcome is $q_{d/2 }q_{d/2 - 1} =\{11 \}$,
    feed forward $\prod_{i=d/2+1}^{d-1}\hat{Z}_{i}$. If the measurement outcome is $q_{d/2 }q_{d/2 - 1} =\{10 \}$, do nothing. 
    \item[3.] Reset $q_{d/2 }q_{d/2 - 1}$ to $|00\rangle$.
    \item[4.]
    Apply a controlled-$R_Y$ CNOT sequence between $q_{d/2-2}q_{d/2-1}$ and $q_{d/2+1}q_{d/2}$. 
    The $R_Y$ rotation angles are given by 
    \begin{equation}
        \tan\left (\frac{\theta_{\frac{d}{2}-1}}{2} \right ) =\frac{c_{\frac{d}{2}-1}}{c_{\frac{d}{2}-2}} \ \ , \ \ \tan\left (\frac{\theta_{\frac{d}{2}}}{2} \right ) =\frac{c_{\frac{d}{2}}}{c_{\frac{d}{2}+1}} \ .
    \end{equation}
\end{itemize}

Table~\ref{tab:summary_W} compares the scaling of the resources required to prepare W states using the methods in this paper to other methods in the literature.

\begin{table}
\renewcommand{\arraystretch}{1.4}
\begin{tabular}{| c | | c |c | c | c |  c | c | c |}
\hline
     Reference & Qubits & Connectivity & CNOT count & CNOT depth & MCM & $p_{\text{success}}$ & ${\cal I}$ \\
     \hline
     \hline
      Ref.~\cite{Cruz_2019} & $d$ & linear & $2d$ & $2d$ & 0 & 1 & 0 \\
     \hline
     Fig.~\ref{fig:IsingWPCircs} & $d$ & linear & $2d$ & $d$ & 0 & 1 & 0 \\ \hline
     Fig.~\ref{fig:WP0Prep}c) & $d$ & linear & $2d$& $d/2$ & 2 & 1/2 & 0 \\ \hline
     Ref.~\cite{Piroli:2024ckr} & $2d$ & linear & $13d/4$ & 11 & $3d/4$ & $\geq0.43 \,\delta$ & ${\cal O}(\delta^2)$ \\ 
     \hline
     Fig.~\ref{fig:ConstantDepth} & $d$ & linear & $17d/4$& $13$ & $3d/4$ & $\geq0.43\delta$ & ${\cal O}(\delta^2)$ \\ 
     \hline
    Fig.~\ref{fig:WP0Prep}a) & $d+1$ & heavy-hex & $2d$ &  $d/2$ & 0 & 1 & 0 \\ 
     \hline
    Ref.~\cite{Cruz_2019}, Fig.~\ref{fig:WP0Prep}b) & $d$ & all-to-all & $2d$ & $2\log_2(d)$ & 0 & 1 & 0 \\ \hline
\end{tabular}
\caption{Summary of the scaling of resources (qubits, connectivity, CNOT gates, mid-circuit measurements, success probability $p_\text{success}$ and infidelity $\mathcal{I}$) needed to prepare the W state on $d$ qubits.
The values for the constant-depth circuits have been adjusted for a device with linear connectivity.}
\label{tab:summary_W}
\end{table}
%

\subsection{More details on using ADAPT-VQE to prepare wavepackets and vacua}
\label{app:57Adapt}
\noindent
The algorithm developed in this work for preparing wavepackets uses ADAPT-VQE to minimize the energy.
Our implementation of ADAPT-VQE is the following:
\begin{itemize}
    \item[1.] Define a pool of operators $\{ \hat{O} \}$ that are translationally invariant and imaginary (so that $e^{i \theta \hat{O}}$ is real).
    They should also respect the other symmetries of the Hamiltonian.
    An operator pool inspired by the Lie algebra of the Hamiltonian is found to be effective.
    \item[2.] Initialize a state with the correct amplitude and phase in each momentum block of the Hamiltonian.
    This state should also have the other relevant quantum numbers of the desired wavepacket, e.g., $Z_2$ parity in scalar field theory or electric charge in the Schwinger model.
    \item[3.] For each operator in the pool $\hat{O}_i$, optimize the variational parameters in $ e^{i \theta_i \hat{O}_i}\lvert \psi_{{\rm ansatz}} \rangle$ to minimize the energy.
    The previously optimized values for $\theta_{1,\ldots,i-1}$ and $\theta_i=0$, are used as initial conditions. 
    \item[4.] Identify the operator $\hat{O}_n$ and variational parameters $\vec{\theta}$ that had the lowest energy in the previous step.
    Update the ansatz $\lvert \psi_{{\rm ansatz}} \rangle \to e^{i \theta_n \hat{O}_n}\lvert \psi_{{\rm ansatz}} \rangle$ using the optimal parameters.
    \item[5.] Return to step 3 until the desired tolerance is achieved.
\end{itemize}
Previous applications of ADAPT-VQE have selected the operator $\hat{O}_n$ in step 4 with the largest energy gradient~\cite{Grimsley:2018wnd,Farrell:2023fgd,Gustafson:2024bww}.
We found that using the gradient leads to worse convergence compared to optimizing every operator in the pool, and choosing the operator that produces the lowest energy. The resulting algorithm is less greedy as it explores more circuits, and is seen to perform better.
This has a computational overhead that scales with the number of operators in the operator pool.
However, our operator pools are very small due to symmetry constraints, and this overhead was not a limitation.

\begin{figure}
    \centering
    \includegraphics[width=0.6\linewidth]{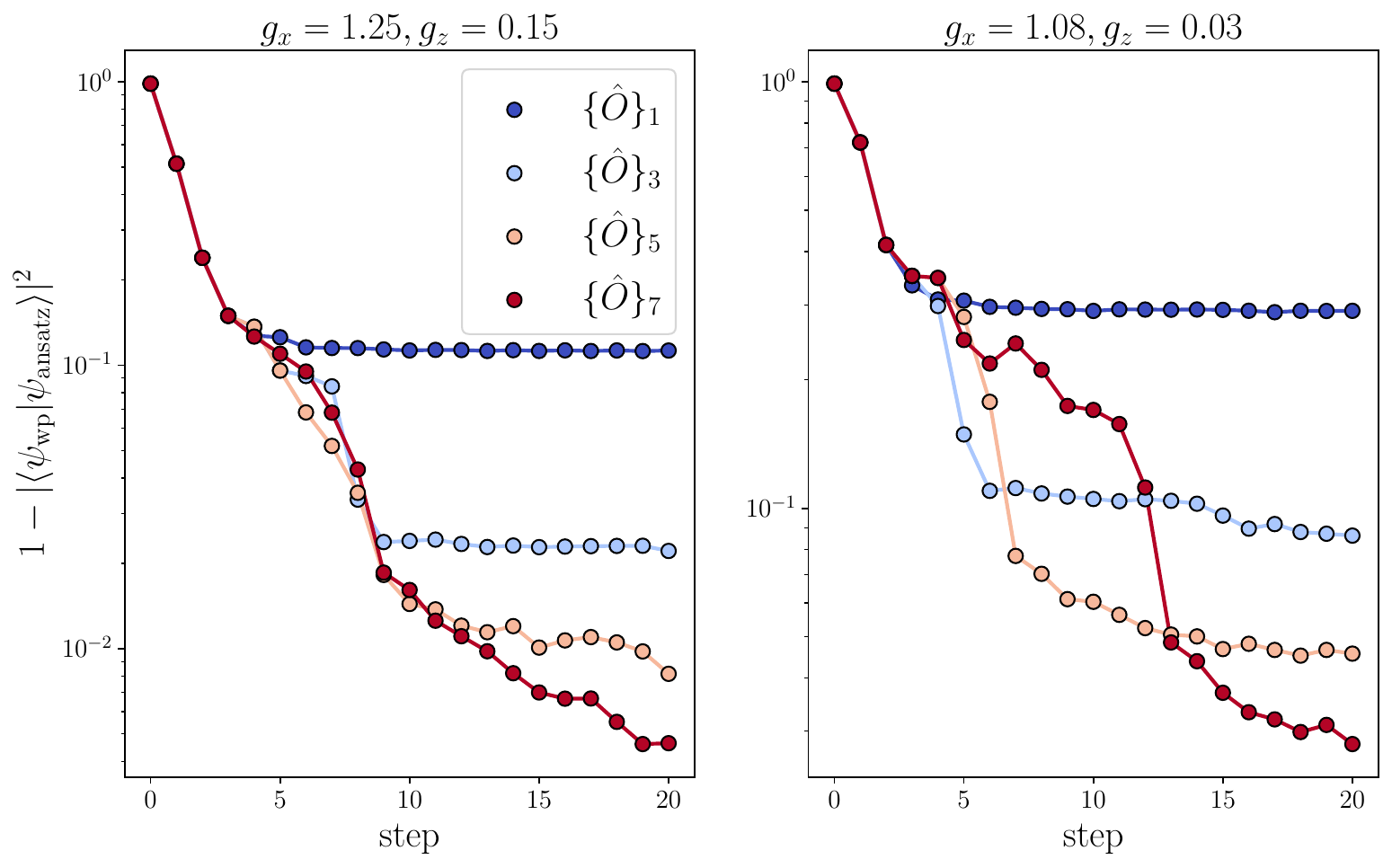}
    \caption{{\it The effect of larger ADAPT-VQE operator pools on wavepacket quality in one-dimensional Ising field theory.}
    The infidelity of wavepackets prepared with up to 20 steps of ADAPT-VQE and parameters $\sigma_{}=0.15,\,k_0=0.36\pi$.
    Results are shown for $L=24$ and using four different operators pools $\{\hat{O}\}_1,\{\hat{O}\}_3,\{\hat{O}\}_5,\{\hat{O}\}_7$, defined in Eq.~\eqref{eq:opPool1357}.
    The left (right) plot is for $g_x=1.25,\,g_z=0.15$ ($g_x=1.08,\,g_z=0.03$).}
    \label{fig:ADAPT1357}
\end{figure}
A successful application of ADAPT-VQE requires using an effective operator pool.
A pool that is too large will be difficult to optimize, partly due to the presence of many local minima in the optimization landscape.
On the other hand, a pool that is too small does not have sufficient expressivity to prepare the desired wavefunction.\footnote{This tradeoff was discussed in the context of brickwall circuits in Ref.~\cite{Zemlevskiy:2024vxt}.}
This tradeoff is explored by preparing wavepackets in one-dimensional Ising field theory using ADAPT-VQE equipped with an operator pool consisting of the unique operators in the Hamiltonian algebra up to 1st, 3rd (used in the main text), 5th and 7th order commutators,
\begin{align}
&\{ {\hat O}\}_1  =  \sum_{n=0}^{L-1} \bigg \{ \hat{Y}_n \,  , \  \left (\hat{Y}_n\hat{Z}_{n+1}+ \hat{Z}_n\hat{Y}_{n+1}\right )   \bigg \} \ , \nonumber \\
&\{ {\hat O}\}_3  =  \{ {\hat O}\}_1 \ \bigcup \ \sum_{n=0}^{L-1} \bigg \{ \hat{Z}_n\hat{Y}_{n+1}\hat{Z}_{n+2}  ,  \  \left (\hat{Y}_n\hat{X}_{n+1}+ \hat{X}_n\hat{Y}_{n+1}\right )  ,  \ \left (\hat{Z}_n\hat{X}_{n+1}\hat{Y}_{n+2}+ \hat{Y}_n\hat{X}_{n+1}\hat{Z}_{n+2}\right )  \bigg \} \ , \nonumber \\
 &\{ {\hat O}\}_5 \ = \  \{\hat{O}\}_3 \ \bigcup \ \sum_{n=0}^{L-1}\Bigg \{ \left (\hat{Z}_n\hat{Y}_{n+1}\hat{X}_{n+2}+\hat{X}_n\hat{Y}_{n+1}\hat{Z}_{n+2} \right ),
\left (\hat{Y}_n\hat{X}_{n+1}\hat{X}_{n+2}+\hat{X}_n\hat{X}_{n+1}\hat{Y}_{n+2} \right ),  \hat{Y}_n\hat{Y}_{n+1}\hat{Y}_{n+2} , \nonumber \\[4pt]
&\left (\hat{Z}_n\hat{X}_{n+1}\hat{Y}_{n+2}\hat{Z}_{n+3}+\hat{Z}_n\hat{Y}_{n+1}\hat{X}_{n+2}\hat{Z}_{n+3} \right ), 
\left (\hat{Y}_n\hat{Z}_{n+1}\hat{Z}_{n+2}+\hat{Z}_n\hat{Z}_{n+1}\hat{Y}_{n+2} \right ), \nonumber \\
&\left (\hat{Z}_n\hat{X}_{n+1}\hat{X}_{n+2}\hat{Y}_{n+3}+\hat{Y}_n\hat{X}_{n+1}\hat{X}_{n+2}\hat{Z}_{n+3} \right ) \Bigg \} \ , \nonumber \\
&\{ {\hat O}\}_7 \ = \  \{\hat{O}\}_5 \ \bigcup \ \sum_{n=0}^{L-1} \Bigg \{ \left (\hat{Z}_n\hat{I}_{n+1}\hat{Y}_{n+2}+\hat{Y}_n\hat{I}_{n+1}\hat{Z}_{n+2} \right ),
\left (\hat{Y}_n\hat{Z}_{n+1}\hat{X}_{n+2}+\hat{X}_n\hat{Z}_{n+1}\hat{Y}_{n+2} \right )\ , \nonumber \\
&\left (\hat{Z}_n\hat{Z}_{n+1}\hat{Y}_{n+2}\hat{Z}_{n+3}+\hat{Z}_n\hat{Y}_{n+1}\hat{Z}_{n+2}\hat{Z}_{n+3} \right ), \left (\hat{Z}_n\hat{X}_{n+1}\hat{Y}_{n+2}\hat{X}_{n+3}+\hat{X}_n\hat{Y}_{n+1}\hat{X}_{n+2}\hat{Z}_{n+3} \right ), \hat{X}_n\hat{Y}_{n+1}\hat{X}_{n+2}, \nonumber \\
&\left (\hat{Y}_n\hat{Y}_{n+1}\hat{Y}_{n+2}\hat{Z}_{n+3}+\hat{Z}_n\hat{Y}_{n+1}\hat{Y}_{n+2}\hat{Y}_{n+3} \right ),
\left (\hat{Z}_n\hat{Y}_{n+1}\hat{X}_{n+2}\hat{X}_{n+3}+\hat{X}_n\hat{X}_{n+1}\hat{Y}_{n+2}\hat{Z}_{n+3} \right ), \nonumber \\
&\left (\hat{Y}_n\hat{X}_{n+1}\hat{X}_{n+2}\hat{X}_{n+3}+\hat{X}_n\hat{X}_{n+1}\hat{X}_{n+2}\hat{Y}_{n+3} \right ),
\left (\hat{Z}_n\hat{X}_{n+1}\hat{X}_{n+2}\hat{Y}_{n+3}\hat{Z}_{n+4}+\hat{Z}_n\hat{Y}_{n+1}\hat{X}_{n+2}\hat{X}_{n+3}\hat{Z}_{n+4} \right ), \nonumber \\
&\left (\hat{Y}_n\hat{X}_{n+1}\hat{Z}_{n+2}\hat{Z}_{n+3}+\hat{Z}_n\hat{Z}_{n+1}\hat{X}_{n+2}\hat{Y}_{n+3} \right ),
\left (\hat{Y}_n\hat{X}_{n+1}\hat{Y}_{n+2}\hat{Y}_{n+3}+\hat{Y}_n\hat{Y}_{n+1}\hat{X}_{n+2}\hat{Y}_{n+3} \right ), \nonumber \\
&\left (\hat{Z}_n\hat{Y}_{n+1}\hat{I}_{n+2}\hat{Z}_{n+3}+\hat{Z}_n\hat{I}_{n+1}\hat{Y}_{n+2}\hat{Z}_{n+3} \right ),\left (\hat{Z}_n\hat{X}_{n+1}\hat{Z}_{n+2}\hat{Y}_{n+3}+\hat{Y}_n\hat{Z}_{n+1}\hat{X}_{n+2}\hat{Z}_{n+3} \right ),
\hat{Z}_n\hat{X}_{n+1}\hat{Y}_{n+2}\hat{X}_{n+3}\hat{Z}_{n+4} , \nonumber \\
&\left (\hat{X}_n\hat{I}_{n+1}\hat{Y}_{n+2}+\hat{Y}_n\hat{I}_{n+1}\hat{X}_{n+2} \right ),\left (\hat{Z}_n\hat{X}_{n+1}\hat{X}_{n+2}\hat{X}_{n+3}\hat{Y}_{n+4}+\hat{Y}_n\hat{X}_{n+1}\hat{X}_{n+2}\hat{X}_{n+3}\hat{Z}_{n+4} \right )\Bigg \} \ .
\label{eq:opPool1357}
\end{align}

The performance of ADAPT-VQE using these different operator pools is shown in Figure~\ref{fig:ADAPT1357}.
The infidelity is given as a function of ADAPT-VQE step for $L=24$, wavepacket parameters $\sigma_{}=0.15,\,k_0=0.36\pi$ and two sets of couplings, $g_x=1.25,\,g_z=0.15$ and $g_x=1.08,\,g_z=1.03$, with mass gaps of $m=1.6$ and $m=0.7$, respectively.
For early ADAPT-VQE steps, the best performing operator pool changes from step to step.
However, for later ADAPT-VQE steps, the largest operator pool $\{\hat{O}\}_7$ always performs the best.
This is more pronounced for the couplings with a smaller mass gap, $g_x=1.08,\,g_z=0.03$, likely because the longer Pauli strings in $\{\hat{O}\}_7$ are able to more effectively build out long-range correlations.
Note that these results do not show the circuit depths, 
which will generally be larger for the higher-order commutator operator pools due to the longer Pauli strings.
We identified $\{\hat{O}\}_3$ as the best balance between convergence and circuit depth, and it was therefore used for the quantum simulations performed in this work.

\begin{figure}
    \centering
    \includegraphics[width=\linewidth]{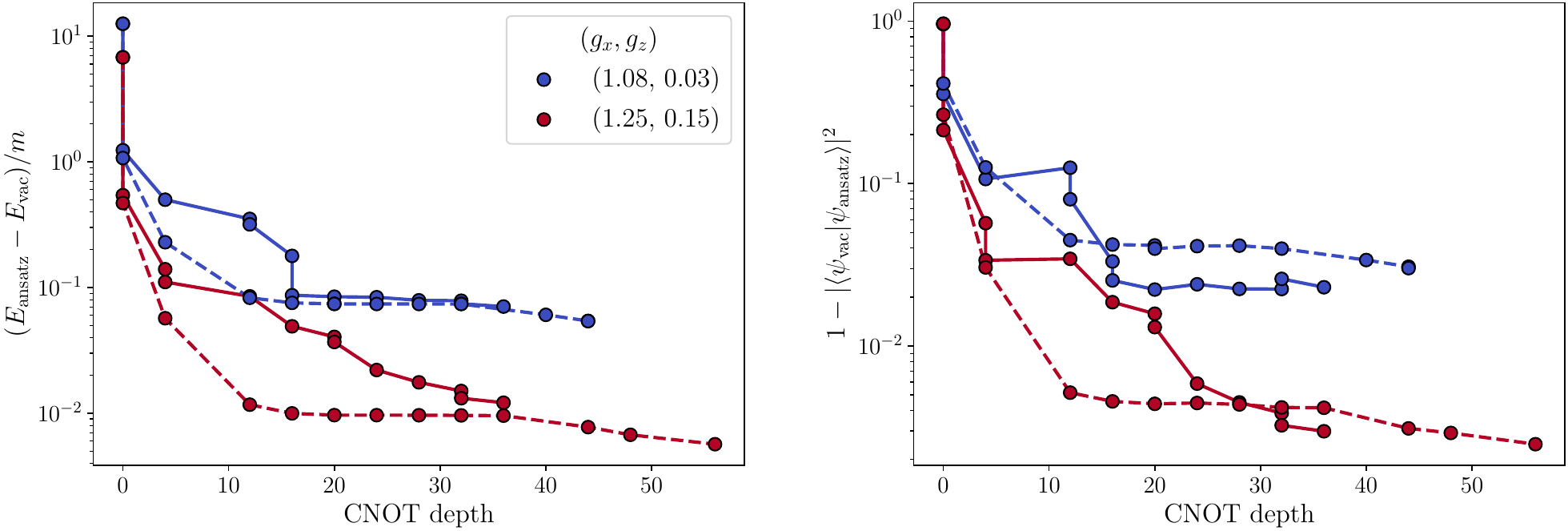}
    \caption{{\it The quality of the one-dimensional Ising field theory vacuum prepared using ADAPT-VQE.}
    The deviation in the energy (left) and infidelity (right) are shown for $L=28$ and up to 12 steps of ADAPT-VQE.
    The dashed line corresponds to the vacuum prepared using $\hat{U}_{\text{vac}}(\vec{\theta}_\star)$ in Eq.~\eqref{eq:ADAPTvac} and the solid line corresponds to $\hat{U}(\vec{\theta}_\star)$  in Eq.~\eqref{eq:ADAPTWPvac2}.
    The $\hat{U}(\vec{\theta}_\star)$ circuits are the same ones that were used in Fig.~\ref{fig:IsingADADPT_results} to prepare wavepackets with $k_0=0.36\pi,\,\sigma_{}=0.13$.}
\label{fig:IsingADADPT_VACresults}
\end{figure}

In the main text, vacuum-subtracted quantities were computed by time evolving an approximate vacuum constructed from the {\it wavepacket} ADAPT-VQE circuit,
\begin{align}
\hat{U}(\vec{\theta}_\star) |0\rangle^{\otimes L}  \ \approx \ |\psi_{\text{vac}}\rangle \ .
\label{eq:ADAPTWPvac2}
\end{align}
Alternatively, ADAPT-VQE can be performed with the goal of preparing the vacuum by starting from the state $|\psi_{\text{ansatz}}\rangle=|00 \ldots 0\rangle$.
This defines the {\it vacuum} ADAPT-VQE circuit,
\begin{align}
\hat{U}_{\text{vac}}(\vec{\theta}_\star) |0\rangle^{\otimes L}  \ \approx \ |\psi_{\text{vac}}\rangle \ .
\label{eq:ADAPTvac}
\end{align}
The results from preparing the vacuum in these two ways are shown in Fig.~\ref{fig:IsingADADPT_VACresults}.
The vacuum ADAPT-VQE circuits $\hat{U}_{\text{vac}}(\vec{\theta})$ are constructed from the operator pool in Eq.~\eqref{eq:opPool}.
Comparing the infidelities with Fig.~\ref{fig:IsingADADPT_results}, it is seen that the vacuum prepared by $\hat{U}(\vec{\theta}_\star)$ (solid line) is of higher quality than the corresponding wavepacket.
The vacuum prepared with $\hat{U}_{\text{vac}}(\vec{\theta}_\star)$ (dashed line)  initially converges quite quickly, but then plateaus after a few steps.
In comparison, the vacuum prepared with $\hat{U}(\vec{\theta}_\star)$ converges slower, but does not feature as prominent of a plateau.
Notice that the vacuum prepared with $\hat{U}(\vec{\theta}_\star)$ is, in some cases, better than with $\hat{U}_{\text{vac}}(\vec{\theta}_\star)$.
This is a result of the greedy nature of ADAPT-VQE that does not always pick the optimal operator ordering.
Comparing the two sets of couplings $(g_x,g_z,m)=\{(1.25,0.15,1.6),(1.08,0.03,0.7)\}$, it is seen that convergence to the vacuum is slower for a smaller mass gap, as expected.
The ADAPT-VQE operator ordering and variational parameters that prepare the vacuum are given in App.~\ref{app:ADAPTparam}.
%

\subsection{State preparation and Trotter errors}
\label{app:systematics}
\begin{figure}
    \centering
    \includegraphics[width=0.5\linewidth]{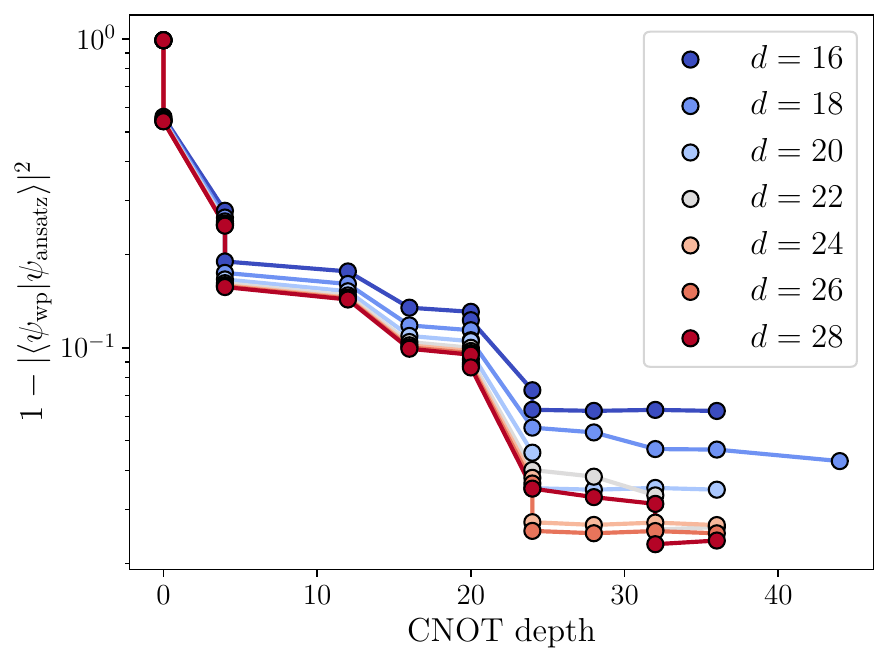}
    \caption{{\it The effect of truncating the spatial extent of $|W(k_0)\rangle$ on the quality of the prepared wavepacket in one-dimensional Ising field theory.
    }
    The infidelity of the prepared wavepacket for a selection of truncations $d$, defined in Eq.~\eqref{eq:WPd}.
    Results are shown for up to 12 steps of ADAPT-VQE with parameters $g_x=1.25,\,g_z=0.15,\,L=28,\,k_0=0.36\pi,\,\sigma_{}=0.13$.}
    \label{fig:dError}
\end{figure}
\noindent
As explained in Methods~\ref{sec:qcirc_scatt}, when the lattice size is much larger than the wavepacket width, $L\gg \sigma_{}^{-1}$, the spatial support of the
initial wavepacket $|W(k_0)\rangle$ is truncated to a spatial interval $d$.
Explicitly, for a wavepacket centered at $x_0$,
\begin{align}
|W(k_0)\rangle \ = \ \sum_{n=0}^{L-1} e^{i\phi_n}c_n |2^n\rangle \ \to  \ {\cal N}\sum_{|n-x_0|\leq d/2} e^{i\phi_n}c_n |2^n\rangle \ . 
\label{eq:WPd}
\end{align}
Results from preparing wavepackets in Ising field theory with different $d$ are shown in Fig.~\ref{fig:dError} for $L=28,\,g_x=1.25,\,g_z=0.15$ and wavepacket parameters $\sigma_{}=0.13,\,k_0=0.36\pi$.
Increasing $d$ decreases the infidelity, as expected, and for these wavepacket parameters, there are essentially no truncation effects for $d\geq20$.
This informs the choice of $d=21$ used in the quantum simulations presented in Sec.~\ref{sec:qsim}.

\begin{figure}
    \centering
    \includegraphics[width=\linewidth]{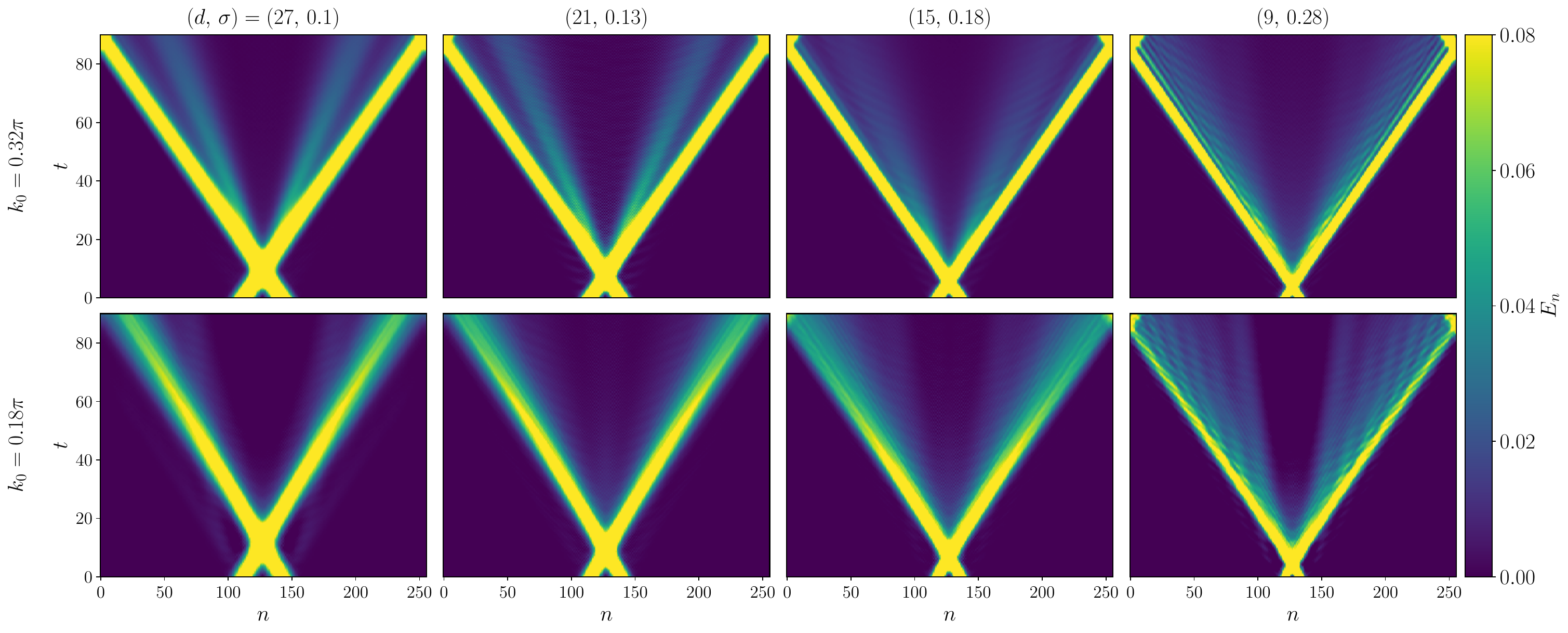}
    \caption{\textit{The effect of wavepacket size on scattering in one-dimensional Ising field theory.}
    The vacuum-subtracted energy density $E_n$ throughout MPS simulations of inelastic (top row) and elastic (bottom row) scattering is shown for an $L=256$ system with PBCs and a time step of $\delta t=1/16$. 
    The spread in momentum $\sigma_{}$ is chosen such that the wavepacket quality is the same after truncation to $d$ sites.}
\label{fig:wp_size_comparison}
\end{figure}

Larger wavepackets take longer to overlap and scatter but spread less under time evolution.
The latter is a result of variations in the group velocity, e.g., see the middle plot in Fig.~\ref{fig:dispersion}. 
At a fixed wavepacket quality, increasing $d$ allows the spread in momentum space $\sigma_{}$ to decrease.
The effect of varying $d$ on scattering dynamics is shown in Fig.~\ref{fig:wp_size_comparison}.
In all plots, $\sigma_{}$ is chosen to maintain a constant wavepacket amplitude after truncating $|W(k_0)\rangle$ to $d$ lattice sites as in Eq.~\eqref{eq:WPd}.
The top row shows the gradual washing out of the inelastic signal for scattering at $k_0=0.32\pi$ as $d$ is decreased from 27 to 9, where they are completely absent.
The bottom row shows elastic scattering for wavepackets with $k_0=0.18\pi$. 
For small $d$, interference effects from negative momenta and higher energies become significant, resulting in a checkerboard pattern in the energy density.
This prevents the elastic and inelastic channels from being reliably distinguished.
The wavepacket size $d=21$ is chosen for the quantum simulations in Sec.~\ref{sec:qsim} as a balance between resolution in momentum space and time to collision.
It is surprising that such a large hierarchy between $d$ and the correlation length $\xi\propto1/m=0.6$ is required to distinguish elastic and inelastic scattering.

The Trotter step size $\delta t$ is chosen to balance Trotter errors and circuit depth.
Smaller $\delta t$ approximates the exact time evolution more closely, but requires deeper circuits to reach a target $t_{\text{max}}$, thus incurring more device errors. 
In principle, if the device noise is characterized well, there exists an optimal $\delta t$ for each target simulation time $t$.
Examples of such exploration can be found for the digital case in Refs.~\cite{Knee:2015,Clinton:2021,Haghshenas:2025euj} and for the analog setting in Ref.~\cite{Zemlevskiy:2023eyw}.
When such knowledge is not readily available, as in this work, $\delta t$ must be chosen heuristically.
\begin{figure}
    \centering
    \includegraphics[width=\linewidth]{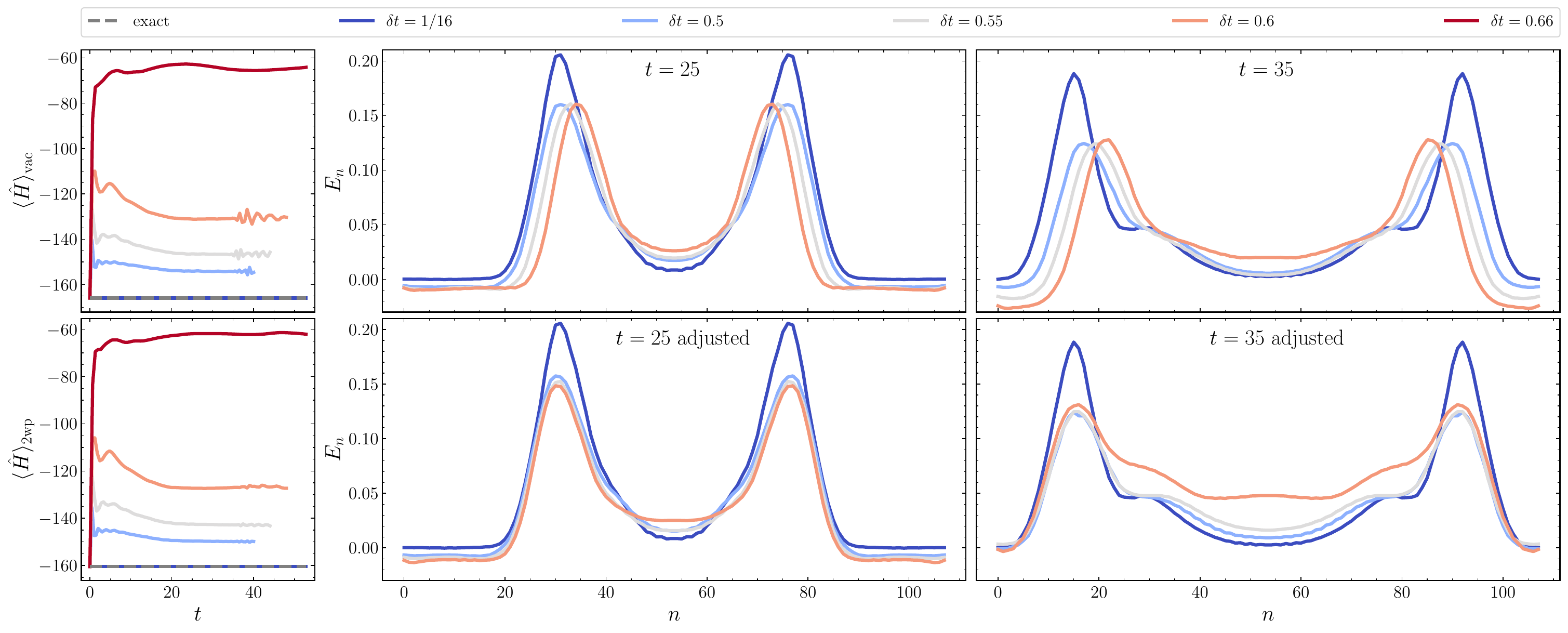}
    \caption{\textit{The effect of Trotter step size $\delta t$ on scattering in one-dimensional Ising field theory.} 
    Results are shown for a $L=108$ system with PBCs simulated using MPS with $k_0=0.32\pi$ and $\sigma_{}=0.13$. 
    Left column: the total energy of $|\psi_\text{vac}\rangle$ (top) and $|\psi_\text{2wp}\rangle$ (bottom) as a function of time $t$ for various $\delta t$. 
    Right: the top plots show the vacuum-subtracted energy density $E_n$ at $t=25$ and $t=35$. 
    The bottom plots show $E_n$ for $\delta t=1/16$  at $t=25,35$, as well as larger $\delta t$ with $t$ chosen so that the wavepacket peaks are aligned.}
\label{fig:trotter_errors}
\end{figure}

Figure~\ref{fig:trotter_errors} shows the effects of Trotter errors on the scattering process through MPS simulations.
The plots in the left column give the total energy of the vacuum and the two-wavepacket state as a function of time $t$ for various $\delta t$. 
A large $\delta t$ effectively quenches the state at $t=0$, causing a sharp change in the energy.
The system is heated to the chaotic regime for $\delta t=0.66$, where the dynamics no longer resemble those of the exact time evolution.
A sharp transition to chaotic dynamics is a generic feature of Trotterized time evolution, e.g., see Refs.~\cite{Heyl_2019,Sieberer_2019,Kargi:2021rww}.
The quench effects are also seen in the four plots on the right, where the vacuum-subtracted energy density $E_n$ is shown for $t=25,35$. 
The height of the wavepacket peaks is reduced for $\delta t > 1/16$ suggesting evolution toward a thermal state~\cite{DAlessio:2014rzv,Lazarides:2014loq,Yang:2023nak}. 

The Trotters errors cause the group velocity to be modified, and the wavepackets travel slower for larger $\delta t$.
This is seen in the top right plots of Fig.~\ref{fig:trotter_errors} for $t=25,35$, where the wavepacket peaks are closer together for larger $\delta t$, indicating slower propagation.
To compare the state when the wavepackets have propagated equal distances, the simulation time is adjusted so that the wavepacket peaks align (bottom right plots of Fig.~\ref{fig:trotter_errors}). 
The best approximation to the exact time evolution is given by the smallest Trotter step $\delta t=1/16$.
A larger $\delta t$ tends to make the energy density more uniform, and washes out the small ``bumps" that are the signatures of inelastic particle production.
The bottom right plots of Fig.~\ref{fig:trotter_errors} show the bump from the emergence of a second pair of of tracks vanishing as $\delta t$ is increased.
In addition, $E_n$ near the point of the collision does not return to the vacuum after the wavepackets have left the region.
Furthermore, a systematic offset from $E_n=0$ is seen to develop in the regions outside of the wavepackets for $\delta t>1/16$. 
This effect increases with Trotter step size and is persistent across all bond dimensions.
We attribute this to the quench from Trotterized time evolution creating a ``negative energy excitation" that travels in the opposite direction from each wavepacket starting at $t=0$.
A Trotter step size of $\delta t=0.55$ is chosen for the quantum simulations presented in Sec.~\ref{sec:qsim} as it retains signatures of inelasticities while minimizing circuit depth and trotter errors.

\subsection{Determining wavepackets with exact diagonalization}
\label{app:ExactWP}
\noindent
In Methods~\ref{sec:WPQFTcircs}, the wavepackets prepared with ADAPT-VQE are compared to the exact wavepacket determined from exact diagonalization.
The exact wavepacket is the superposition of single particle eigenstates $|\psi_k\rangle$ as in Eq.~\eqref{eq:psiWPFull}.
These $|\psi_k\rangle$ are obtained by finding the lowest energy state of the Hamiltonian projected onto the desired $k$ block.
This appendix outlines how this projection is done.

The columns of the projection matrix consist of orthonormal basis vectors that span a given $k$ block. 
The task is to construct a basis for a given $k$ block.
A generic state with momentum $\tilde{k}$, $|\tilde{k}\rangle$, is an eigenstate under translations, $e^{i \hat{k} \Delta}|\tilde{k}\rangle \ = \ e^{i \tilde{k} \Delta}|\tilde{k}\rangle$.
A basis for these states can be built by grouping together all states that are related by translation with the appropriate phase.
As an example, consider a $L=4$ lattice in Ising field theory.
A basis for each momentum sector is,
\begin{align}
\{|k=0\rangle \} \ = \ & \Big \{ |0000\rangle \ , \ \frac{1}{2}\left (|0001\rangle +|0010\rangle + |0100\rangle +|1000\rangle \right ) \ ,  \frac{1}{2}\left (|0011\rangle +|0110\rangle + |1100\rangle +|1001\rangle \right ) \nonumber \\
& \frac{1}{\sqrt{2}}\left (|0101\rangle +|1010\rangle \right ) \ , \ \frac{1}{2}\left (|0111\rangle +|1110\rangle + |1101\rangle +|1011\rangle \right )\ , \ |1111\rangle \Big \} \ , \nonumber \\[4pt]
\{|k=\pi/2\rangle \} \ = \ 
& \Big \{ \frac{1}{2}\left (\vert0001 \rangle + i \vert0010 \rangle - \vert0100 \rangle -i \vert1000 \rangle\right ) \ , \  \frac{1}{2}\left (\vert0011 \rangle + i \vert0110 \rangle - \vert1100 \rangle -i \vert1001 \rangle\right ) \ , \nonumber \\
&\frac{1}{2}\left (\vert1110 \rangle + i \vert1101 \rangle - \vert1011 \rangle -i \vert0111 \rangle\right ) \Big \} \ , \nonumber \\[4pt]
\{|k=-\pi/2\rangle \} \ = \ 
& \Big \{ \frac{1}{2}\left (\vert0001 \rangle - i \vert0010 \rangle - \vert0100 \rangle +i \vert1000 \rangle\right ) \ , \  \frac{1}{2}\left (\vert0011 \rangle - i \vert0110 \rangle - \vert1100 \rangle +i \vert1001 \rangle\right ) \ , \nonumber \\
&\frac{1}{2}\left (\vert1110 \rangle - i \vert1101 \rangle - \vert1011 \rangle +i \vert0111 \rangle\right ) \Big \} \ , \nonumber \\[4pt]
\{|k=\pi\rangle \} \ = \ 
& \Big \{ \frac{1}{2}\left (\vert0001 \rangle - \vert0010 \rangle + \vert0100 \rangle - \vert1000 \rangle\right ) \ , \  \frac{1}{2}\left (\vert0011 \rangle - \vert0110 \rangle + \vert1100 \rangle - \vert1001 \rangle\right ) \ , \nonumber \\
&\frac{1}{2}\left (\vert1110 \rangle -  \vert1101 \rangle + \vert1011 \rangle - \vert0111 \rangle\right ) \ , \  \frac{1}{\sqrt{2}}\left (|0101\rangle -|1010\rangle \right )\Big \} \ .
\label{eq:L4kbasis}
\end{align}
All the above states are orthonormal and together span the $2^4$ dimensional Hilbert space.
Note that the state $\vert 0101 \rangle$ is not included in the $k=\pm\pi/2$ set as it maps onto itself with a minus sign under a translation by two sites.
Also notice that the states in $k=+\pi/2$ and $k=-\pi/2$ are related by complex conjugation as required by time reversal symmetry.

The basis vectors are then collected into the columns of the projection matrix $\hat{V}_k$, and the Hamiltonian is projected into a $k$ block $\hat{H}_k$ by
\begin{equation}
\hat{H}_k \ = \ \hat{V}^{\dagger}_{k} . \hat{H} . \hat{V}_k \ .
\end{equation}
For example, projecting onto $k=0$ using the basis in Eq.~\eqref{eq:L4kbasis} gives a $\hat{V}_0$ that is a $16\times 6$ matrix and a $\hat{H}_0$ that is a $6\times 6$ matrix.
Once the Hamiltonian has been projected into the desired $k$ block, the single particle eigenstates $|\psi_k\rangle$ are determined by diagonalizing $\hat{H}_k$.

The $|\psi_k\rangle$ can now be added in superposition to form the desired wavepacket (after using $\hat{V}_k$ to map them back to the full Hilbert space).
Here, a subtlety occurs. 
The eigenstates of a matrix are only determined up to an overall complex phase.
Therefore, the $|\psi_k\rangle$ that are obtained will generically have an overall complex phase that changes every time $\hat{H}_k$ is diagonalized.
This is a problem when constructing $|\psi_{\text{wp}}\rangle$ because the relative phases are set by the $e^{-i kx_0}$ in Eq.~\eqref{eq:psiWPFull}.
A consistent phase convention can be established by choosing a basis state that has a non-zero amplitude in every $|\psi_k\rangle$, and then multiplying each $|\psi_k\rangle$ by an overall phase to make that amplitude real.
For example, in Ising field theory, every $|\psi_k\rangle$ will contain the state $|0\rangle^{\otimes L-1}|1\rangle$, and $|\psi_k\rangle$ can be multiplied by an overall phase that makes the amplitude of $|0\rangle^{\otimes L-1}|1\rangle$ real.

\subsection{Useful circuit identities}
\label{app:qcircs}
\begin{figure}
    \centering
    \includegraphics[width=0.8\linewidth]{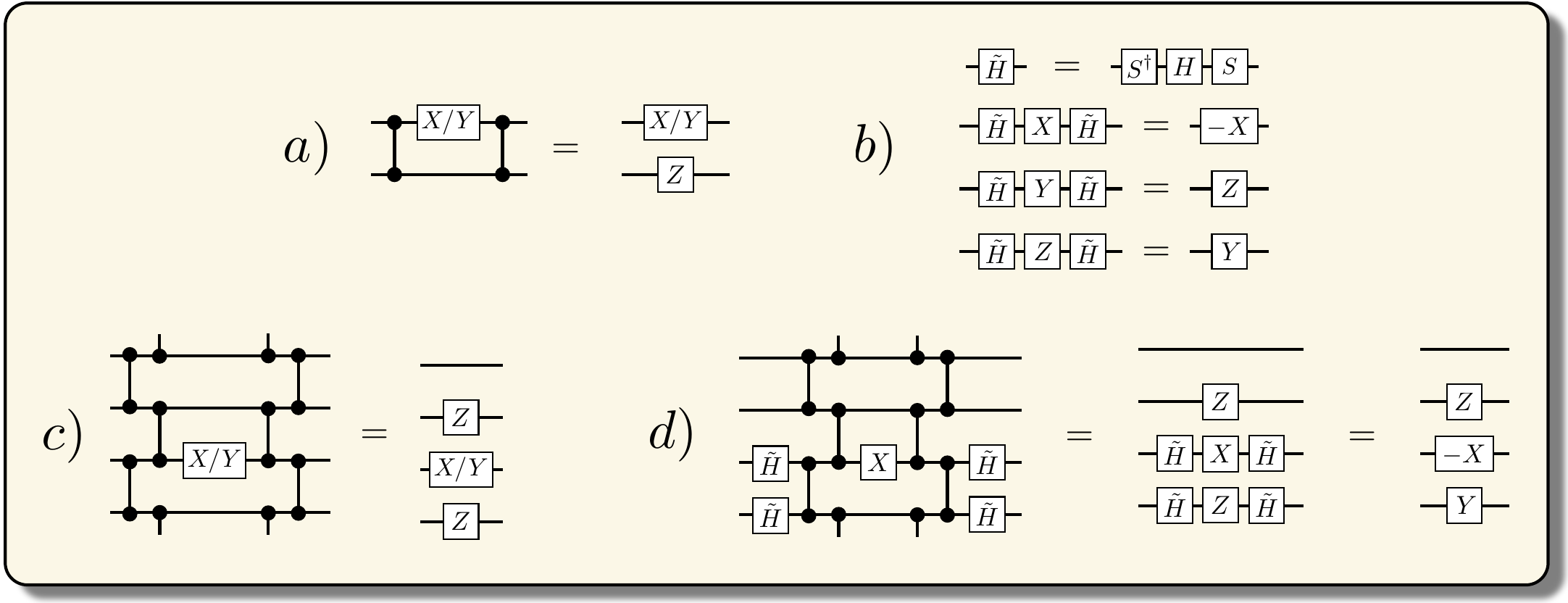}
    \caption{Circuit identities used to construct the circuits in Fig.~\ref{fig:IsingWPCircs} that prepare wavepackets in one-dimensional Ising field theory.}
    \label{fig:circIdentities}
\end{figure}

\noindent
The quantum circuits used to implement the unitary evolution with respect to the operator pool in Eq.~\eqref{eq:opPool} are inspired by the techniques in Ref.~\cite{Chernyshev:2025jyw}.
First, consider the circuit in Fig.~\ref{fig:IsingWPCircs}c) that implements unitary evolution of $\sum_n\hat{Z}_n\hat{Y}_{n+1}\hat{Z}_{n+2}$.
The building block for this circuit is shown in Fig.~\ref{fig:circIdentities}c).
Defining a brickwall sequence of $CZ$ as $\overline{CZ}$, this circuit implements the following unitary,
\begin{equation}
\left (\overline{CZ}\right )\hat{X}_n\left (\overline{CZ}\right ) \to \hat{Z}_{n-1}\hat{X}_n\hat{Z}_{n+1} \ \ , \ \ \left (\overline{CZ}\right )\hat{Y}_n \left (\overline{CZ}\right )\to \hat{Z}_{n-1}\hat{Y}_n\hat{Z}_{n+1}
\end{equation}
These relations can be derived by applying the identity shown in Fig.~\ref{fig:circIdentities}a).
The circuit in Fig.~\ref{fig:IsingWPCircs}c) immediately follows 
\begin{equation}
\left (\overline{CZ}\right )\prod_{n=0}^{L-1} R_Y(\theta)_n \left (\overline{CZ}\right )\to \exp \left [-i \frac{\theta}{2}\left (\sum_{n=0}^{L-1}\hat{Z}_{n-1}\hat{Y}_n\hat{Z}_{n+1}\right )\right ] \ . 
\end{equation}

Next, consider the circuit in Fig.~\ref{fig:IsingWPCircs}d) that implements unitary evolution of $\sum_n\left (\hat{Z}_n\hat{X}_{n+1}\hat{Y}_{n+2} + \hat{Y}_n\hat{X}_{n+1}\hat{Z}_{n+2}\right )$.
A useful gate is $\tilde{H} \equiv S.H.S^{\dagger}$, which acts like a Hadamard between the $Y$- and $Z$-bases, as shown in Fig.~\ref{fig:circIdentities}b). 
In Fig.~\ref{fig:circIdentities}d) this change of basis is combined with Fig.~\ref{fig:circIdentities}c) to show that
\begin{equation}
\tilde{H}_n\tilde{H}_{n+1}\left (\overline{CZ}\right ) \hat{X}_n\left (\overline{CZ}\right )\tilde{H}_n\tilde{H}_{n+1} \ = \ -\hat{Z}_{n-1}\hat{X}_n \hat{Y}_{n+1} \ . 
\end{equation}
Similarly it can be shown that 
\begin{align}
&\tilde{H}_n\tilde{H}_{n+1}\left (\overline{CZ}\right ) \hat{X}_{n+1}\left (\overline{CZ}\right )\tilde{H}_n\tilde{H}_{n+1} \ = \ -\hat{Y}_{n}\hat{X}_{n+1} \hat{Z}_{n+2} \ ,\nonumber \\[4pt]
&\tilde{H}_n\tilde{H}_{n+1}\left (\overline{CZ}\right ) \hat{X}_{n-1}\left (\overline{CZ}\right )\tilde{H}_n\tilde{H}_{n+1} \ = \ \hat{Z}_{n-2}\hat{X}_{n-1}\hat{Y}_n \ .
\end{align}
Next, split the Pauli strings in $\sum_n\left (\hat{Z}_n\hat{X}_{n+1}\hat{Y}_{n+2} + \hat{Y}_n\hat{X}_{n+1}\hat{Z}_{n+2}\right )$ into two commuting sets, 
\begin{align}
&\{\hat{O}_1\} \ = \ \left \{\sum_{n=0}^{\lfloor (L-1)/3\rfloor}\left (\hat{Z}_{3n}\hat{X}_{3n+1}\hat{Y}_{3n+2}  +\delta_{3n+1<L}\hat{Z}_{3n+1}\hat{X}_{3n+2}\hat{Y}_{3n+3} +\delta_{3n+2<L}\hat{Y}_{3n+2}\hat{X}_{3n+3}\hat{Z}_{3n+4} \right ) \right \} \ , \nonumber \\[4pt]
&\{\hat{O}_2\} \ = \ \left \{\sum_{n=0}^{\lfloor (L-1)/3\rfloor}\left (\hat{Y}_{3n}\hat{X}_{3n+1}\hat{Z}_{3n+2}  +\delta_{3n+1<L}\hat{Y}_{3n+1}\hat{X}_{3n+2}\hat{Z}_{3n+3} +\delta_{3n+2<L}\hat{Z}_{3n+2}\hat{X}_{3n+3}\hat{Y}_{3n+4}  \right ) \right \} \ .
\end{align}
The unitary evolution of the exponential of each set is implemented separately using the construction in Fig.~\ref{fig:circIdentities}d).
The circuit in Fig.~\ref{fig:IsingWPCircs}d) first implements the unitary of evolution of $\{ \hat{O}_1\}$ and then $\{\hat{O}_2\}$.
Half of the rotations are $R_X(-\theta)$ due to the minus sign in $\tilde{H}\hat{X}\tilde{H}=-\hat{X}$.

%

\subsection{Operators and variational parameters for preparing wavepackets in one-dimensional Ising field theory}
\label{app:ADAPTparam}
\noindent
The variational parameters and operator ordering used to prepare wavepackets in one-dimensional Ising field theory with ADAPT-VQE for the quantum simulations performed in Sec.~\ref{sec:qsim} are given in Table~\ref{tab:qsimIsingAdaptAng},
for the {\tt qiskit} statevector simulations performed in Methods~\ref{sec:IsingStatevector} are given in Table~\ref{tab:IsingAdaptAngL28},
for the MPS simulations performed in Methods~\ref{sec:csimscatt} are given in  Tables~\ref{tab:IsingAdaptAngL256} and~\ref{tab:IsingAdaptAngL256_2}
and for the vacuum preparation in App.~\ref{app:57Adapt} are given in Table~\ref{tab:VacIsingAdaptAng}.

As in Refs.~\cite{Farrell:2024fit,Zemlevskiy:2024vxt},
the variational parameters that prepare wavepackets converge exponentially with increasing system size.
For example, the parameters used to prepare wavepackets with $g_x=1.25,\,g_z=0.15$ and $k_0=0.36\pi$ for $L=28$, Table~\ref{tab:IsingAdaptAngL28}, and for $L=256$, Table~\ref{tab:IsingAdaptAngL256} agree to 4 decimal places. 
This is due to the exponential convergence of the mass gap with increasing system size, and the associated exponential suppression of finite size effects.
This convergence motivated the SC-ADAPT-VQE workflow~\cite{Farrell:2023fgd} (see also Refs.~\cite{Klco:2019yrb,Klco:2020aud}) where quantum circuits are optimized on small lattices and then systematically extrapolated to prepare states on large lattices.
For preparing wavepackets, this approach has the limitation that it can only prepare wavepackets of a fixed size, i.e., $d/L$ is a constant.
Indeed, in this work we found that the operators and variational parameters that prepare small wavepackets were not optimal for preparing larger wavepackets.
In this work no parameter extrapolations were necessary as our state preparation circuits for large $L$ could be directly simulated using MPS.
However, circuit extrapolations and the SC-ADAPT-VQE workflow will likely be helpful for preparing wavepackets in higher dimensions where classical computing techniques are not able to faithfully simulate large circuits.
\begin{table}[!t]
\renewcommand{\arraystretch}{1.4}
\begin{tabularx}{\textwidth}{|c || Y | Y | Y | Y | Y | Y  | Y |}
  \hline
 \diagbox[height=23pt]{$k_0$}{$\theta_i$} & 
 $\hat{Y}$&
 $\hat{Y}\hat{Z}$ & 
 $\hat{Y}$& $\hat{Z}\hat{X}\hat{Y}$& $\hat{Y}\hat{Z}$&$\hat{Z}\hat{Y}\hat{Z}$&
 $\hat{Y}$

 \\
 \hline\hline
$0.32\pi$ &  0.0191&  0.0276& -0.4497&  0.0226&  0.0618&  0.0900&   -0.2238\\
 \hline
\end{tabularx}
\caption{
Operators/parameters used to prepare wavepackets in the quantum simulations of scattering in one-dimensional Ising field theory presented in Sec.~\ref{sec:qsim}.
These parameters correspond to the 7th step of ADAPT-VQE for $L=104$, $g_x=1.25,\,g_z=0.15$ and $k_0=0.32\pi,\, \sigma_{}=0.13$.
For clarity, each operator is labeled by one term, e.g., $\hat{Y}\hat{Z}$ corresponds to $\hat{Y}\hat{Z} + \hat{Z}\hat{Y}$ in Eq.~\eqref{eq:opPool}. }
 \label{tab:qsimIsingAdaptAng}
\end{table}
\begin{table}[!t]
\renewcommand{\arraystretch}{1.4}
\begin{tabularx}{\textwidth}{|c || Y | Y | Y | Y | Y | Y | Y | Y | Y |Y|Y|Y|}
  \hline
 \diagbox[height=23pt]{$(g_x,g_z)$}{$\theta_i$} & 
 $\hat{Y}$&
 $\hat{Y}\hat{Z}$ &  $\hat{Z}\hat{X}\hat{Y}$ & 
 $\hat{Y}$ &  
 $\hat{Z}\hat{X}\hat{Y}$&  $\hat{Z}\hat{Y}\hat{Z}$&  
 $\hat{Y}\hat{Z}$&  
 $\hat{Y}$&
 $\hat{Y}\hat{Z}$
 &
 $\hat{Y}$
&
 $\hat{Y}\hat{X}$
 &
 $\hat{Z}\hat{Y}\hat{Z}$

 \\
 \hline\hline
$(1.25,0.15)$ &  0.1212 & 0.0185 & -- & -0.5452 &  0.0397 & --&0.0599& --&0.0556 &-0.2637& --&0.0566\\
 \hline
 $(1.18,0.08)$ & -0.3517 & 0.0610 &  0.0477 & -0.1425 &--& 0.1107&--& -0.2030 &  0.0310&--&   0.0176&--\\
 \hline
  $(1.15,0.05)$ &-0.3866 & 0.0700&    0.0473&  0.0201&--&  0.0705&--& -0.3268 & 0.0288&--&  --&0.0368\\
   \hline
  $(1.08,0.03)$ & -0.3653 & 0.0754 & 0.0539 &-0.1910&--&   0.1234&--& -0.1772&  0.0313&--&  0.0266&--\\
 \hline
\end{tabularx}
\caption{
Operators/parameters used to prepare wavepackets in one-dimensional Ising field theory for $L=28$ and $k_0=0.36\pi,\, \sigma_{}=0.13$.
These parameters correspond to the 8th step in Fig.~\ref{fig:IsingADADPT_results}.
For clarity, each operator is labeled by one term, e.g., $\hat{Y}\hat{Z}$ corresponds to $\hat{Y}\hat{Z} + \hat{Z}\hat{Y}$ in Eq.~\eqref{eq:opPool}.}
 \label{tab:IsingAdaptAngL28}
\end{table}
\begin{table}[!t]
\renewcommand{\arraystretch}{1.4}
\begin{tabularx}{\textwidth}{|c || Y | Y | Y | Y | Y | Y | Y | Y | Y |Y|Y|Y|Y|Y|}
  \hline
 \diagbox[height=23pt]{$k_0$}{$\theta_i$} & 
 $\hat{Y}$&
 $\hat{Y}\hat{Z}$ & $\hat{Y}$& 
 $\hat{Z}\hat{X}\hat{Y}$&  $\hat{Y}$&$\hat{Y}\hat{Z}$
 &$\hat{Z}\hat{Y}\hat{Z}$&  
 $\hat{Y}\hat{Z}$&  
 $\hat{Y}$&
 $\hat{Y}\hat{Z}$
 &$\hat{Z}\hat{Y}\hat{Z}$&
 $\hat{Y}$
 &
 $\hat{Z}\hat{X}\hat{Y}$

 \\
 \hline\hline
$0.36\pi$ &  0.1212&  0.0185& -0.5452&  0.0397&--&  0.0599&--&  0.0556& -0.2637&--&0.0566&--&--\\
\hline
$0.32\pi$ &  0.0505&  0.0006&-0.3983&  0.0316&--&  0.0750 & 0.0868& --&-0.3029&--& 0.0349&--&--\\
\hline
$0.28\pi$ &  0.0499&  0.0314& -0.5092&  0.0231&--&  0.0401&--&  0.0325&--&--&  0.0739&-0.2064&--\\
\hline
$0.18\pi$ &  -0.2663&  0.0566&--&  0.0470& -0.1871&--&  0.0806&--& -0.2214& 0.0524&--&--&0.0127\\
 \hline
\end{tabularx}
\caption{
Operators/parameters used to prepare wavepackets in one-dimensional Ising field theory for $L=256$,\, $g_x=1.25,\,g_z=0.15$ and $\sigma_{}=0.13$.
These parameters were used to prepare the initial state of the scattering simulations in Fig.~\ref{fig:2WP_scattering_MPS}.
For clarity, each operator is labeled by one term, e.g., $\hat{Y}\hat{Z}$ corresponds to $\hat{Y}\hat{Z} + \hat{Z}\hat{Y}$ in Eq.~\eqref{eq:opPool}.}
 \label{tab:IsingAdaptAngL256}
\end{table}
\begin{table}[!t]
\renewcommand{\arraystretch}{1.4}
\begin{tabularx}{\textwidth}{|c || Y | Y | Y | Y | Y | Y  | Y |Y|Y|Y|Y|Y|}
  \hline
 \diagbox[height=23pt]{$k_0$}{$\theta_i$} & 
 $\hat{Y}$&
 $\hat{Y}\hat{Z}$ & 
 $\hat{Z}\hat{X}\hat{Y}$& $\hat{Y}$& $\hat{Z}\hat{Y}\hat{Z}$&$\hat{Y}$&
 $\hat{Y}\hat{Z}$&  
 $\hat{Z}\hat{Y}\hat{Z}$&
 $\hat{Y}$
 &
 $\hat{Z}\hat{X}\hat{Y}$

 \\
 \hline\hline
$0.20\pi$ &  -0.3124&  0.0412&  0.0511&  0.1360 &  0.0235& -0.3669&  0.0579&
        0.0828& -0.1367&  0.0009\\
 \hline
\end{tabularx}
\caption{
Same as Table~\ref{tab:IsingAdaptAngL256} but for $k_0=0.20\pi$ and with 10 ADAPT-VQE steps.
Operators/parameters used to prepare wavepackets in one-dimensional Ising field theory for $L=256$,\, $g_x=1.25,\,g_z=0.15$ and $\sigma_{}=0.13$.
These parameters were used to prepare the initial state of the scattering simulations in Fig.~\ref{fig:2WP_scattering_MPS}.
For clarity, each operator is labeled by one term, e.g., $\hat{Y}\hat{Z}$ corresponds to $\hat{Y}\hat{Z} + \hat{Z}\hat{Y}$ in Eq.~\eqref{eq:opPool}.}
 \label{tab:IsingAdaptAngL256_2}
\end{table}
\begin{table}[!t]
\renewcommand{\arraystretch}{1.4}
\begin{tabularx}{\textwidth}{|c || Y | Y | Y | Y | Y | Y | Y |Y|Y|Y|}
  \hline
 \diagbox[height=23pt]{$(g_x,g_z)$}{$\theta_i$} & 
 $\hat{Y}$&
 $\hat{Y}\hat{Z}$ &  $\hat{Z}\hat{X}\hat{Y}$ &  
 $\hat{Y}\hat{Z}$&
 $\hat{Y}\hat{Z}$&
 $\hat{Z}\hat{Y}\hat{Z}$&
 $\hat{Y}$&$\hat{Y}\hat{X}$
 &$\hat{Z}\hat{Y}\hat{Z}$&
 $\hat{Y}\hat{X}$
 \\
 \hline\hline
$(1.25,0.15)$ & -0.4735 & 0.0244 & 0.0145 & 0.0209 & 0.0195 & 0.0008 &--&-0.0024& -0.0029&-- \\
   \hline
  $(1.08,0.03)$ &-0.4885 & 0.0288 & 0.0194 & 0.0248 & 0.0235 & --&-0.0048 & 0.0021&--& -0.0026 \\
 \hline
\end{tabularx}
\caption{
Operators/parameters used to prepare the vacuum in one-dimensional Ising field theory for $L=28$.
These parameters correspond to the 8th step in Fig.~\ref{fig:IsingADADPT_VACresults}.
For clarity, each operator is labeled by one term, e.g., $\hat{Y}\hat{Z}$ corresponds to $\hat{Y}\hat{Z} + \hat{Z}\hat{Y}$ in Eq.~\eqref{eq:opPool}.}
 \label{tab:VacIsingAdaptAng}
\end{table}
%

\subsection{Error mitigation with energy rescaling}
\label{app:energy_rescale}
\noindent
An ideal simulation would conserve the total energy $E_{\text{tot}}=\sum_nE_n$.
In a quantum simulation, the combination of Trotter errors and the imperfect mitigation of device noise leads to a violation of energy conservation.
We find that a new error mitigation step that restores energy conservation in post-processing can reduce systematic errors.
This is done by applying a multiplicative rescaling of the vacuum-subtracted energy density predicted by ODR,
\begin{align}
    E_n = E_n^{(\text{ODR})} \frac{E_{\text{tot}}}{\sum_{j=0}^{L-1} E_j^{(\text{ODR})}}\ ,
\end{align}
where $E_{\text{tot}}$ is the energy of the initial state and $E_j^{(\text{ODR})}$  is the energy density predicted after ODR.
This rescaling corrects Trotter and device errors that lead to violations of energy conservation.
It also reduces the classical computing overhead required in ODR.
Our implementation of ODR, described in Methods~\ref{sec:qsimDetails}, assumed that observables could be efficiently computed in the time-evolved vacuum $U_2(t) |\psi_{\text{vac}}\rangle$.
This observable $\langle \hat{O}\rangle_\text{pred}$ is used to compute the signal strength in Eq.~\eqref{eq:odr_p_j}.
Efficient classical computation of observables in the time-evolved vacuum assumes that the time-evolved state remains close to the true vacuum, which can be efficiently represented by an MPS in one dimension.
However, the large Trotter step size of $\delta t = 0.55$ used in this work effectively quenches the vacuum, and the bond dimension required to compute observables in $U_2(t) |\psi_{\text{vac}}\rangle$ can be quite large (although not as large as in the scattering simulations).
This challenge is overcome if the energy conservation is enforced in post-processing because the {\it unevolved} vacuum can be used to compute $\langle \hat{O}\rangle_\text{pred}$ and determine the signal strength,
\begin{align}
    p_{\hat{O}} \ = \ \frac{\langle \hat{O}\rangle_\text{meas}}{\langle \hat{O}\rangle_\text{pred}} \ = \ \frac{\langle \psi_\text{vac}|\hat{U}^{\dagger}_2(t) \hat{O} \hat{U}_2(t)|\psi_\text{vac}\rangle_\text{meas}}{\langle \psi_\text{vac}| \hat{O} |\psi_\text{vac}\rangle_\text{pred}} \ .
    \label{eq:odr_p_j_vac}
\end{align}
The $\sim\!15\%$ ``error" incurred by using the static vacuum to compute $\langle \hat{O}\rangle_\text{pred}$ instead of the time-evolved vacuum is fixed when energy conservation is enforced.

\begin{figure}[t!]
    \centering
    \includegraphics[width=0.75\linewidth]{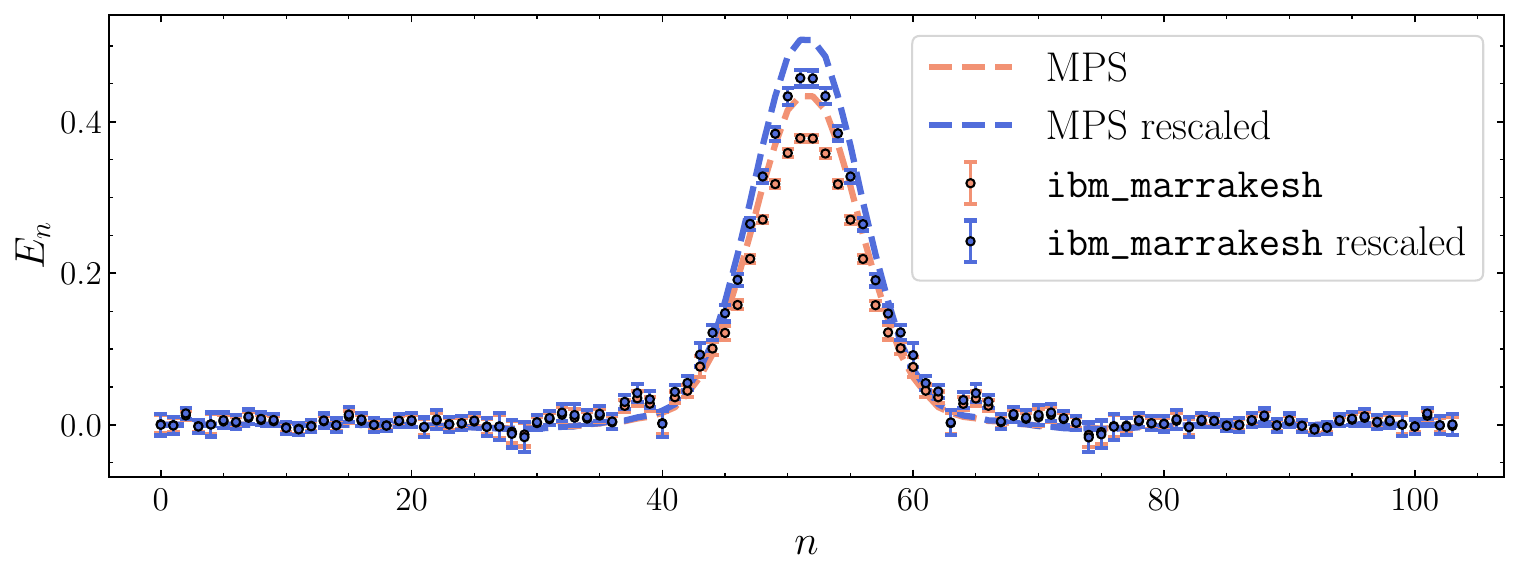}
    \caption{{\it The effect of energy rescaling.}
    MPS predictions (dashed lines) are compared with data obtained from {\tt ibm\_marrakesh} (points) for $t_1=8.25$.
    The results without energy rescaling presented in Fig.~\ref{fig:ibm_results}b) are shown in orange while the results with energy conservation enforced are shown in blue.}
    \label{fig:Energy_rescale}
\end{figure}

The effect of enforcing energy conservation on the results is shown in Fig.~\ref{fig:Energy_rescale} for $t_1 = 8.25$.
Without energy conservation, the energy density determined from {\tt ibm\_marrakesh} (orange points) is significantly lower than the MPS predictions in the middle of the collision (orange dashed line).
Enforcing energy conservation has the largest impact on the energy density in the middle of the lattice, and the rescaled energy density (blue points) perfectly aligns with MPS.
However, for a fair a comparison, the rescaled energy density should be compared to the rescaled MPS predictions (blue dashed line).
The energy in the MPS simulations changes because of Trotter errors.
The systematic error in the energy density near the point of collision is seen to be reduced, although it is still present.
This is primarily because there is a slight positive-energy bias compared with MPS in the quantum results away from the middle of the lattice. 
When summed over many sites, this small bias becomes significant and reduces the energy rescale value.
For later simulations times, the positive energy bias in the quantum results becomes more significant and the energy rescaling becomes ineffective.
Techniques that use the light-cone to zero the energy density outside of the scattering region would remove the positive bias, and could improve the efficacy of this error mitigation strategy.
%

\subsection{Table of results}
\noindent
Table~\ref{tab:results_t} gives data for the vacuum-subtracted energy density $E_n$ throughout the quantum simulations of inelastic scattering presented in Sec.~\ref{sec:qsim}, and Table~\ref{tab:results_t_1wp} gives data for $E_n$ for the single wavepacket simulations.
An $L=104$ system with OBCs and $g_x=1.25,\,g_z=1.5$ is simulated using a Trotter step size $\delta t=0.55$.
The wavepacket parameters $k_0=0.32\pi,\,\sigma_{}=0.13$ are used. 
For each time $t$, the left column gives the MPS data and the right column shows the results from {\tt ibm\_marrakesh} after error mitigation.
Parity symmetry about the collision point equates $E_n$ and $E_{L-1-n}$, and the energy density is not shown for sites related by this symmetry. 
The MPS values of $E_n$ are not exactly 0 outside of the simulation light cone due to finite Trotter step size. 
\begin{table}[h]
\renewcommand{\arraystretch}{1.1}
\begin{tabularx}{\linewidth}{|c||Y|Y||Y|Y||Y|Y||Y|Y|}
\hline
\rule{0pt}{10pt} & \multicolumn{8}{c|}{\large$E_n$}\\\hline\hline
\rule{0pt}{10pt} \multirow{2}{*}{\makecell{Spatial site\\$n$}} &  \multicolumn{2}{c||}{$t_0=0.0$}  &  \multicolumn{2}{c||}{$t_1=8.25$} & \multicolumn{2}{c||}{$t_2=16.5$} & \multicolumn{2}{c|}{$t_3=24.75$} \\\cline{2-9}
\rule{0pt}{10pt} & MPS & QC & MPS & QC & MPS & QC & MPS & QC\\
\hline\hline
  0 & 0.000 & 0.005(10)  &  0.000 & 0.001(12)  & -0.002 & -0.001(10) & -0.023 & 0.001(08)  \\
  1 & 0.000 & -0.009(10) &  0.000 & -0.000(09) & -0.001 & -0.009(07) & -0.014 & 0.002(06)  \\
  2 & 0.000 & -0.016(11) &  0.000 & 0.012(06)  & -0.001 & 0.007(06)  & -0.015 & -0.006(05) \\
  3 & 0.000 & -0.003(12) &  0.000 & -0.002(08) & -0.001 & -0.009(07) & -0.015 & -0.011(06) \\
  4 & 0.000 & 0.007(11)  &  0.000 & 0.001(13)  & -0.001 & 0.005(11)  & -0.015 & -0.008(08) \\
  5 & 0.000 & 0.006(11)  &  0.000 & 0.004(09)  & -0.001 & -0.002(07) & -0.015 & 0.003(06)  \\
  6 & 0.000 & -0.002(09) &  0.000 & 0.003(07)  & -0.001 & -0.008(06) & -0.015 & -0.002(07) \\
  7 & 0.000 & 0.001(10)  &  0.000 & 0.009(08)  & -0.002 & -0.001(06) & -0.015 & -0.010(06) \\
  8 & 0.000 & 0.001(09)  &  0.000 & 0.006(07)  & -0.002 & -0.012(06) & -0.014 & -0.006(06) \\
  9 & 0.000 & 0.000(08)  &  0.000 & 0.005(06)  & -0.003 & -0.010(06) & -0.014 & -0.004(05) \\
 10 & 0.000 & -0.001(11) &  0.000 & -0.003(06) & -0.003 & -0.012(06) & -0.014 & 0.012(06)  \\
 11 & 0.000 & -0.002(16) &  0.000 & -0.005(06) & -0.004 & -0.002(06) & -0.014 & -0.008(06) \\
 12 & 0.000 & -0.000(15) &  0.000 & -0.001(07) & -0.004 & 0.003(06)  & -0.014 & -0.003(06) \\
 13 & 0.000 & -0.001(13) &  0.000 & 0.005(07)  & -0.005 & -0.009(07) & -0.014 & 0.009(08)  \\
 14 & 0.000 & -0.021(22) &  0.000 & -0.000(07) & -0.006 & -0.004(08) & -0.014 & -0.002(07) \\
 15 & 0.000 & -0.026(18) &  0.000 & 0.011(08)  & -0.006 & -0.004(08) & -0.015 & 0.003(06)  \\
 16 & 0.000 & 0.002(10)  &  0.000 & 0.005(07)  & -0.006 & 0.008(07)  & -0.014 & 0.005(06)  \\
 17 & 0.000 & -0.002(09) &  0.000 & -0.000(07) & -0.007 & 0.007(07)  & -0.014 & 0.006(06)  \\
 18 & 0.000 & 0.000(05)  &  0.000 & -0.001(06) & -0.007 & -0.000(07) & -0.012 & 0.003(06)  \\
 19 & 0.000 & -0.002(08) &  0.000 & 0.005(07)  & -0.007 & 0.005(07)  & -0.010 & 0.008(06)  \\
 20 & 0.000 & 0.007(08)  &  0.000 & 0.005(08)  & -0.006 & 0.009(08)  & -0.005 & 0.023(08)  \\
 21 & 0.000 & -0.017(10) & -0.001 & -0.003(11) & -0.006 & -0.003(12) &  0.002 & 0.030(09)  \\
 22 & 0.000 & -0.017(13) & -0.001 & 0.006(10)  & -0.006 & -0.009(11) &  0.014 & 0.039(07)  \\
 23 & 0.000 & 0.003(12)  & -0.001 & 0.001(08)  & -0.006 & 0.008(07)  &  0.031 & 0.061(07)  \\
 24 & 0.000 & -0.000(07) & -0.002 & 0.002(08)  & -0.005 & -0.004(07) &  0.050 & 0.068(08)  \\
 25 & 0.000 & 0.001(09)  & -0.003 & 0.004(08)  & -0.006 & 0.009(07)  &  0.070 & 0.103(07)  \\
 26 & 0.000 & -0.011(17) & -0.004 & -0.002(10) & -0.007 & 0.015(08)  &  0.094 & 0.121(07)  \\
 27 & 0.000 & -0.001(12) & -0.004 & -0.002(15) & -0.006 & 0.005(15)  &  0.117 & 0.130(09)  \\
 28 & 0.001 & -0.006(08) & -0.005 & -0.009(15) & -0.007 & 0.018(15)  &  0.132 & 0.139(12)  \\
 29 & 0.002 & 0.001(09)  & -0.005 & -0.013(16) & -0.005 & 0.031(15)  &  0.143 & 0.127(08)  \\
 30 & 0.012 & 0.019(12)  & -0.004 & 0.002(08)  & -0.003 & 0.022(08)  &  0.148 & 0.120(06)  \\
 31 & 0.020 & 0.015(10)  & -0.003 & 0.008(08)  &  0.002 & 0.021(07)  &  0.144 & 0.110(06)  \\
 32 & 0.036 & 0.019(12)  & -0.002 & 0.013(10)  &  0.008 & 0.038(07)  &  0.139 & 0.095(06)  \\
 33 & 0.057 & 0.094(26)  & -0.002 & 0.010(12)  &  0.020 & 0.042(11)  &  0.127 & 0.084(08)  \\
 34 & 0.088 & 0.072(07)  &  0.001 & 0.008(08)  &  0.041 & 0.062(08)  &  0.123 & 0.088(08)  \\
 35 & 0.124 & 0.108(09)  &  0.002 & 0.012(08)  &  0.065 & 0.080(07)  &  0.106 & 0.075(06)  \\
 36 & 0.168 & 0.150(11)  &  0.005 & 0.003(09)  &  0.092 & 0.100(07)  &  0.088 & 0.074(06)  \\
 37 & 0.209 & 0.183(09)  &  0.005 & 0.025(08)  &  0.120 & 0.146(06)  &  0.083 & 0.064(06)  \\
 38 & 0.246 & 0.236(09)  &  0.008 & 0.035(10)  &  0.152 & 0.132(08)  &  0.066 & 0.056(07)  \\
 39 & 0.271 & 0.253(08)  &  0.011 & 0.028(09)  &  0.174 & 0.172(08)  &  0.054 & 0.056(06)  \\
 40 & 0.280 & 0.292(08)  &  0.016 & 0.002(15)  &  0.187 & 0.177(09)  &  0.049 & 0.064(06)  \\
 41 & 0.271 & 0.284(06)  &  0.023 & 0.036(07)  &  0.193 & 0.170(07)  &  0.044 & 0.053(06)  \\
 42 & 0.245 & 0.239(07)  &  0.041 & 0.045(08)  &  0.191 & 0.137(08)  &  0.038 & 0.055(06)  \\
 43 & 0.210 & 0.187(07)  &  0.063 & 0.077(14)  &  0.176 & 0.172(14)  &  0.034 & 0.055(06)  \\
 44 & 0.166 & 0.140(10)  &  0.093 & 0.101(08)  &  0.158 & 0.136(08)  &  0.029 & 0.050(06)  \\
 45 & 0.126 & 0.124(09)  &  0.139 & 0.121(09)  &  0.138 & 0.135(09)  &  0.027 & 0.040(08)  \\
 46 & 0.087 & 0.074(06)  &  0.192 & 0.158(06)  &  0.116 & 0.098(07)  &  0.023 & 0.057(05)  \\
 47 & 0.058 & 0.048(06)  &  0.251 & 0.219(05)  &  0.096 & 0.090(06)  &  0.021 & 0.055(05)  \\
 48 & 0.035 & 0.021(07)  &  0.315 & 0.271(05)  &  0.080 & 0.091(06)  &  0.019 & 0.055(05)  \\
 49 & 0.021 & 0.022(09)  &  0.370 & 0.318(05)  &  0.066 & 0.083(06)  &  0.018 & 0.042(06)  \\
 50 & 0.012 & -0.003(16) &  0.415 & 0.359(05)  &  0.058 & 0.063(06)  &  0.017 & 0.052(07)  \\
 51 & 0.003 & 0.004(09)  &  0.434 & 0.378(05)  &  0.055 & 0.065(06)  &  0.017 & 0.052(05)  \\
 \hline
\end{tabularx}
\caption{The vacuum-subtracted energy density $E_n$ for $t=0,\,8.25,\,16.5,\,24.75$ corresponding to Fig.~\ref{fig:ibm_results}. 
Results are shown from MPS simulations and from quantum simulations using {\tt ibm\_marrakesh} after error mitigation (columns labeled QC). 
The energy density of the sites not shown can be obtained by the parity symmetry that relates $E_n = E_{L-1-n}$.}
\label{tab:results_t}
\end{table}

\begin{table}[h]
\renewcommand{\arraystretch}{1.1}
\begin{tabularx}{\linewidth}{|c||Y|Y||Y|Y||c||Y|Y||Y|Y|}
\hline
\rule{0pt}{10pt} & \multicolumn{4}{c||}{\large$E_n$} & \rule{0pt}{10pt} & \multicolumn{4}{c|}{\large$E_n$}\\\hline\hline
\rule{0pt}{10pt} \multirow{2}{*}{\makecell{Spatial site\\$n$}} &  \multicolumn{2}{c||}{$t_2=16.5$} & \multicolumn{2}{c||}{$t_3=24.75$} & \rule{0pt}{10pt} \multirow{2}{*}{\makecell{Spatial site\\$n$}} & \multicolumn{2}{c||}{$t_2=16.5$} & \multicolumn{2}{c|}{$t_3=24.75$} 
\\\cline{2-5}
\cline{7-10}
\rule{0pt}{10pt} & MPS & QC & MPS & QC & \rule{0pt}{10pt} & MPS & QC & MPS & QC\\
\hline\hline
0 & -0.001 & 0.008(11)  & -0.021 & -0.002(09) &  52 &  0.020 & 0.043(06)  &  0.001 & 0.027(06)  \\
  1 & -0.001 & 0.006(09)  & -0.015 & 0.008(07)  &  53 &  0.018 & 0.024(07)  &  0.000 & -      \\
  2 & -0.001 & -0.013(07) & -0.016 & -0.002(06) &  54 &  0.015 & 0.034(08)  & -0.001 & 0.025(07)  \\
  3 & -0.001 & -0.003(07) & -0.016 & -0.005(06) &  55 &  0.013 & 0.041(05)  & -0.002 & 0.036(05)  \\
  4 & -0.001 & 0.005(09)  & -0.016 & -0.002(07) &  56 &  0.010 & 0.025(06)  & -0.003 & 0.028(05)  \\
  5 & -0.001 & 0.004(07)  & -0.016 & 0.010(07)  &  57 &  0.009 & 0.030(07)  & -0.004 & 0.019(06)  \\
  6 & -0.001 & 0.004(06)  & -0.016 & -0.006(07) &  58 &  0.007 & 0.011(09)  & -0.004 & 0.031(08)  \\
  7 & -0.001 & 0.005(05)  & -0.016 & -0.001(06) &  59 &  0.006 & 0.036(06)  & -0.005 & 0.027(06)  \\
  8 & -0.001 & -0.000(06) & -0.015 & -0.002(06) &  60 &  0.004 & 0.031(06)  & -0.005 & 0.021(06)  \\
  9 & -0.001 & 0.014(06)  & -0.015 & -0.001(06) &  61 &  0.004 & 0.036(06)  & -0.006 & 0.030(06)  \\
 10 & -0.001 & 0.012(06)  & -0.015 & 0.004(06)  &  62 &  0.003 & 0.028(05)  & -0.006 & 0.015(06)  \\
 11 & -0.001 & 0.008(06)  & -0.015 & -0.002(06) &  63 &  0.002 & 0.028(08)  & -0.007 & 0.029(06)  \\
 12 & -0.001 & -0.001(07) & -0.015 & -0.002(07) &  64 &  0.001 & 0.015(06)  & -0.006 & 0.021(06)  \\
 13 & -0.001 & 0.006(07)  & -0.015 & -0.004(08) &  65 &  0.001 & 0.005(08)  & -0.007 & 0.028(07)  \\
 14 & -0.001 & 0.000(06)  & -0.015 & 0.003(08)  &  66 & -0.001 & 0.015(06)  & -0.008 & 0.019(06)  \\
 15 & -0.002 & 0.007(07)  & -0.014 & 0.004(07)  &  67 & -0.001 & 0.012(07)  & -0.006 & 0.010(06)  \\
 16 & -0.002 & 0.009(06)  & -0.013 & -0.001(07) &  68 & -0.003 & 0.008(06)  & -0.005 & 0.005(06)  \\
 17 & -0.002 & 0.014(05)  & -0.012 & 0.001(06)  &  69 & -0.003 & 0.005(07)  & -0.009 & 0.001(07)  \\
 18 & -0.002 & -0.001(06) & -0.010 & 0.011(06)  &  70 & -0.004 & -0.005(07) & -0.006 & 0.008(07)  \\
 19 & -0.002 & 0.001(05)  & -0.007 & 0.013(06)  &  71 & -0.004 & 0.015(06)  & -0.005 & 0.013(06)  \\
 20 & -0.002 & 0.006(06)  & -0.003 & 0.026(06)  &  72 & -0.004 & 0.021(06)  & -0.007 & 0.015(05)  \\
 21 & -0.002 & 0.013(08)  &  0.004 & 0.021(06)  &  73 & -0.005 & 0.009(07)  & -0.006 & 0.011(07)  \\
 22 & -0.002 & -0.000(07) &  0.018 & 0.038(06)  &  74 & -0.004 & 0.011(08)  & -0.007 & -0.001(12) \\
 23 & -0.002 & -0.016(07) &  0.037 & 0.073(06)  &  75 & -0.004 & -      & -0.007 & -      \\
 24 & -0.002 & -0.021(08) &  0.059 & 0.079(07)  &  76 & -0.003 & 0.019(12)  & -0.006 & -0.000(10) \\
 25 & -0.002 & -0.006(07) &  0.082 & 0.082(06)  &  77 & -0.004 & 0.019(08)  & -0.008 & 0.008(07)  \\
 26 & -0.002 & -0.011(08) &  0.113 & 0.100(07)  &  78 & -0.003 & 0.009(07)  & -0.008 & 0.018(06)  \\
 27 & -0.002 & 0.005(10)  &  0.142 & 0.117(10)  &  79 & -0.002 & 0.008(08)  & -0.009 & -0.010(07) \\
 28 & -0.001 & -      &  0.161 & 0.117(13)  &  80 & -0.003 & 0.013(07)  & -0.010 & 0.006(07)  \\
 29 &  0.001 & 0.007(09)  &  0.175 & 0.112(07)  &  81 & -0.003 & 0.014(08)  & -0.010 & 0.007(07)  \\
 30 &  0.004 & 0.010(07)  &  0.179 & 0.093(06)  &  82 & -0.003 & 0.023(09)  & -0.011 & 0.014(10)  \\
 31 &  0.008 & 0.005(07)  &  0.169 & 0.085(06)  &  83 & -0.003 & 0.002(07)  & -0.011 & 0.006(11)  \\
 32 &  0.015 & 0.019(06)  &  0.155 & 0.087(06)  &  84 & -0.004 & 0.001(06)  & -0.011 & 0.012(06)  \\
 33 &  0.027 & 0.024(07)  &  0.133 & 0.092(08)  &  85 & -0.004 & -0.011(06) & -0.011 & 0.003(06)  \\
 34 &  0.047 & 0.031(07)  &  0.120 & 0.061(07)  &  86 & -0.004 & -0.005(06) & -0.011 & 0.010(06)  \\
 35 &  0.072 & 0.045(06)  &  0.094 & 0.064(06)  &  87 & -0.004 & -0.003(06) & -0.010 & 0.014(06)  \\
 36 &  0.098 & 0.082(06)  &  0.069 & 0.057(06)  &  88 & -0.004 & 0.004(06)  & -0.011 & 0.008(07)  \\
 37 &  0.128 & 0.111(06)  &  0.060 & 0.036(06)  &  89 & -0.003 & -0.003(07) & -0.010 & 0.019(06)  \\
 38 &  0.160 & 0.114(07)  &  0.040 & 0.041(08)  &  90 & -0.003 & -0.003(07) & -0.010 & -0.002(08) \\
 39 &  0.182 & 0.136(06)  &  0.028 & 0.029(08)  &  91 & -0.002 & 0.011(06)  & -0.009 & 0.000(06)  \\
 40 &  0.193 & 0.124(07)  &  0.025 & 0.040(06)  &  92 & -0.002 & -0.003(06) & -0.009 & 0.003(05)  \\
 41 &  0.198 & 0.144(06)  &  0.022 & 0.038(06)  &  93 & -0.001 & -0.008(06) & -0.010 & 0.009(05)  \\
 42 &  0.192 & 0.135(06)  &  0.018 & 0.027(06)  &  94 & -0.001 & -0.006(06) & -0.009 & 0.012(05)  \\
 43 &  0.172 & 0.125(06)  &  0.015 & 0.038(06)  &  95 & -0.001 & -0.002(06) & -0.010 & 0.006(05)  \\
 44 &  0.152 & 0.106(06)  &  0.013 & 0.032(06)  &  96 &  0.000 & 0.006(06)  & -0.010 & 0.008(05)  \\
 45 &  0.129 & 0.110(09)  &  0.011 & 0.020(05)  &  97 &  0.000 & 0.009(07)  & -0.011 & 0.008(05)  \\
 46 &  0.103 & 0.086(05)  &  0.009 & 0.023(06)  &  98 &  0.000 & 0.006(07)  & -0.010 & 0.010(06)  \\
 47 &  0.081 & 0.072(05)  &  0.007 & 0.023(05)  &  99 &  0.000 & 0.002(09)  & -0.010 & 0.003(07)  \\
 48 &  0.062 & 0.052(06)  &  0.006 & 0.026(06)  & 100 &  0.000 & -0.002(07) & -0.010 & 0.006(06)  \\
 49 &  0.044 & 0.060(08)  &  0.004 & 0.024(08)  & 101 &  0.000 & 0.003(07)  & -0.010 & 0.007(06)  \\
 50 &  0.031 & 0.037(07)  &  0.003 & 0.029(06)  & 102 &  0.000 & 0.001(10)  & -0.009 & 0.011(06)  \\
 51 &  0.024 & 0.035(06)  &  0.002 & 0.033(05)  & 103 &  0.000 & 0.007(10)  & -0.016 & -0.002(08) \\
 \hline
\end{tabularx}
\caption{The vacuum-subtracted single wavepacket energy density $E_n$ for  $t=16.5,\,24.75$ corresponding to Fig.~\ref{fig:1wp_vs_2wp}. 
Results are shown from MPS simulations and from quantum simulations using {\tt ibm\_marrakesh} after error mitigation (columns labeled QC). 
Several entries are missing data as a result of filtering out qubits with noisy readout.}
\label{tab:results_t_1wp}
\end{table}

\clearpage
\newpage
    \bibliography{bibi}
\end{document}